\def\figsize{9.5cm}

\def\rn{}
\def\nn#1 #2{#2. #1}				% Name with 1 initial
\def\nnn#1 #2 #3{#2. #3. #1}			% Name with 2 initials
\def\nnnn#1 #2 #3 #4{#2. #3. #4 #1}		% Name with 3 initials
\def\nnnnn#1 #2 #3 #4 #5{#2. #3. #4 #5. #1}	% Name with 4 initials
\def\dualand{ and\hbox{ }}				
\def\multiand{, and\hbox{ }}				
\def\rf#1;#2;#3;#4;#5 {{\frenchspacing\par\rn#1, #3 {\bf #4}, #5 (#2). \par}}
\def\rg#1;#2;#3;#4;#5;#6 {{\frenchspacing\par\rn#1, #3 {\bf #4}, #5 (#2). \par}}
\def\rfbook#1;#2;#3;#4;#5 {{\frenchspacing\par\rn#1, {\it #3} (#5, #4, #2).\par}}
\def\rfprep#1;#2;#3 {{\par\frenchspacing\rn#1, #3 (#2).\par}}
\def\rfproc#1;#2;#3;#4;#5;#6 {{\frenchspacing\par\rn#1 #2, in {\it #3}, ed. #4 (#5: #6)\par}}
\def\rfprocp#1;#2;#3;#4;#5;#6;#7 {{\frenchspacing\par\rn#1 #2, in {\it #3}, ed. #4 (#5: #6), p#7\par}}
\def\rg#1;#2;#3;#4;#5;#6 {\par\rn#1 #2, {\it #3}, {\bf #4}, #5 (``#6'') \par}
\def\rf#1;#2;#3;#4;#5 {\par\rn#1, {\it #3}, {\bf #4}, #5 (#2)\par}
\def\rfbook#1;#2;#3;#4;#5 {{\frenchspacing\par\rn#1, {\it #3} (#4: #5, #2)\par}}
\def\rfproc#1;#2;#3;#4;#5;#6 {{\frenchspacing\par\rn#1 #2, in {\it #3}, ed. #4 (#5: #6)\par}}
\def\rfprocp#1;#2;#3;#4;#5;#6;#7 {{\frenchspacing\par\rn#1 #2, in {\it #3}, ed. #4 (#5: #6), p#7\par}}
\def\rfprep#1;#2;#3  {{\par\rn#1, #3, #2\par}}
\def\rfprepp#1;#2;#3 {{\par\rn#1 #2, #3\par}}

\def\K{{\rm K}}

\def\GeV{{\rm GeV}}
\def\TeV{{\rm TeV}}

\def\expec#1{\langle#1\rangle}

\def\etal{{\frenchspacing\it et al.}}
\def\ie{{\frenchspacing\it i.e.}}
\def\eg{{\frenchspacing\it e.g.}}
\def\cf{{\frenchspacing\it c.f.}}
\def\etc{{\frenchspacing\it etc.}}
\def\beq#1{\begin{equation}\label{#1}}
\def\eeq{\end{equation}}
\def\beqa#1{\begin{eqnarray}\label{#1}}
\def\eeqa{\end{eqnarray}}
\def\eq#1{equation~(\ref{#1})}
\def\Eq#1{Equation~(\ref{#1})}
\def\eqn#1{~(\ref{#1})}

\def\fig#1{Figure~\ref{#1}}
\def\Fig#1{Figure~\ref{#1}}

\def\Sec#1{Section~\ref{#1}}

\def\nskip{\hskip-2mm}

\def\spose#1{\hbox to 0pt{#1\hss}}
\def\simlt{\mathrel{\spose{\lower 3pt\hbox{$\mathchar"218$}}
     \raise 2.0pt\hbox{$\mathchar"13C$}}}
\def\simgt{\mathrel{\spose{\lower 3pt\hbox{$\mathchar"218$}}
     \raise 2.0pt\hbox{$\mathchar"13E$}}}
\def\simpropto{\mathrel{\spose{\lower 3pt\hbox{$\mathchar"218$}}
     \raise 2.0pt\hbox{$\propto$}}}

\def\ed{\end{document}}

\documentclass[twocolumn,amsmath,nofootinbib]{revtex4} % For astro-ph
\usepackage{url}
\begin{document}
% Include Rokicki's epsf.sty file for Encapsulated PostScript graphics
\input{epsf.sty}

% FOR APJ:
%\documentstyle[emulateapj,danonecolfloat]{article}
%%\documentstyle[aasms4]{article}
%\def\NoApjSectionMarkInTitle#1{#1.\ }
%%\draft
%\begin{document}
%\twocolumn[%%% Begin front material

%%%%%%%%%%%%%%%%%%%%%%%%%%%%%

%\tighten
%\eqsecnum
%\received{4 August 1988}
%\accepted{23 September 1988}
%\journalid{337}{15 January 1989}
%\articleid{11}{14}

\def\affilmrk#1{$^{#1}$}
\def\affilmk#1#2{$^{#1}$#2;}

% COSMOLOGICAL PARAMETERS:
\def\Otot{\Omega_{\rm tot}}
\def\Ob{\Omega_b}
\def\Oc{\Omega_{\rm cdm}}
\def\Ok{\Omega_k}
\def\Ol{\Omega_\Lambda}
\def\Om{\Omega_m}
\def\Od{\Omega_d}
\def\On{\Omega_\nu}
\def\ob{\omega_b}
\def\oc{\omega_{\rm cdm}}
\def\od{\omega_d}
\def\ok{\omega_k}
\def\ol{\omega_\Lambda}
\def\om{\omega_{\rm m}}
\def\on{\omega_\nu}
\def\fb{f_b}
\def\fn{f_\nu}
\def\ns{{n_s}}
\def\nt{{n_t}}
\def\al{\alpha}
\def\Ot{\Omega_{\rm tot}}
\def\As{A_s}
\def\At{A_t}
\def\Ap{A_p}
\def\dH{Q}
\def\dHt{Q_t}
\def\Qstar{Q_*}
\def\fppmin{f''_{\rm min}}
\def\Ntot{N_{\rm tot}}
\def\Nbar{{\bar N}}
\def\delt{\delta_H}
\def\deltt{\delta_H^T}
\def\nb{n_{\rm b}}
\def\ng{n_\gamma}
\def\nnu{n_\nu}
\def\rhob{\rho_{\rm b}}
\def\rhoc{\rho_{\rm c}}
\def\rhod{\rho_{\rm d}}
\def\rhon{\rho_\nu}
\def\rhol{\rho_\Lambda}
\def\rholgr{\rho_\Lambda^{\rm GR}}
\def\rhog{\rho_\gamma}
\def\rhom{\rho_{\rm m}}
\def\rhok{\rho_{\rm k}}
\def\rhogmk{\rho_{\gamma{\rm mk}}}
\def\rhostar{\rho_*}
\def\astart{a_{\rm start}}
\def\aexit{a_{\rm exit}}
\def\ak{a_{\rm k}}
\def\aend{a_{\rm end}}
\def\areheat{a_{\rm reheat}}
\def\aenter{a_{\rm enter}}
\def\aeq{a_{\rm eq}}
\def\anow{a_0}
\def\abh{a_{\rm bh}}
\def\rhostart{\rho_{\rm start}}
\def\rhoexit{\rho_{\rm exit}}
\def\rhoend{\rho_{\rm end}}
\def\rhoreheat{\rho_{\rm reheat}}
\def\rhoenter{\rho_{\rm enter}}
\def\rhoeq{\rho_{\rm eq}}
\def\rhonow{\rho_0}
\def\Hexit{H_{\rm exit}}
\def\Nbefore{N_{\rm before}}
\def\Nexit{N_{\rm exit}}
\def\Neq{N_{\rm eq}}
\def\Nexit{N_{\rm exit}}
\def\Ngal{N_{\rm gal}}
\def\Nnow{N_0}
\def\zgal{z_{\rm gal}}
\def\Phorizon{P_{\rm horizon}}
\def\Mnu{M_\nu}
\def\xig{\xi_\gamma}
\def\xib{\xi_{\rm b}}
\def\xic{\xi_{\rm c}}
\def\xid{\xi_{\rm d}}
\def\xin{\xi_\nu}
\def\phiend{{\phi_{\rm end}}}
\def\phistop{{\phi_{\rm stop}}}
\def\aend{{a_{\rm end}}}
\def\xend{{x_{\rm end}}}
\def\rhoend{{\rho_{\rm end}}}
\def\rhoreh{{\rho_{\rm reh}}}
\def\dN{\partial_N}
\def\dt{\partial_t}
\def\dphi{\partial_\phi}
\def\mbar{{\bar m}}
\def\mh{{m_h}}
\def\mv{{m_v}}
\def\vphi{{\mbox{\boldmath$\phi$}}}
\def\vPhi{{\mbox{\boldmath$\Phi$}}}
\def\finf{f^{\rm inf}}
\def\fhalo{f^{\rm halo}}
\def\fhat{{\hat f}}
\def\nskip{\hskip-2mm}
\def\Dx{\Delta x}
\def\trig{{\rm trig}\>}
\def\C{{\bf C}}
\def\l{\ell}
\def\erfc{{\rm erfc}\,}
\def\Qmin{{Q_{\rm min}}}
\def\Qmax{{Q_{\rm max}}}
\def\QQQmin{{Q^3_{\rm min}}}
\def\QQQmax{{Q^3_{\rm max}}}
\def\Lmin{{\rhol^{\rm min}}}
\def\Lmax{{\rhol^{\rm max}}}
\def\tento#1{\times 10^{#1}}

\def\p{{\bf p}}
\def\pSRA{\p_{\rm SRA}}
\def\r{{\bf r}}
\def\lnr{R}
\def\lnV{W}
\def\fgal{f_{\rm gal}}
\def\fphi{f_\phi}
\def\Vphi{V_\phi}
\def\Vinf{V_{\rm infl}}
\def\Vcom{V_{\rm c}}
\def\Vreheat{V_{\rm reh}}
\def\Treheat{T_{\rm reh}}
\def\rreheat{\rho_{\rm reh}}
\def\rinf{\rho_{\rm infl}}
\def\Vtherm{V_{\rm therm}}
\def\ttherm{t_{\rm therm}}
\def\atherm{a_{\rm therm}}
\def\Tcmb{T_{\rm CMB}}
\def\Ttherm{T_{\rm therm}}
\def\wvol{w_{\rm vol}}
\def\wgal{w_{\rm gal}}
\def\whalo{w_{\rm halo}}
\def\wN{w_N}
\def\wQ{w_Q}
\def\x{{\bf x}}
\def\rhomeq{{\rhom^{\rm eq}}}
\def\Teq{{T_{\rm eq}}}
\def\Gl{{G_\Lambda}}

%\submitted{Submitted to ApJ July 2 2001, accepted January 15 2002}
%\submitted{\today. To be submitted to ApJ.}
%\submitted{Submitted to ApJL September 16; accepted February 2}

\def\mpl{m_{\rm Pl}}
\def\mp{m_{\rm p}}

\title{What does inflation really predict?}

\author{Max Tegmark}
\address{
Dept. of Physics, Massachusetts Institute of Technology, Cambridge, MA 02139, USA\\
Department of Physics, University of Pennsylvania, Philadelphia, PA 19104, USA
}

\date{\today.}
%\date{Submitted to Phys. Rev. D October 27 2003, revised December 22, accepted January 4.}

%
\begin{abstract}
If the inflaton potential has multiple minima, as may be expected in, \eg, the 
string theory ``landscape'', inflation predicts a probability distribution for 
the cosmological parameters describing spatial 
curvature ($\Otot$), dark energy ($\rho_\Lambda$, $w$, \etc), 
the primordial density fluctuations ($\dH$, $\ns$, $d\ns/d\ln k$, \etc)
and primordial gravitational waves 
%$\delta_H^T$, 
($r$, $\nt$, \etc).
We compute this multivariate probability distribution for various classes of 
single-field slow-roll
models, % , modeling the selection effect by number den
exploring its dependence on the characteristic inflationary energy scales,
the shape of the potential $V$ and 
and the choice of measure underlying the calculation.
We find that unless the characteristic scale $\Delta\phi$ on which $V$ % the inflaton potential $V$
varies happens to be near the Planck scale, the only aspect of $V$ that matters observationally 
is the statistical distribution of its peaks and troughs.
For all energy scales and plausible measures considered, 
we obtain the predictions
$\Otot\approx 1\pm 10^{-5}$, $w=-1$ and $\rhol$ in the observed ballpark but uncomfortably high.
%$0$-$4$ magnitudes larger than observed.
The high energy limit predicts
$\ns\approx 0.96$, $d\ns/d\ln k\approx -0.0006$, 
$r\approx 0.15$ and $\nt\approx -0.02$, consistent with observational data and
indistinguishable from % single-field 
eternal 
$V\propto\phi^2$ inflation.
%Surprisingly, $\dH$ is minimized at the Planck scale and 
%{\it increases} towards lower energies as tiny $\epsilon$-values 
%dominate the statistics, with no energy scale naturally favoring 
%the observed fluctuation level $\dH\sim 10^{-5}$.
The low-energy limit predicts 5 parameters but prefers larger $Q$ and redder $\ns$ than observed.
We discuss the {\it coolness problem}, the {\it smoothness problem} and the {\it pothole paradox},
which severely limit the viable class of models and measures. 
Predictions insensitive to pre-inflationary conditions can arise either from 
eternal inflation attractor behavior or from anthropic 
selection effects
probing merely a tiny non-special part of the initial distribution.
We argue that these two mechanisms are severely challenged by the 
coolness problem and the smoothness problem, respectively.
%Our findings are encouraging for the prospects of detecting an inflationary gravitational wave signature
%with future CMB polarization experiments, with generic single-field models arguably favoring 
%the readily detectable level $r\sim 0.03$.
%Our findings bode well for the search for an inflationary gravitational wave signature,
%since the arguably best motivated single-field models predict a level $r\sim 0.03$
%well within reach of future CMB polarization experiments.
Our findings bode well for detecting an inflationary gravitational wave signature
with future CMB polarization experiments, with the arguably best-motivated single-field models 
favoring the detectable level $r\sim 0.03$.

\end{abstract}

%\keywords{large-scale structure of universe 
%--- galaxies: statistics 
%--- methods: data analysis}
% ]%%% End front material

% PACS, the Physics and Astronomy Classification Scheme
% http://www.aip.org/pacs/pacs03/pacs0390.html
%\pacs{98.80.Es}
%\keywords{cosmic microwave background  -- diffuse radiation}
%%-- radiation mechanisms: thermal and non-thermal
%%-- methods: data analysis}
%]%%% End front material
  
\maketitle

%%%%%%%%%%%%%%%%%%%%%%%%%%%%%%%%%%%%%%%%%%%%%%
%%%%%%%%%%%%%%%%%%%%%%%%%%%%%%%%%%%%%%%%%%%%%%

\setcounter{footnote}{0}

\section{Introduction}
\label{IntroSec}

\begin{table}
\noindent 
{\footnotesize
Table~1: Constraints on fundamental cosmological parameters computed from WMAP+SDSS as in \cite{sdsspars,sdsslyaf}.
Inflation may predict the first 8.
%The CMB temperature $T$ is of course not a parameter computable from fundamental theory, merely a time variable 
%specifying the current epoch. 
% Note that other popular parameters such as $\Omega_b$, $\omega_b$ and $\Omega_\Lambda$ are
%time-dep. But so is $\dH$, since $H$ changes!
We have replaced the conventional parameters $(\Ol,\Ob,\Oc,\On)$ by the
less anthropocentric ones $(\rhol,\xib,\xic,\xin)$ since the latter do not depend on the observing epoch. 
\tiny\begin{center}
\begin{tabular}{|l|ll|}
\hline			
p			&Meaning							&Measured value\nskip\\
\hline
% Densities:
$\Otot       	$	&Spatial curvature						&$1.01\pm 0.02$\\ % sdsspars
$\rhol		$	&Dark energy density						&$(9.3\pm 2.5)\tento{-124}\mpl^4$\\ % sdsspars (anthroneutrino), not quited by sdsslyaf
$w		$	&Dark energy equation of state					&$-1\pm 0.1$\\ % sdsspower, also sdsslyaf
$\dH       	$	&Scalar fluctuation amplitude $\delta_H$ on horizon		&$(2.0\pm 0.2)\times 10^{-5}$\\ 
$\ns	        $	&Scalar spectral index					        &$0.98\pm 0.02$\\ % sdsslyaf.	sdsspower: 0.98\pm 0.03
$\alpha 	$       &Running of spectral index $d\ns/d\ln k$		        &$|\alpha|\simlt 0.01$\\ % sdsslyaf
$r           	$	&Tensor-to-scalar ratio	$(\dHt/\dH)^2$      	        	&$\simlt 0.36$\\ % sdsspars <0.47, sdsslyaf <0.36
$\nt           	$ 	&Tensor spectral index						&Unconstrained\\
\hline
$\xib		$	&Baryon mass per photon $\rhob/\ng=\mp\eta$			&$(0.60\pm 0.03)$ eV\\
$\xic		$	&CDM mass per photon $\rhoc/\ng$				&$(3.3\pm 0.2)$ eV\\
$\xin		$	&Neutrino mass per photon $\rhon/\ng={3\over 11}\sum m_{\nu_i}$	&$<0.11$ eV\\ % sdsspars 0.5 eV, sdsslyaf (3/11)*0.42eV
% Anthroneutrono notes.txt:
% xib27       0.046874000    0.049097000    0.052066000
% xic27       0.249420000    0.269030000    0.289690000
% xin27       0.004483900    0.013385000    0.026653000
% 1eV/Eplanck ~ 8.1897d-29
% set eV = 0.081897 # 1eV*10e27 = 	
% set p={0.046874000    0.049097000    0.052066000}
% set p={0.249420000    0.269030000    0.289690000}
% echo $(p[1]/eV) +- $(0.5*(p[2]-p[0])/eV)
% xib = 0.5994968414 +- 0.03169835359
% xic = 3.28497982 +- 0.2458575964
%\hline
%$T		$	&CMB temperature (merely a time variable)			&$2.725$K (today)\\
\hline   
\end{tabular}
\end{center}     
} 
%\vskip-0.5cm
\end{table}

\begin{figure} 
\vskip-4.4cm
\centerline{\epsfxsize=\figsize\epsffile{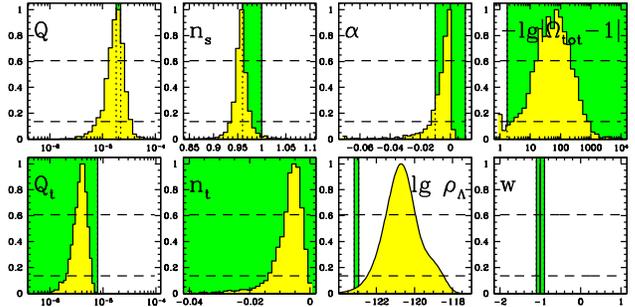}}
\vskip-1cm
\caption[1]{\label{1DconditionedFig}\footnotesize%
Yellow/light grey cosmological parameter distributions show inflationary predictions 
for our \S 5 example with characteristic $\phi$-scale $\mh\approx 2\mpl$,
$V$-scale $\mv=0.002\mpl$ and uniform measure over initial $\phi$-values.
Green/dark grey regions show observational constraints ($1\sigma$) from Table 1 and \cite{sdsspars,sdsslyaf}.
$\rhol$ is in Planck units.
}
\end{figure}

The spectacular recent progress in observational cosmology has 
produced measurements of about a dozen cosmological parameters
\cite{Spergel03,sdsspars,sdsslyaf} as
summarized in Table 1 and \fig{1DconditionedFig} (vertical bands). 
% Otot, eta, xi, xin, rhol, w, As, ns, al, r, nt
This observational progress is in stark contrast to the theoretical 
situation: we still lack a firm {\it a priori} prediction for any one of them.
Inflation is promising in potentially being able to explain a strikingly large
fraction of these numbers: 8 of the 11, including (as we will argue) $\rhol$ and $w$, 
the only holdouts being the baryon parameter 
$\xib$ (awaiting an understanding of baryogenesis),
the dark matter parameter $\xic$ 
(perhaps awaiting an understanding of supersymmetry parameters governing 
dark matter freezeout) and
the neutrino parameter $\xin$ (awaiting an understanding of the origin of neutrino masses).

So what does inflation predict for the parameter vector 
\beq{pEq}
\p\equiv(\Otot,\rhol,w,\dH,\ns,\al,r,\nt)?
\eeq
The rhetorical answer is that the answer is determined by the inflaton 
potential $V(\vphi)$, where $\vphi$ is a vector of one or more scalar fields, 
and that the only fairly robust predictions are $\Otot\approx 1$ and the slow-roll 
consistency relation $r=-8\nt$ --- with known exceptions even to 
these, {\eg} \cite{LindeOpen98,LindeClosed04,LindeFastroll01,AlbrechtOpen04}.
Since a bewilderingly large number of potentials have been studied in the literature
(see \cite{LindeBook,DodelKinneyKolb97,LythRiotto99,LiddleLythBook,Peiris03,Kinney03,LiddleSmith03,Wands03}
for recent reviews), producing predictions in pretty much ever corner of the currently allowed
parameter space, it is easy to come away with the impression that inflation 
is not very predictive and hence difficult to test or rule out.

\begin{figure} 
%\vskip\smtopskip
\centerline{\epsfxsize=\figsize\epsffile{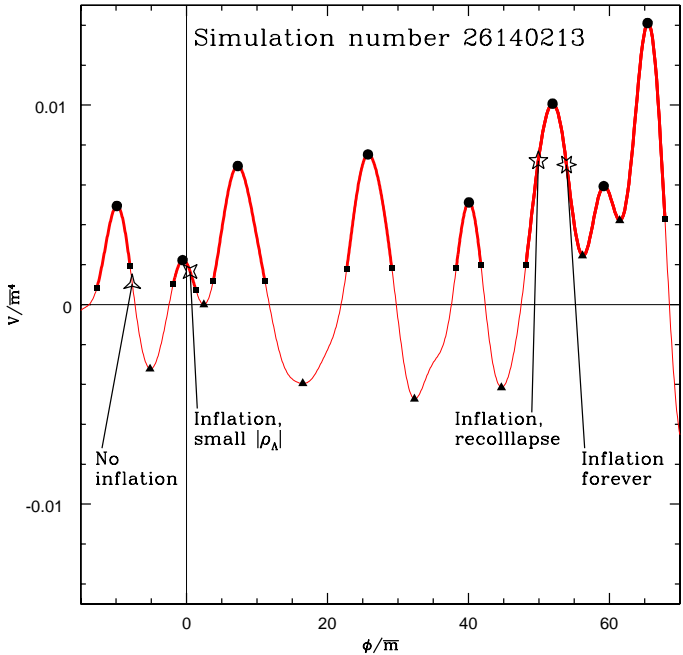}}
%\vskip\smbotskip
\caption[1]{\label{phiExampleFig}\footnotesize%
A small segment of the inflaton potential $V(\phi)$ for simulation number 26140213 from \S 5,
with 
%$\gamma=0$ and 
vertical and horizontal energy scales $\mh=3\mbar$ and $\mh=0.007\mbar$,
respectively ($\mbar=\mpl/\sqrt{8\pi}$).
A half-basin of attraction stretches from a local maximum (circle) to a local minimum (triangle).
The thicker curve indicates regions where the slow-roll approximation is valid, with squares indicating
inflation endpoints. Many starting points $\phi$ fail to produce galaxies, 
either because the slow-roll approximation in invalid and there is 
no inflation (three-pointed star), because inflation is rapidly followed by 
$\rhol<0$ recollapse (five-pointed star) or because inflation never ends (six-pointed star).
Only a small fraction of starting points (like the four-pointed star) give inflation followed
by a vacuum density $|\rhol|$ near zero.
}
\end{figure}

However, this pessimistic conclusion rests on a rather dubious premise: that the way inflation 
occurred in our region of space is the only way that inflation occurred anywhere. 
This is the antithesis of the ``landscape'' paradigm now emerging from string theory, with an 
extremely complicated potential with perhaps $10^{300}$ different minima 
\cite{Bousso00,Feng00,KKLT03,Susskind03,AshikDouglas04,KKLMMT03,WKV04,Denef04,GK04,Freivogel04,Conlon04,DeWolfe04,Nima05,Acharya05},
{\cf} \cite{BanksDineGorbatov04,Banks04}. 

Whatever the correct fundamental theory may turn out to be, 
one can easily imagine it giving rise to an inflaton potential $V(\vphi)$ with multiple minima,
as in the example shown in \fig{phiExampleFig}.
As long as $\vphi$ takes on a range of values early on\footnote{Having $\vphi$ take on a range of values early on is easy to arrange.
The same quantum fluctuation process responsible for generating our observed density fluctuations 
gives rise to the familiar diffusion process that can drive $\vphi$ uphill as well as downhill
\cite{LindeDiffusion86}, 
effectively populating all $\vphi$-values in a quantum superposition. Decoherence \cite{ZehBook} then insures
that we can for all practical purposes treat $\vphi$ as having different classical values in different spatial regions 
\cite{PolarskiStarobinsky96,KieferPolarski98}.
Alternatively, an early hot state with energy in excess of typical $V$-values would naturally populate
all $\vphi$-values.}, 
different post-inflationary domains will correspond
to $\vphi$ having rolled down into different minima, and so 
the parameter values we measure will depend on which domain we inhabit.
In other words, what is traditionally referred to as {\it the} inflaton potential is  
merely the basin of attraction around our particular local minimum $\vphi_0$.
Our observed cosmological constant will be $\rhol=V(\vphi_0)$, 
our gravity wave amplitude will be given by $V(\vphi_1)$ at the point $\vphi_1$ about
55 $e$-foldings before our inflation ended \cite{LiddleLeach03,HuiDodel03}, 
$\ns$ will be determined by the first and second derivatives of $\ln V$ at at $\vphi_1$, and so on.
Even if the potential has only one minimum, parameters such as $\ns$ will depend
on the direction from which one rolls down to this minimum 
(from the left or from the right for single-field inflation like in \fig{phiExampleFig}, or from a continuum of directions
for multi-field inflation where $\vphi$ is a vector). 

If one or more explicit effective potentials $V(\vphi)$ emerge from string theory or some other
fundamental theory \cite{Hossain04}, then it is clearly worthwhile computing or estimating the 
corresponding 8-dimensional parameter probability distribution $f_p(\p)$ (\fig{1DconditionedFig} shows an example), 
since this will provide a powerful 
observational test of the theory.
The purpose of the present paper is to lay the groundwork for such a calculation 
in the context of classic slow-roll inflation, computing 
$f_p(\p)$ for various classes of randomly generated potentials and 
exploring its dependence both on the potential $V(\vphi)$
% characteristic inflationary energy scale
and on the choice of measure underlying the calculation.
We will see that a number of general 
% and at times disturbing 
conclusions can be drawn rather independently of the detailed nature of the potential $V(\vphi)$, depending mainly on
energy scales.
For related discussions of such issues in the context in inflationary flow equations, 
see \cite{HoffmanTurner01,Kinney02,Liddle03}.

We find that the issue of what measure to use when sampling 4D spacetime is no less important than 
the issue of what potential $V(\vphi)$ to use.  
This measure problem cannot be brushed aside as irrelevant philosophy, since the measure dramatically affects the
testable prediction for $f_p(\p)$, and we will argue that some plausible-sounding and oft-discussed 
measures like synchronous volume-weighting are already ruled out by observation.
There is some correct measure that nature 
subscribes to, and we need to figure out which one it is, just as was successfully done in 
the past for the measures allowing us to compute probabilities in statistical mechanics and quantum physics.

The rest of this paper is organized as follows.
\Sec{Vsec} briefly summarizes how the potential $V(\vphi)$ affects what we observe,
emphasizing that modulo a slight caveat in the multidimensional case, 
all inflationally observables (including the eight inflation-related parameters in Table 1) 
are determined by the cosmic density history $\rho(a)$ alone, independent of the details of $V(\vphi)$.
Sections~\ref{MeasureSec} and~\ref{OrderingSolutionSec}
discuss how the measure affects the parameter probability distribution $f_p(\p)$,
\ie, what we should expect to observe given $V(\vphi)$, classifying various plausible and ruled-out measure candidates.
\Sec{MonteSec} presents a suite of numerical experiments exploring the effect of inflationary energy scales,
and \Sec{AnthroSec} shows how to compute $f_p(\p)$ from these results when including selection effects.
\Sec{AnalyticSec} complements our numerical exploration with a suite of analytic calculations,
and \Sec{ConcSec} summarizes our conclusions. Appendix A and B give technical details of how we
compute $\p$ from $V$.

%\notyet
% Comment on 1D and nD?

\section{How the potential affects what we observe}
\label{Vsec}

Given the measure subtleties discussed in the next section, it is crucial to be clear on what question we are asking.
In this section, we address the following question:
If we live in a part of spacetime where the inflaton has evolved as $\vphi(a)$ 
and the density has evolved as $\rho(a)$, then what do we observe?
This is mainly a review of standard material, organized to clarify how
in many cases, all observables can be calculated directly from $\rho(a)$ alone, 
without needing to know $\vphi(a)$ or the inflaton potential.

We wish to make predictions for four basic observables:
\begin{enumerate}
\item The curvature of space given by $\Otot$.
\item The dark energy density as a function of cosmic scale factor, $\rhol(a)$.
\item The primordial power spectrum of gravitational waves $\deltt(k)$.
\item The primordial power spectrum of adiabatic density fluctuations $\delt(k)$.
\end{enumerate}
It is popular to approximate the three cosmological functions 
$\rhol(a)$, $\deltt(k)^2$ and $\delt(k)^2$ 
by linear and quadratic functions in log-log space, parametrized by parameters
2-8 from Table 1:
\beqa{ParametrizationEq}
\rhol(a) 	&=& \rhol\times (a/a_0)^{-3(1+w)},\\
\deltt(k)^2 	&=& \dH^2\times r (k/H)^\nt,\\
\delt(k)^2 	&=& \dH^2\times (k/H)^{\ns-1+\alpha\ln(k/H)/2},   % power ~ k**(ns - 1 + al*log(kk/0.05)/2)
\eeqa 
where $H\equiv\dot a/a$ is the current Hubble parameter, so that the wavenumber $k=H$ corresponds to a fluctuation on the
current horizon scale\footnote{The horizon is $c/H$ --- throughout this paper, we use standard particle
physics units where $\hbar=c=k_b=1$, so the Planck mass is simply $\mpl=G^{-1/2}$.}\footnote{For simplicity, we have defined
the normalizations ($Q$, $rQ$) and tilts ($\ns,\nt$) on the horizon scale. 
A popular alternative convention \cite{cmbfast,Spergel03,sdsspars,sdsslyaf}
is to specify them on the comoving scale corresponding to $k=0.05/$Mpc at the present 
epoch --- this corresponds to shifting the emphasis from, say,
scales that are $N=55$ e-foldings before the end of inflation to $N\approx 50$, making only a minor difference in $\p$ that
is irrelevant for the purposes of this paper.}.
Here and throughout, we will assume the standard inflationary fluctuation-generation mechanism which produces adiabatic fluctuations --- for a 
review of alternatives, see \cite{LindeBook,DodelKinneyKolb97,LythRiotto99,LiddleLythBook,Peiris03,Kinney03,LiddleSmith03,Wands03,ZaldaGhost}.

\begin{figure} 
%\vskip\smtopskip
\centerline{\epsfxsize=\figsize\epsffile{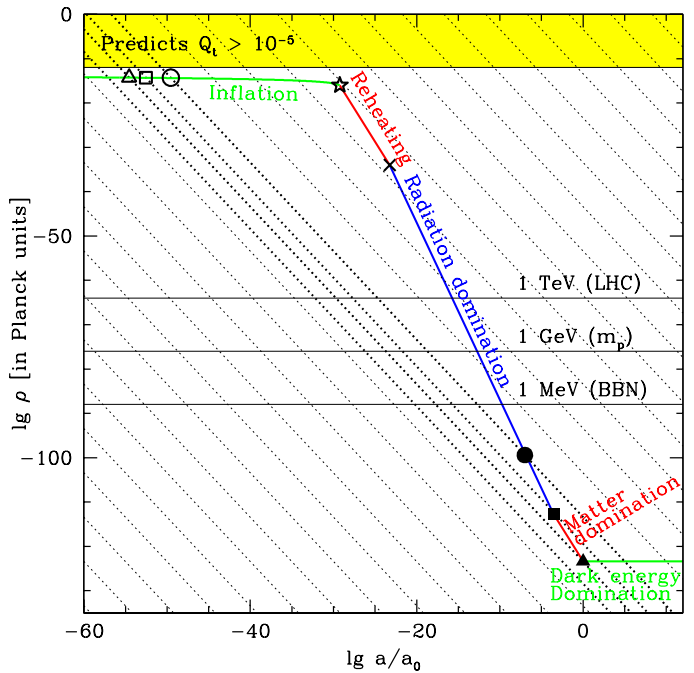}}
%\vskip\smbotskip
\caption[1]{\label{arhoFig}\footnotesize%
The spatial curvature parameter $\Otot$, the primordial gravitational wave power spectrum $\deltt(k)$ and the dark energy 
function $\rhol(a)$ are all determined by the single curve $\rho(a)$ above, as is the primordial
density fluctuation spectrum $\delt(k)$ in many cases. 
Dotted diagonals of slope $-2$ are lines of constant comoving horizon size $a/H^{-1}=\dot a$, 
governing when fluctuations exit end reenter the horizon.
Inflation $(\ddot a>0)$ is when the cosmic density history $\rho(a)$ has a shallower slope 
than these diagonals, $d\ln\rho/d\ln a>-2$, and this logarithmic slope normally becomes $-3$ during reheating,
$-4$ during radiation domination, $-3$ during matter domination and $>-2$ now during dark energy domination.
When two points lie on the same diagonal (like the two
triangles), it means that that the horizon volume at the two epochs is the same comoving spatial region. 
$\Otot$ is constant on these diagonals, approaching unity towards the upper right,
and unless the $\rho(a)$ curve crosses the leftmost heavy diagonal,
we should observe curvature $|\Otot-1|>10^{-5}$.
The dark energy function $\rhol(a)$ is simply the curve $\rho(a)$ at late times minus the matter contribution.
The primordial gravitational wave power spectrum is simply the curve $\rho(a)$ at early
times, rescaled vertically (by a factor of two since $\deltt(k)\propto\rho^{1/2}$) 
and horizontally (mapping $a$ into $k$ matching horizon exit and reentry with the diagonal lines,
as with the pairs of triangles, squares and circles).
For single-field inflation and many multi-field cases,
the primordial density fluctuation spectrum is determined from the curve $\rho(a)$ and its 
derivative by $\delt(k)\propto\rho^{1/2} (-d\ln\rho/d\ln a)^{-1/2}$.
}
\end{figure}

\Fig{arhoFig} is an attempt to capture all the inflationary observables in one plot.
It illustrates how the late-time behavior of $\rho(a)$ encodes the dark energy function $\rhol(a)$
whereas its early-time behavior determines the curvature parameter $\Otot$, the 
gravitational wave power spectrum $\deltt(k)$ and,
in many cases, the power spectrum of density fluctuation $\delt(k)$.
Key epochs in \fig{arhoFig} are 
$\astart$ (when inflation began; $\astart=0$ if there was no beginning),
$\aexit$ (open triangle, when our current Hubble volume left the horizon),
$\ak$ (when a fluctuation of wavenumber $k$ leaves the horizon; open square or circle, say),
$\aend$ (star, when inflation ended),
$\areheat$ (cross, when reheating ended),
$\aeq$ (filled square, matter-radiation equality) and
$\anow$ (filled triangle, now).
\Fig{arhoFig}
also provides a simple graphical way of reading off bounds 
on the number of e-foldings $N$ between the creation of our observed 
horizon-scale fluctuations (open triangle) and the end of inflation 
(star) --- most popular models give $N\sim 55\pm 5$ \cite{LiddleLeach03,HuiDodel03}.

For comparison, the ekpyrotic model \cite{Ekpyrotic01,Ekpyrotic03} has  
fluctuations leave the horizon during a contracting phase when we move 
up to the left in \Fig{arhoFig} with steeper slope than $-2$.

%For the number of e-foldings $N=\ln(\aend/aexit)$ between the creation of our observed 
%horizon-scale fluctuations (open triangle) and the end of inflation (star), most popular models give $N\sim 55\pm 5$ .
%We will now review the calculation of these various quantities.

%Given the cosmic density history $\rho(a)$, the 
%the Friedman equation
%\beq{FriedmanEq}
%H^2 = {8\pi\over 3\mpl^2}\rho = {1\over 3\mbar^2}\rho,
%\eeq
%where the Hubble parameter $H\equiv d\ln a/dt=\dot a/a$ and we have defined the 

%\beq{OtotEq3}
%|\Otot-1|_{\rm observed}\sim  10^{-5} + e^{-2\Nbefore}.
%\eeq

%\beq{rholEq}
%\rhol=\lim_{a\to\infty}\rho(a)
%\eeq

%\beq{dHtEq2}
%\deltt(k)\approx \sqrt{8\rho(\ak)\over 75\pi^2\mbar^4}.
%\eeq
%\beq{nDdHeq2}
%\delt(k) \ge \sqrt{\rho(\ak)\over 150\pi^2\mbar^4\epsilon},
%\eeq

We will now summarize the calculation of the cosmological functions 
$\rhol(a)$, $\deltt(k)$, $\deltt(k)$ and the eight inflationary parameters from Table 1.
As detailed in Appendix A,
our three cosmological functions and the Hubble parameter are given in terms of $\rho(a)$ by 
\beqa{funcSummaryEq}
\rhol(a)	&=&\rho(a)-\rhogmk(a),\\
\deltt(k)	&=&\sqrt{8\rho(\ak)\over 75\pi^2\mbar^4},\label{delttEq}\\
\delt(k)	&\ge&\sqrt{\rho(\ak)\over 75\pi^2\mbar^4(\ln\rho)'},\label{deltEq}\\
H(a)		&\equiv&{d\ln a\over dt}= \sqrt{8\pi\rho(a)\over 3\mpl^2} = {\rho(a)^{1/2}\over\sqrt{3}\mbar},
\eeqa
where $'\equiv d/dN=-d/d\ln a$,
$\rhogmk(a)$ is the density contribution from photons, matter and curvature,
the mode exit scale is
\beq{akEq}
a_k \approx {\aexit\over H(\aexit)} k = {\sqrt{3}\mbar\aexit\over\rho(\aexit)^{1/2}} k,
\eeq
$\aexit$ is the solution to the equation $\aexit^2\rho(\aexit)=\anow^2\rho(\anow)$, and 
the {\it reduced Planck mass}
\beq{mbarDefEq}
\mbar\equiv{\mpl\over\sqrt{8\pi}}\approx {\mpl\over 5}.
\eeq
is defined so as to minimize the number of $8\pi$-factors in this paper. 
The corresponding eight observable cosmological parameters in Table 1 are given in terms of $\rho(a)$ by (see Appendix A)
\beqa{parSummaryEq}
%|\Otot-1|	&\sim&10^{-5} + e^{-2\Nbefore},\\
|\Otot-1|	&\sim&10^{-5} + \left[{\aexit\rho(\aexit)\over\astart\rho(\astart)}\right]^2,\label{OtotSummaryEq}\\
\rhol		&=&\lim_{a\to\infty}\rho(a),\\
w		&=&-1,\\
\dH		&\ge&\sqrt{\rho(\aexit)\over 75\pi^2\mbar^4(\ln\rho)'},\label{SRA_Qeq}\\
\ns		&=&1-(\ln\rho)'+{(\ln\rho)''\over(\ln\rho)'}\\
\al		&=&(\ln\rho)'+{(\ln\rho)''^2\over(\ln\rho)'^2}-{(\ln\rho)'''\over(\ln\rho)'},\\
r		&\le&8(\ln\rho)',\label{rEq}\\
\nt		&=&-(\ln\rho)',\label{ntSummaryEq}
\eeqa
where $'\equiv {d\over d\ln a}$.
The usual slow-roll parameters are
\beqa{epsilonEq2}
\epsilon	&=&{1\over 2}(\ln\rho)',\\
\eta		&=&(\ln\rho)' + {(\ln\rho)''\over 2(\ln\rho)'},\label{etaEq2}\\
%zzz \xi_2		&=&...,\label{xi2Eq2}
\eeqa
in terms of which we have the standard relations 
\beqa{SRA_nsEq}
\ns&=&1-6\epsilon+2\eta,\\
\nt&=&-2\epsilon,\label{SRA_ntEq}\\
r&\le&16\epsilon,\label{SRA_rEq}\\
\al&=&16\epsilon\eta-24\epsilon^2-2\xi_2.\label{SRA_alEq}
\eeqa
%$\ns=1-6\epsilon+2\eta$, 
%$\nt=-2\epsilon$, 
%$r\le 16\epsilon$ and
%$\al=16\epsilon\eta-24\epsilon^2-2\xi_2$.
%for the single-field case.
As detailed in Appendix A, the inequalities\eqn{SRA_Qeq},\eqn{rEq} and\eqn{SRA_rEq} become equalities for 
single-field inflation and also to good approximation for most multi-field models.
Similarly, the expressions for $\ns$ and $\al$ are exact for 
single-field models and usually good approximations for multi-field models. 

In addition to determining these numerical parameters, the $\rho(a)$-curve in \fig{arhoFig} readily allows
qualitative conclusions to be read off: 
To solve the horizon problem, the curve must cross the dotted diagonal that runs through the
filled triangle, and 
\eq{OtotSummaryEq} shows that to solve the flatness problem, it must also cross the diagonal line lying about 
a factor of 10 further to the left.

\section{How the measure affects what we observe}
\label{MeasureSec}

In this Section, we discuss how to confront an inflation model with observational data, \ie, 
how to compute the theoretically predicted probability distribution $f_p(\p)$ for the cosmological parameters
once $V(\vphi)$ is known.

\subsection{Overview of the measure problem}

Inflation helps create an interestingly complex spacetime, one or more parts of which 
contain observers like us.
Loosely speaking, the probability distribution $f_p(\p)$ for the cosmological parameters
is simply an eight-dimensional histogram, reflecting the distribution of parameters measured
by the different observers. Let us now make this more rigorous.
To make inflation a fully specified and testable theory from which $f_p(\p)$ can be computed,
three things need to be specified: 
\begin{enumerate}
\item the choice of {\it reference objects} (say points, protons, planets or observers),
\item the {\it order} in which to count them (this matters when there are infinitely many),
\item the {\it initial conditions} (irrelevant in some cases).
\end{enumerate}
%For reasons that will soon become clear, 
We will refer to the specification of these three things as a choice of measure.
We will first summarize these three issues briefly, then revisit them in greater detail.

\subsubsection{Reference objects}

The first of the three choices is the most obvious and straightforward to deal with,
so we will not dwell on it here --- see \cite{conditionalization,BostromBook} for a detailed discussion.
The resulting parameter probability distribution $f_p(\p)$ depends strongly on whether the reference objects
are points, protons, dark matter halos, galaxies, stars or 
self-aware observers (whatever that means), so the 
key point is simply to be explicit about what reference objects are used and to quantify the sensitivity of
the results to this choice --- we will explore various choices below,
in \Sec{AnthroSec}.
Mathematically, this is simply standard use of conditional probabilities:
we compute the probability distribution for the parameters $\p$ measured from a spacetime point given
various constraints, say that it is the location of a galaxy.\footnote{Discussions of selection effects often turn heated when somebody mentions the
``A-word'', {\it anthropic}. The author feels that discussions of the so-called anthropic 
principle \cite{Carter74,BarrowTipler,Balashov,toe} have generated more heat than light, largely 
because of a preponderance of different and mutually
incompatible definitions and interpretations of what it means.
The author is not aware of anybody disagreeing with 
what might be termed MAP, the ``minimalistic anthropic principle'':
{\it When testing fundamental theories with observational data, 
ignoring selection effects can give incorrect conclusions.} 
This is all we subscribe to in the present paper.
We wish to test any fundamental theory that predicts $V(\vphi)$ by calculating its 
cosmological parameter predictions $f_p(\p)$, and including selection effects is clearly not optional.
The question of precisely which selection effects to use (\ie, what reference objects to condition on) is difficult
and not settled. The appropriate response to this is clearly not to 
% to start foaming at the mouth and 
give up and ignore selection effects altogether, but 
rather to explore a range of options and quantify the extent to which the choice affects the results.
}

\begin{figure} 
%\vskip\smtopskip
\centerline{\epsfxsize=\figsize\epsffile{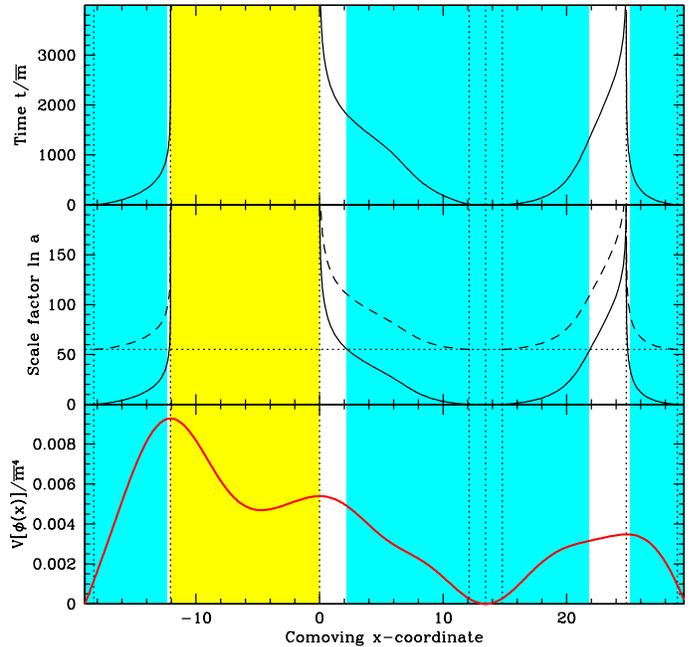}}
%\vskip\smbotskip
\caption[1]{\label{OrderingFig}\footnotesize%
Spacetime foliation examples.
Solid curves (top and middle panels)
show infinite hypersurfaces corresponding to the end of inflation,
starting at $t=0$, $a=1$ with an inflaton field $\phi(\r)=\varepsilon x$ 
with $\varepsilon$ so small that spatial gradients are negligible.
Dashed curve (middle panel) corresponds to the present density.
The four shaded regions of comoving space fail to produce any Hubble volumes 
resembling ours, either because inflation never ends (yellow/light grey)
or because inflation never starts or lasts less than 55 e-foldings 
(cyan/grey). 
Lower panel shows the inflaton potential used, with 
vertical dotted lines showing maxima, minima and 
inflation endpoints.
}
\end{figure}

\subsubsection{Ordering}

\Fig{OrderingFig} illustrates 
what we will call the {\it ordering problem}.
We have chosen our reference objects (protons, say), and 
need to locate all of them in the spacetime manifold.
Inflation creates photons and other particles on spatial hypersurfaces corresponding to the end of inflation,
illustrated by the U-shaped curves in the figure. 
As detailed in \Sec{AnalyticSec}, each of these 
hypersurfaces typically has infinite physical volume, producing infinitely many reference objects. Such a hypersurface
corresponds to a particular basin of attraction in $V(\vphi)$, \ie, everywhere in this comoving volume, 
the inflaton field $\vphi$ rolls down to the same local minimum $\vphi_0$.
We will usually refer to such spatial domains as 
thermalized regions, pocket universes \cite{Guth00,Guth00b,Guth04} or ``pockets'' for brevity --- other 
terms
% names 
for them include bubble universes
% O-regions \cite{GarrigaVilenkin01} - NOT
and Level I multiverses \cite{multiverse}.\footnote{Note that there is not a one-to-one correspondence
between basins and pockets. Each pocket corresponds to exactly one basin, but a basin can correspond to 
multiple pockets, either because of initial conditions (in \fig{OrderingFig}, imagine, say, 
an initial inflaton field $\phi(\r)=\phi_{\rm peak} + \epsilon\sin x$ that is draped over the maximum
at $\phi_{\rm peak}$ %$\approx 0$ 
multiple times) 
or because of quantum diffusion (which can drape the inflaton field repeatedly over the maximum 
even starting with the simple initial condition $\phi(\r)=\varepsilon x$ where the field crosses the
maximum only once) \cite{LindeDiffusion86}).
}
If we order the reference objects 1, 2, 3, ... and let $\p_i$ denote the parameters observed from the $i^{th}$ object, then
we can compute the probability distribution as
\beq{OrderingEq}
f_p(\p)\equiv \lim_{n\to\infty} {1\over n}\sum_{i=1}^n \delta(\p-\p_i).
\eeq
The crux is that when there are infinitely many objects \cite{Guth00,Guth00b,Guth04}, the ordering 
affects the answer! 
For example, suppose we order the objects by increasing physical distance from some given point $\r$
at fixed time (defined as fixed cosmic density).
This may sound like an innocent and reasonable algorithm, since it is equivalent to the familiar procedure of 
averaging in a spherical volume and then letting the sphere radius approach infinity.
However, since the pockets (thermalized regions) are infinite, this ordering will only include objects within the same
pocket that contains the reference point $\r$, giving zero statistical weight to the reference objects in
all other pockets.

This illustrates a difficult aspect of the ordering problem: how to decide the relative order of 
reference objects in different pockets.
This is equivalent to assigning statistical weights to
the different pockets once the ordering and hence the $\p$-distribution has been determined within each pocket.
As detailed below, many attempts to solve the ordering problem have implicitly 
involved grouping the reference objects by 
the value of some physical parameter at their spacetime location
\cite{LindeMezhlumian93,LindeLindeMezhlumian94,Garcia94,Garcia95,LindeLindeMezhlumian,Vilenkin95,WinitzkiVilenkin96,LindeMezhlumian96,LindeLindeMezhlumian96,VilenkinWinitzki97}, 
say by 
% increasing 
proper time $t$ 
(\fig{OrderingFig}, top panel), 
% increasing 
scale factor $a$ (\fig{OrderingFig}, middle panel) or 
% decreasing 
density $\rho$ 
(roughly the black curves in top two panels). 
%, which is in some case equivalent to ordering them by the time, scale factor or density at which they were formed. 
%For the time example, say, 
%this is equivalent to foliating spacetime into three-dimensional constant-time spatial hypersurfaces, averaging over each 
%hypersurface and (hopefully) finding that the result converges to some well-defined distribution $f_p(\p)$ as $t\to\infty$.

In \Sec{OrderingSolutionSec}, we will review and discuss various ordering choices and their implications. 
This is a severe problem in the sense that the ordering strongly affects the testable prediction $f_p(\p)$, yet 
we currently lack a compelling argument leading to a unique choice.
As detailed below, it even affects the sense in which we should think of inflation as eternal.
This embarrassing problem is clearly one of physics rather than philosophy, since its resolution affects the
testable inflationary predictions for $f_p(\p)$.
On an optimistic note, the measure problem (how to compute probabilities) plagued both both statistical mechanics 
and quantum physics early on, so there is real hope that inflation too can overcome its birth pains and become
a testable theory whose probability predictions are unique.

%Be careful not to steal the thunder from Aguirre \& Tegmark (2004).

%selection effects can obviously be important.
%temperature distribution:
%for a point, peaked at cold IGM.
%for a proton, more weighted towards stars.
%for an observer, much narrower range.

\subsubsection{Initial conditions and predictions}

Let us cast the above issues in mathematical form, including the initial conditions $\vPhi$ which are the inflaton field 
and its first derivative early on:
\beq{PhiEq}
\vPhi\equiv\left(\begin{tabular}{c}
$\vphi$\\
$\dot\vphi$\\
\end{tabular}\right).
\eeq
Given any particular initial conditions $\vPhi$ and inflaton potential $V(\vphi)$,
the standard inflation formalism allows us to compute the probability distribution $f_p(\p;\vPhi,V)$ 
for the parameters $\p$.
If quantum diffusion is negligible, then the parameters are uniquely determined by $\vPhi$ and $V$ and we have simply
\beq{fpApproxEq}
f_p(\p;\vPhi,V) \approx \delta[\p-\p(\vPhi,V)],
\eeq
where $\p(\vPhi,V)$ is computed as in \Sec{Vsec} or by going beyond the slow-roll approximation if 
necessary \cite{LiddleLythBook}.
When quantum diffusion is important, a given starting point $\vPhi$ no longer gives a unique classical time-evolution 
$\vphi(t)$ and so the probability distribution $f_p(\p;\vPhi,V)$ widens --- for this case, $f_p(\p;\vPhi,V)$ can be
evaluated either by solving a stochastic ordinary differential equation or, equivalently, 
by integrating a Fokker-Planck equation 
\cite{LindeBook,Vilenkin83,Starobinsky84,Starobinsky86,Goncharov86,SalopekBond91,LindeLindeMezhlumian94}.
As will be discussed in 
\Sec{OrderingSolutionSec}, there are even some choices of measure 
% weighting and ordering: say volume weighting + time ordering,
% (so reference object, ordering) = (cubic meter of space, time ordering)
where attractor dynamics 
\cite{LindeBook,Vilenkin83,Starobinsky84,Starobinsky86,Goncharov86,SalopekBond91,LindeLindeMezhlumian94}
causes
$f_p(\p;\vPhi,V)$ to be completely independent of the initial conditions $\vPhi$.

Quite generally, we can now write the theoretical prediction for the parameter probability distribution as 
\beq{fpEq}
f_p(\p) = w(\p)\expec{\expec{f_p(\p;\vPhi,V)}_\vPhi}_V,
\eeq
where $w(\p)$ is a weight function reflecting
the choice and ordering of reference objects. 
Here $\expec{~}_\vPhi$ denotes averaging over initial conditions $\vPhi$, 
to cover cases where the theory predicts a distribution of initial conditions rather than some particular
initial conditions. Analogously, $\expec{~}_V$ denotes averaging over inflaton potentials $V$ 
to cover cases where this is necessary --- on could, for instance, envision theories
predicting a quantum superposition of many different effective inflaton potentials corresponding to different 
false vacua of the underlying theory.

For notational convenience, we augment our parameter vector from \eq{pEq} to 
\beq{pEq2}
\p\equiv(\Otot,\rhol,w,\dH,\ns,\al,r,\nt,\Ntot),
\eeq
making it include also $\Ntot$, the total number of e-foldings during inflation. 
Although, unlike the other parameters, $\Ntot$ is not directly observable,
it enters in many interesting choices of the weight function $w(\p)$. 
% that involve the volume expansion factor $e^{3\Ntot}$.
In \Sec{AnthroSec} we will explore a variety of weight functions $w(\p)$
including factors such as $e^{3\Ntot}$ (weighting by the thermalized volume produced),
$\theta(\Ntot-55)$ (a Heaviside step function giving weight only to regions inflating 
by at least 55 e-foldings, a requirement linked to galaxy formation), 
$\fhalo(Q,\rhol,\Otot)$ (the fraction of protons ending up in 
dark matter halos) and $f_Q(Q)$ (a factor related to galaxy formation and planetary stability).

%As discussed in \Sec{Vsec}, it is occasionally useful to think of $\p$ as 
%simply encoding the cosmic expansion history $\rho(a)$.

We will find that the choice of measure makes a dramatic difference, and leads to one of 
three qualitatively different situations for the 
theoretical prediction $f_p(\p)$:
\begin{enumerate}
\item Predictions independent of initial conditions because of attractor dynamics.
\item Predictions independent of initial conditions because selection effects in $w(\p)$ probe
      only tiny non-special part of initial distribution.
\item Predictions dependent on initial conditions, \ie, on pre-inflationary physics.
\end{enumerate}
Case 1 is discussed in \Sec{AnalyticSec}, 
with the conclusion that some implementations of it are observationally ruled out.
% with the conclusion that it is observationally ruled out (modulo some caveats).
Case 2 is discussed in \Sec{MonteSec} and is a multidimensional analogy of the classic example of 
anthropic constraints on 
$\rhol$ \cite{BarrowTipler,LindeLambda,Weinberg87,Efstathiou95,Vilenkin95,Martel98,GarrigaVilenkin03}\footnote{Weinberg's 
argument \cite{Weinberg87} went as follows: 
Since galaxies only form for $|\rhol|\simlt 10^{-123}$, all models predicting 
an a priori probability distribution $f(\rhol)$ are equivalent as long as the 
characteristic width of the distribution is $\gg 10^{-123}$ 
and the value $\rhol=0$ is not in any way special.
The probability distribution for $\rhol$ seen from a reference object (in this case a galaxy)
can therefore be computed independently of the detailed shape of 
the initial condition distribution $f(\rhol)$, simply treating $f(\rhol)$
as constant.}.
For some cases explored in \Sec{MonteSec}, we will find the predictions to be independent of the
details of the inflaton potential $V(\vphi)$ as well.
Cases 1 and 2 are nice in the sense that we can make quantitative inflationary predictions from $V(\vphi)$ alone,
without having a theory of pre-inflationary physics. 
On the other hand, case 3 offers an opportunity to learn about pre-inflationary physics from 
cosmological observations of $\p$.

In summary, we have seen that the when computing inflationary predictions, 
the choice of measure discussed in this section has just as important observational consequences as 
the choice of inflaton potential discussed in \Sec{Vsec}. Moreover, 
we found that the measure problem splits into three parts: 
one straightforward (chosing reference objects), one hard (chosing an ordering)
and one perhaps irrelevant (chosing initial conditions).

 \section{Possible solutions to the ordering problem}
\label{OrderingSolutionSec}

Above we saw that to make inflation a testable physical theory, the ordering problem must be solved.
This is a wide open problem in dire need of further work --- the purpose of this section is not to give a
solution, merely to describe some possible approaches to solving it. We will compare their strengths and shortcomings
in the remainder of the paper, first using numerical computations in \Sec{MonteSec} and then using analytic calculations
in \Sec{AnalyticSec}.
We will see that all orderings can be viewed as regularization techniques to deal with infinities, and that many are 
observationally ruled out.

%\subsection{Overview of ordering options}

A variety of solutions to the ordering problem have been discussed in the literature, 
albeit
% implicitly, 
using different terminology.
We will consider two broad classes of orderings:
\begin{enumerate}
\item Global orderings based on some time variable (say $t$ or $a$).
\item Pocket-based orderings (ordering separately within each thermalized region (pocket), 
then averaging the pocket results with some weighting (say equally, by $\vphi$-volume or by $\r$-volume).
\end{enumerate}

\subsection{Global time-based orderings}

The first class of orderings foliates spacetime into a sequence of three-dimensional spatial hypersurfaces,
each corresponding to a fixed ``time'', and computes the parameter probability distribution $f_p(\p)$ separately on each 
hypersurface. We will see that this is in many cases equivalent to simply ordering the reference objects by increasing formation ``time''.
The utility of this approach stems from two facts:
\begin{itemize}
\item In many cases, the ordering problem vanishes on each individual hypersurface. With time variables
like $t$ or $a$, a finite initial volume will give a finite hypersurface volume for any fixed future time,
so that all orderings of the (finitely many) reference objects give the exact same answer $f_p(\p)$.
\item In many cases, $f_p(\p)$ converges to a well-defined distribution as this ``time'' approaches infinity,
eliminating the need to select a particular hypersurface.
\end{itemize}
As detailed below, this first class of orderings is intimately linked to the 
infamous problem of chosing a time variable in the inflationary Fokker-Planck equation,
and suffers from serious problems
\cite{LindeMezhlumian93,LindeLindeMezhlumian94,Garcia94,Garcia95,LindeLindeMezhlumian,Vilenkin95,WinitzkiVilenkin96,LindeMezhlumian96,LindeLindeMezhlumian96,VilenkinWinitzki97,Guth00,Guth00b,Guth04}.

For all cases where the total number of reference objects grows exponentially over time, 
any asymptotic $t\to\infty$ distribution for reference objects at fixed time will be
identical to the asymptotic distribution for reference objects formed at time $\le t$, 
because older objects which formed at time $\ll t$  are exponentially
underrepresented. The latter distribution is the one corresponding to 
simply ordering all reference objects by their formation time, so we will 
use the expressions ``$t$-ordering'' and ``$t$-foliation'' interchangeably.

\subsection{Pocket-based orderings}

In contrast, attempting to use density, temperature or expansion rate (or indeed any quantity which, unlike $t$ or $a$, 
is physically observable) as a time variable to foliate by leads to the second class of orderings --- such ``time variables'' do
{\it not} specify the ordering sufficiently 
for inflation to predict a unique parameter probability distribution, since there will typically be 
infinitely many reference objects at any fixed ``time''.
Specifically, infinite numbers of reference objects at a given density, say, can typically be found in 
each of multiple thermalized pockets (the U-shaped regions in \fig{OrderingFig}, say)
and neither the intra-pocket ordering nor the inter-pocket weighting has been specified 
--- we will discuss multiple options for both.

\subsubsection{Intra-pocket orderings}

For the former, \ie, within a given pocket, all orderings clearly give the same answer if the volume
is finite. We will study the following two orderings for the infinite-volume case:
\begin{enumerate}
%\item {\it Empirical intra-pocket ordering}:
\item {\it Empirical ordering}:
Order the reference objects by increasing physical distance from some given point $\r$.
%\item {\it Volume-based intra-pocket ordering:} 
\item {\it Volume-based ordering:} 
Order the reference objects sampling each spatial region in proportion to its relative volume fraction,
defined in terms of how much it has inflated.
\end{enumerate}
The empirical ordering is equivalent to the procedure that an empirically minded observer at
$\r$ would adopt: compute the parameter probability distribution by 
averaging in a spherical volume centered at $\r$ (on a hypersurface defined by a fixed density, say) 
and then let the sphere radius approach infinity. 
This procedure has been advocated by \cite{Vilenkin98,Vanchurin00}, who applied it on the hypersurface corresponding to
the end of inflation.
It is easy to see that its predictions depend only on which pocket the point $\r$ lies in, not on its
location within that pocket. 

In contrast, a volume-based ordering would sample each spatial region with a frequency proportional to how much is has 
expanded since some early reference instant --- the latter is challenging to define unambiguously, and currently there is no
rigorously defined implementation of a volume-based ordering.
% For a detailed technical definition of the volume-based ordering, proposed by Vilenkin and collaborators,
% see \cite{Vilenkin98,Vanchurin00,Garriga01}.
These two orderings are distinguished by their answer to the question ``are some volumes more infinite than others",
to which they answer ``no'' and ``yes'', respectively. Consider, for example, the spatial region 
with $0\simlt x\simlt 22$ in \fig{OrderingFig}.
This is a single basin of attraction with infinite thermalized volume both to the left and right side of 
the plane $x=14$, corresponding to the inflaton having rolled down to the minimum from the left and from the right, respectively. 
This predicts a cosmological parameter distribution of the form 
\beq{LeftRightfpEq}
f_p(\p)=p_L\delta(\p-\p_L)+(1-p_L)\delta(\p-\p_R),
\eeq
where $\p_L$ and $\p_R$ are the cosmological parameter vectors 
corresponding to $\phi$ rolling down from the left and right sides, respectively,
and $p_L$ is the fraction of the reference objects in regions where $\phi$ rolled down from the left.
The vectors $\p_L$ and $\p_R$ are determined by the potential $V(\phi)$ alone as per \Sec{Vsec}, whereas
the constant $p_L$ depends also on the measure.
As long as both half-basins have infinite volume, 
the empirical ordering will predict $p_L=0.5$, 
whereas a volume-based ordering will generally not.
For instance, if the reference instant corresponds to a given density, then 
volume-based ordering assigns a larger probability to the side where it takes
more e-foldings to roll down. % \cite{Vilenkin98,Vanchurin00,Garriga01}. 
For the multi-field case where $\vphi$ is a $d$-dimensional vector, 
$\p_L$ and $\p_R$ are generically replaced by a continuum of parameter vectors corresponding to the directions in $\vphi$-space
from which the inflaton rolled down, and \eq{LeftRightfpEq} gets generalized to
a weighted angular average in $\vphi$-space.
% EMPIRICAL ORDERING ALWAYS GIVES ANGULAR AVERAGE IN REAL SPACE. NOT IN phi-SPACE!
% To define a volume-based ordering in this case, one must define an initial measure along the boundary of the basin of attraction.

In summary, both the empirical and volume-based intra-basin measures agree within any finite sub-volume
(where they both weight by volume, \ie, reduce to standard Lebesgue measure), but differ in 
how to resolve the ambiguities associated with infinite volumes.

% bounded by two maxima, 
%so that the cosmological parameter
%prediction is $\p_L$ and $\p_R$ depending on whether the inflaton rolls down from the left maximum or from the right one.
%This gives a cosmological parameter distribution of the form 
%\beq{LeftRightfpEq}
%f_\p(p)=p_L\delta(\p-\p_L)+(1-p_L)\delta(\p-\p_R)\right],
%\eeq
%where $p_L$the fraction of the reference objects in regions where the inflaton rolled down from the left.
%As long as both half-basins have infinite volume, 
%the empirical ordering will predict a 50-50 split $f_\p(p)=[\delta(\p-\p_L)+\delta(\p-\p_R)]/2$ 
%whereas the volume-based ordering generally assigns a larger probability to the side where it takes
%more e-foldings to roll down \cite{VilenkinOrdering}. 

\subsubsection{Inter-pocket weightings}
\label{InterPocketSec}

Once an intra-pocket ordering has been prescribed within each pocket, the inter-pocket weighting must specify
how reference objects from the different pockets are to be merged into a single ordered sequence.
For instance, ordering objects $A_1,A_2,...$ from pocket A and objects $B_1,B_2,...$ from pocket B jointly 
as $A_1,B_1,A_2,B_2,...$ corresponds to giving equal weight to the two pockets, whereas 
$A_1,A_2,B_1,A_3,A_4,B_2,A_5,A_6,B_3...$
gives twice as much weight to pocket A as to pocket B.
There are at least five obvious candidates for the inter-pocket weighting:
\begin{enumerate}
\item Equal weighting for each basin of attraction (in $\vphi$-space)
\item Equal weighting for each pocket (in physical space)
\item Weighting by relative $\vphi$-volume
\item Weighting by relative thermalized physical volume
\item Weighting by relative number of reference objects
%\item Weighting by relative number of spatial regions populating the $\phi$-basin
%\item Equal weighting for all basins
%\item Weighting basins by $\vphi$-volume
%\item Weighting basins by relative thermalized physical volume
%\item Weighting basins by relative number of reference objects
\end{enumerate}
1 and 2 generally differ, since many disconnected pockets can have $\vphi$ roll down into the same basin of attraction.
An explicit definition of 4 is given in \cite{Garriga01} for the special case of inflaton potentials
with a single peak.

Since inflation typically produces infinitely many pockets, there is a potential subtlety related to pocket ordering within the weighted average.
We define two pockets to be of the same {\it type} if they are observationally indistinguishable, 
\ie, give the same $f_p(\p)$.
If there are only finitely many types of pockets, the inter-pocket weighting is fully specified by
giving the weight for each type.
If there are infinitely many types, however, a naive weighted average over the types is ambiguous, since
the order in which one adds the types (or pockets) matters. This is why, most generally, one needs to merge  
the reference objects from all pockets into a single ordered sequence as stipulated above.

\subsection{Symmetry arguments}

In the following sections, we will explore the predictions of various orderings listed above, 
and find that they are sufficiently different that it might be possible to determine observationally 
which one, if any, is the correct one.
In the spirit of attempting to predict rather than postdict the correct answer, we note that there are three
heuristic arguments favoring orderings where different pockets are treated symmetrically, \ie, receive equal weight.

The first argument is that, mathematically, two sets contain the same number of objects if
there exists a one-to-one correspondence between them. 
Applying this criterion to the case of a one-dimensional inflaton potential, 
all countable infinities are equal and hence all infinite half-pockets should get equal weight,
which corresponds to empirical intra-pocket weighting and equal inter-pocket weighting.

The second argument involves the ``pothole paradox'' that will be presented in \Sec{potholeSec}, and 
also favors equal intra-pocket weighting.

The third argument involves the following {\it Gedanken} experiment.
Consider two basins of attraction in a one-dimensional inflaton potential $V(\phi)$,
and let A, B, C and D denote four infinite spatial regions where $\phi$ has rolled down from
the left into basin 1, from the right into basin 1, from the left into basin 2 and from 
the right into basin 2, respectively.
In the special case where $V(\phi)$ has identical shape in the left parts of the 
two basins, \ie, where there is some offset $\Delta\phi$ such that 
$V(\phi)=V(\phi+\Delta\phi)$ for all $\phi$ in the first half-basin,
then one can argue that volumes A and C should receive equal statistical weight 
by symmetry.\footnote{This argument fails for measures where 
quantum  diffusion/tunneling from nearby basins is important.}
Since empirical intra-basin weighting equates the weight of A with B and the weight of C with D,
transitivity of equivalence implies that B and D have equal weight, and therefore that 
the two pockets A$+$B and C$+$D have equal weight even though they are causally disconnected. 
Since this argument applies regardless of the potential shape in half-basins C and D,
it suggests that all half-pockets and hence all pockets have equal weight.  
The assumed shape-equivalence of A and C can be eliminated from the argument if we consider a
messy random potential like in \Sec{MonteSec} with infinitely many basins where for any
given half-basin, there will be another one that approximates its shape arbitrarily closely.
In summary, this suggests that empirical intra-basin weighting implies equal inter-pocket weighting.

% {\bf STUFF RELATED TO CONDITIONING}
We close this section by noting that
the ordering problem is closely linked (but not equivalent) to the conditional probability calculation corresponding
to conditioning on a class of reference object.
For example, chosing our reference objects to be ``protons at time $t$'' 
implies using the above-mentioned $t$-foliation. % which can give a well-defined albeit observationally ruled out parameter probability distribution $f_p(\p)$.
In contrast, chosing reference objects to be ``protons at density $\rho$'' (or conditioned on $T$, $H$ or another observable)
does not fully specify an ordering for the reasons given above, merely implying a pocket-based ordering.

%%%%%%%%%%%%%%%%%%%%
% MonteSec goes here 
%%%%%%%%%%%%%%%%%%%%

\section{Monte Carlo analysis}
\label{MonteSec}

Above we discussed how the cosmological parameter probability distribution $f_p(\p)$ predicted by inflation 
depends on the inflaton potential $V(\vphi)$ (\Sec{Vsec}) and on the measure (\Sec{MeasureSec}).
We wish to keep our discussion from becoming overly abstract, so 
% as an illustration, 
let us now compute the predictions for some concrete examples.
We will then quantify the measure dependence for a wider class of orderings below in \Sec{AnalyticSec}.

\subsection{Choice of potentials}

Since the inflaton $\vphi$ and the potential $V$ have units of mass and density (mass$^{4}$), respectively, 
let us write the potential as
\beq{VunitEq}
V(\vphi) = \mv^4 f\left({\vphi\over\mh}\right),
\eeq
where $f$ is a dimensionless function with values and derivatives of order unity, and the
characteristic energy scales of the horizontal and vertical axes in a plot of $V(\vphi)$ are absorbed into 
the two constants $\mh$ and $\mv$.
If $V(\vphi)$ emerges from some fundamental physical theory without fine-tuning, one might naturally
expect $\mh\sim\mv$, but we will also explore the more general case where the horizontal 
and vertical energy scales are different.

Ideally, we would wish to explore generic potentials whose shape $f$ is derived from some fundamental theory. 
Currently lacking this, we will instead explore potentials that are generic in the sense of being randomly generated.
We will explore the case where $f$ is a one-dimensional Gaussian random field
with unit variance and power spectrum $P(q)\propto q^\gamma e^{-q^2/2}$ (see \fig{phiExampleFig}).
In practice, we define 
\beq{RanFieldEq}
%f(x) = \sum_{k=-n}^n a_k e^{kx},
f(x) = {a_0\over\sqrt{2}} + \sum_{k=1}^n a_k\cos\left({kx\over\sqrt{n}}\right)+a_{-k}\sin\left({kx\over\sqrt{n}}\right),
\eeq
where the $(2n+1)$ Fourier coefficients 
%$a_k$
$a_{-n},...,a_n$ 
are independent real-valued Gaussian random variables with zero mean and variance 
\beq{PowerSpectrumEq}
\expec{a_k^2} = A\sum q^\gamma e^{-{q^2\over 2}},
\eeq
$q\equiv k/\sqrt{n}$ and the normalization constant $A$ is chosen so that 
$\expec{f(x)^2}={1\over 2}\sum_{k=-n}^n \expec{a_k^2} = 1$.
We take $n=100$ and $\gamma=0$ as our baseline, but explore a variety of other choices below.
We will find that the resulting parameter probability distribution $f_p(\p)$ is rather insensitive
to the detailed shape of $f$, but depends strongly on the energy scales $\mh$ and $\mv$.
A key goal of this section is to identify which features of a complicated potential
$V(\phi)$ are most important in determining the testable predictions $f_p(\p)$.

% Also mention taylor expansions with random coefficients. Try it? (Very quick.)

\subsection{Choice of measures}

We will compute results for two different measures that we term Measure A and Measure B.
Both weight all half-basins roughly equally, the difference being that A counts only those producing an infinite 
volume. Measure A thus corresponds roughly to the empirical intra-pocket ordering and the equal weighting of basins.
As described below, this also corresponds approximately to a uniform distribution over the
initial conditions $(\phi,\dot\phi)$, including the volume weighting factor $e^{3\Ntot}$ into
the $w(\p)$-factor of \eq{fpEq} for Measure A but excluding it for Measure B.
Both of these two measures may be the wrong choice, but they are least free
from the most egregious paradoxes and observational contradictions that afflict many other measures 
as discussed below in \Sec{AnalyticSec}, and hence instructive to explore quantitatively.

Specifically, we perform the following numerical experiment:
\begin{enumerate}
\item Generate a random potential $V(\phi)$, start at $\phi=0$.
\item Go uphill until a maximum is reached.
\item If $V<0$ (Big Crunch imminent) or the slow-roll approximation (SRA) is invalid ($\epsilon>1$ or $|\eta|>1$), 
count as an $\Ntot=0$ failure and stop, otherwise evolve according to the SRA.
\item If $\phi$ gets stuck in an eternally inflating local minimum, count as a failure and stop (will receive zero weight since
no reference objects produced --- not even protons).
\item When the SRA breaks down (which defines $\aend$), compute $\p$ using equations~(\ref{OtotSummaryEq})-(\ref{ntSummaryEq}),
taking $\aexit$ to be 55 e-foldings earlier ($\aexit=e^{-55}\aend)$ 
and evaluating $\rhol=V(\phi)$ at the subsequent local minimum of the potential.
\item Repeat to obtain a statistically large sample, say $10^9$ times.
\item Repeat the entire experiment for a range of energy scales $\mh$ and power spectrum 
shape parameters $\gamma$.
\end{enumerate}
Step 2 is optional: we include it for Measure A but not for Measure B.
We are thus simulating ``shotgun inflation'' in the sense that we 
spray starting points randomly across the potential surface (Measure B) or at random maxima (Measure A).
%In \Sec{AnthroSec}, we will use either galaxies or objects in galaxies as reference objects,
%so the weight function $w(\p)$ from \eq{fpEq} will 
%eliminate all cases where there is too much curvature ($N\simlt 55$ e-foldings) or too much 
%dark energy ($|\rhol|\gg 10^{-112}Q^3$) for galaxies to form. 
In \Sec{AnthroSec}, we will use either galaxies or objects in galaxies as reference objects,
so the weight function $w(\p)$ from \eq{fpEq} will 
eliminate all cases where there is too much dark energy ($|\rhol|\gg 10^{-112}Q^3$) for galaxies to form
or where black holes prevent the formation of reference objects (for $\Ntot\simlt 55$ as per \Sec{BlackHoleSec}).
As detailed in \Sec{AnalyticSec}, Measure A will automatically give $\Ntot=\infty$, whereas Measure B will give $\Ntot=\infty$
only if the maximum above the starting point satisfies the SRA, $\ie$, has $\eta>-1$.
This $\Ntot$-cutoff thus affects only Measure B.

The two measures defined by this prescription were chosen to maximize computational efficiency, not to
be maximally simple to interpret. However, we will now see that, modulo some 
weight factors or order unity, they correspond to weighting all half-basins
equally and also to uniform initial conditions.

Since the statistical properties of our Gaussian random field $V(\phi)$ are 
translationally invariant, our choice to start at $\phi=0$ is equivalent to starting
at a random $\phi$-value drawn from a uniform distribution.
Apart from the numerical practicality that $n$ is finite so that $V$ is periodic,
we can thus reinterpret our generation of many random potentials as simply trying many different starting 
points in the same infinitely extended potential.
The probability that we roll down into any particular half-basin is therefore proportional 
to its length for Measure A and to the length of the sub-region giving $\Ntot>55$ for Measure B. 
The length of a half-basin (the distance between a maximum and a minimum of $V$)
is of order $\mh$ with random variations of order unity, so the different half-basins 
get weighted roughly equally.% , and so do the resulting spatial pockets --- ignoring quantum diffusion.
%\footnote{
%If we wish to make sure that only half-basins with infinite volume get counted, 
%we can require that the local maximum from which $\phi$ arrived at the starting point
%satisfy the slow-roll condition, guaranteeing a near-infinite number of e-foldings.
%In practice, this is essentially equivalent to using our
%present prescription with a weight function $w(\p)$ requiring $N\gg 60$, which we explore below --- among
%other things, this predicts $\Otot\approx 1$.
%} 
% To be more careful, I could decide to compensate for this. I could also reject half-basins that 
% don't give infinite volume. OK, DONE!

The parameter $\rhol$ of course depends only on which basin we roll down into, since it is the height of its minimum.
Since the potential is one-dimensional, the parameters
$(\dH,\ns,\al,r,\nt)$ depend only on the half-basin (whether we roll down from the left or from the right),
so the only parameters affected by where in the basin we start are 
$\Otot$ and $\Ntot$ --- for Measure A, where we start at a maximum, they are $\Otot=1$ and $\Ntot=\infty$.

% KEEP THESE TWO FIGS TOGETHER ON THE SAME PAGE:  

\begin{figure} 
%\vskip\smtopskip
\epsfxsize=16.5cm\epsffile{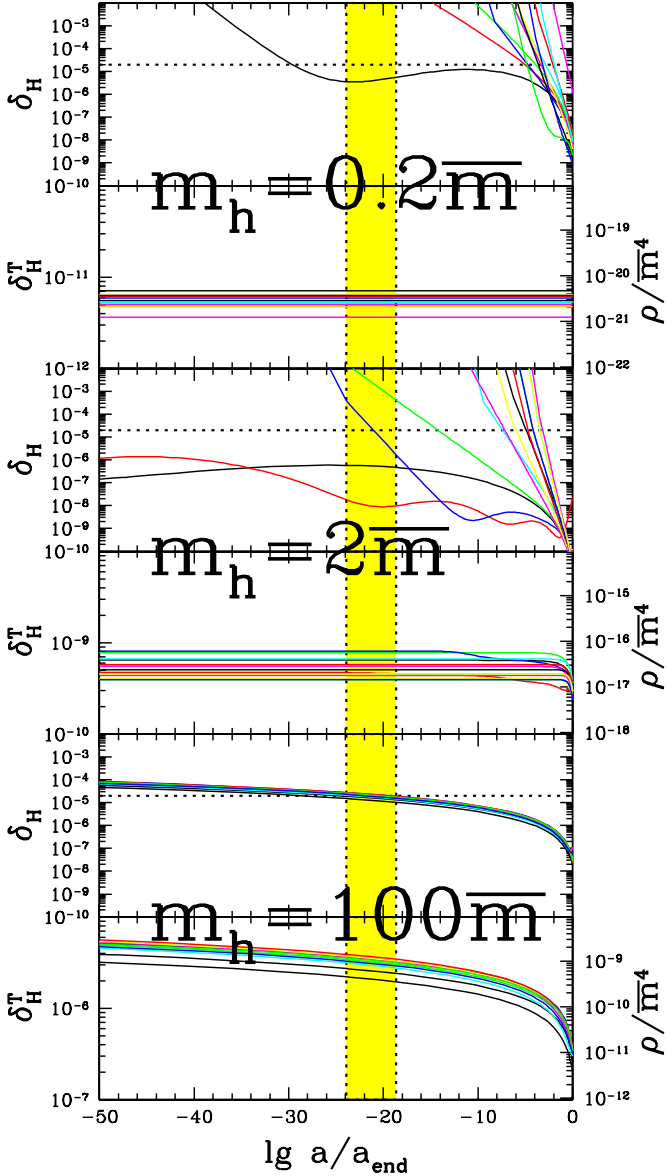}
%\vskip\smbotskip
\caption[1]{\label{PcomboBasinFig}\footnotesize%
The cosmic density history $\rho(a)$ 
is shown in the three panels with labeling on the right side
for a handful of simulations with Measure A and inflation mass scales 
$(\mh,\mv)$=$(0.2\mbar,10^{-6}\mbar)$,
$(2\mbar,0.0004\mbar)$ and
$(100\mbar,0.02\mbar)$.
%
% Qscaling=0.00000003 ~ mv^2
% mv = 0.006741645746*sqrt(0.00000003) ~ 1.16768706e-06
%
%\mh=2\mbar$ and $\mv\sim 0.0004\mbar$.
% mv = 0.006741645746*sqrt(0.003) ~ 0.0003692550818
%
%$\mh=100\mbar$ and $\mv\sim 0.02\mbar$.
%% mv = 0.006741645746*sqrt(10) ~ 0.02131895348
The corresponding scalar and tensor power spectra are shown with labeling on the left side.
The shaded vertical band indicates the
range 43-55 e-foldings from the end of inflation where there is current hope to measure these 
primordial fluctuations\protect\cite{pwindows,spacetime}.
}
\end{figure}

\begin{figure} 
%\vskip\smtopskip
\epsfxsize=16.5cm\epsffile{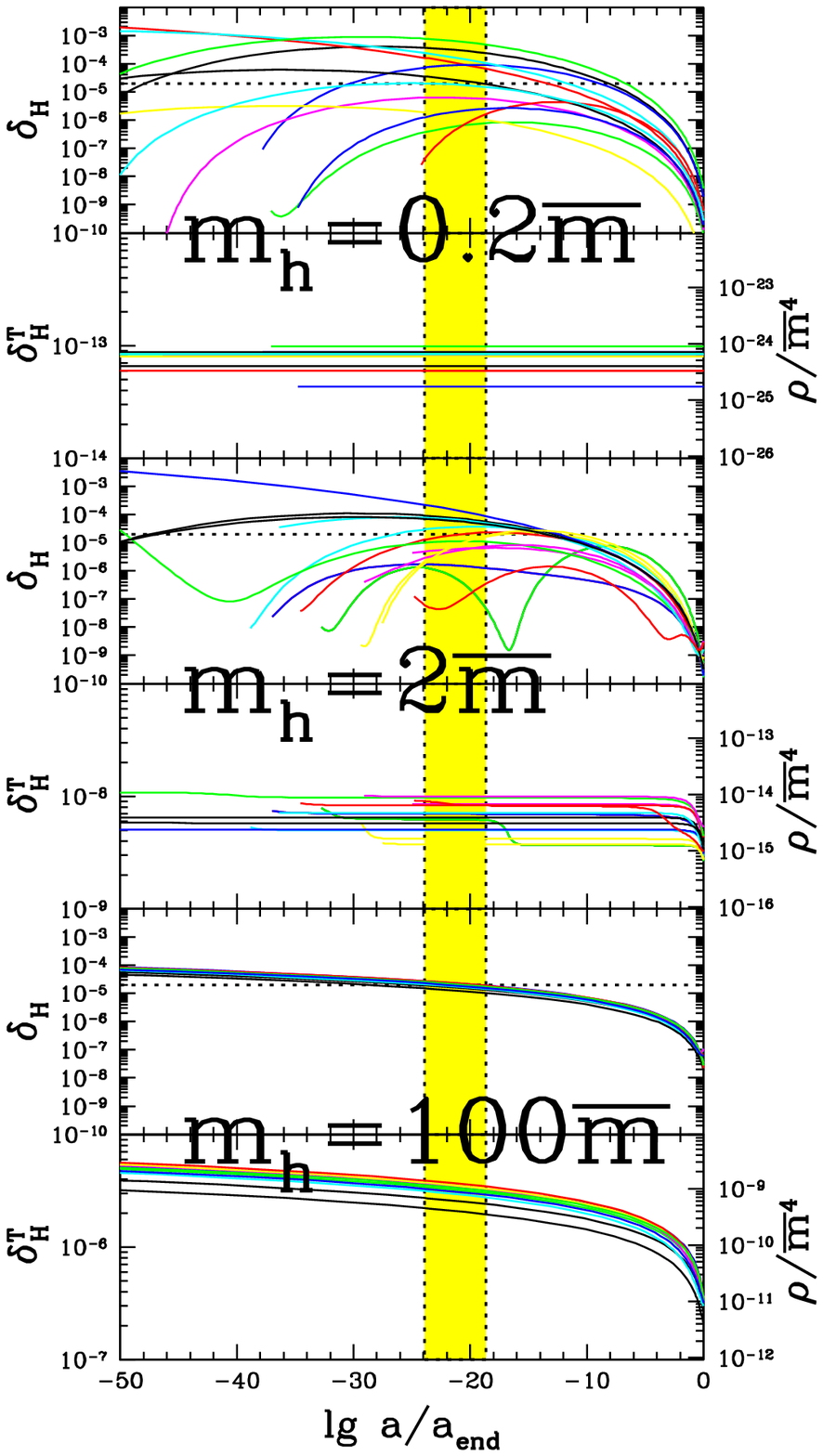}
%\vskip\smbotskip
\caption[1]{\label{PcomboFig}\footnotesize%
Same as \fig{PcomboBasinFig}, but for Measure B, which includes thermalized regions with finite volume.
% measure giving statistical weight only to those thermalized regions that have infinite volume.
}
\end{figure}

Finally, let us discuss the effect of requiring the slow-roll approximation (SRA) to hold.
When instead using the exact inflationary dynamics of \eq{nDphiEq}, we find that
almost all cases giving substantial amounts of inflation do so while the slow-roll approximation (SRA) is
valid. Because our $w(\p)$ gives negligible weight to 
cases with $\simlt 55$ e-foldings (black holes prevent the formation of reference objects as per \Sec{BlackHoleSec}),
our statistics will be dominated by trajectories $\phi(t)$ involving at least one slow-roll period.
Our prescription counts those starting with slow-roll and discards those where the SRA becomes 
valid only later on (for instance, a large $\dot\phi$ can be finely tuned to just barely make 
$\phi$ roll up a hill, so that the SRA becomes valid near the top).
If $\epsilon<1$ and $|\eta|<1$ at a given position $\phi$, then 
the well-known SRA attractor behavior \cite{LiddleLythBook} will 
soon drive the derivative $\dot\phi$ to the SRA value 
$\dot\phi\approx-\mbar V'(\phi)\sqrt{3/V(\phi)}$ if it starts 
out in the slow-roll range  $|\dot\phi|\simlt\sqrt{2V(\phi)}$.
If we began with random and uniformly distributed initial conditions
$\vPhi=(\phi,\dot\phi)$, 
the fraction of the models surviving the $\dot\phi$ cut would thus be $\propto V^{1/2}$,
so that the $\dot\phi$-cut is equivalent to starting with the SRA value of $\dot\phi$
and weighting the initial $\phi$-distribution by $V(\phi)^{1/2}$.
We find that $V^{1/2}$ does not vary strongly between successfully inflating models, 
so our prescription is roughly equivalent to simply starting with uniformly random initial conditions $\vPhi$.

% In practice, to save CPU time, we zap cases with negative lambda.

%Also equivalent to $a$-foliation?
%In this section, I'll foliate using $a$ as the time variable, 
%so it won't matter whether I weight by physical or comoving volume.
%NO, THAT'S WRONG! ONLY TRUE WHILE INFLATION IS ONGOING. BUT I SHOULD COUNT THERMALIZED VOLUME WHEN INFLATION 
%ENDS.

For computational efficiency, we ignore quantum diffusion in our calculations.
We will discuss the effects of diffusion in \Sec{DiffusionSec} and find that they are unimportant
for our qualitative conclusions.
As discussed in \Sec{AnalyticSec}, diffusion can, with some measures, drastically alter the relative
weighting between different basins. Once we have specified the inter-pocket weighting, however,
the only observable effect that diffusion can have is to modify the evolution during the 
last 55 or so e-foldings --- this is important only for models with 
$Q\simgt 1$ and is completely irrelevant for models with $Q\sim 10^{-5}$.

%\begin{figure} 
%%\vskip\smtopskip
%\centerline{\epsfxsize=\figsiz\epsffile{P0.2.ps}}
%%\vskip\smbotskip
%\caption[1]{\label{P0.2_fig}\footnotesize%
%The cosmic density history $\rho(a)$ (bottom panel, right labeling)
%is shown for a handful of simulations with 
%$\mh=0.2\mbar$, $\mv\sim=10^{-6}\mbar$, $\gamma=0$,
%% Qscaling=0.00000003 ~ mv^2
%% mv = 0.006741645746*sqrt(0.00000003) ~ 1.16768706e-06
%together with the corresponding tensor power spectrum (bottom panel, left labeling)
%and scalar power spectrum (top panel). The shaded vertical band indicates the
%range 43-55 e-foldings from the end of inflation where there is hope to measure these 
%primordial fluctuations.
%}
%\end{figure}

%\begin{figure} 
%%\vskip\smtopskip
%\centerline{\epsfxsize=\figsiz\epsffile{P2.ps}}
%%\vskip\smbotskip
%\caption[1]{\label{P2_fig}\footnotesize%
%Same as \fig{P0.2_fig}, but for $\mh=2\mbar$ and $\mv\sim 0.0004\mbar$.
%% mv = 0.006741645746*sqrt(0.003) ~ 0.0003692550818
%}
%\end{figure}

%\begin{figure} 
%%\vskip\smtopskip
%\centerline{\epsfxsize=\figsiz\epsffile{P100.ps}}
%%\vskip\smbotskip
%\caption[1]{\label{P100_fig}\footnotesize%
%Same as \fig{P0.2_fig}, but for $\mh=100\mbar$ and $\mv\sim 0.02\mbar$.
%% mv = 0.006741645746*sqrt(10) ~ 0.02131895348
%}
%\end{figure}

% KEEP THESE TWO FIGS TOGETHER ON THE SAME PAGE:

\begin{figure} 
%\vskip\smtopskip
%\centerline{\epsfxsize=15cm\epsffile{1d_combo.ps}}
\hglue-0.9cm\epsfxsize=15.2cm\epsffile{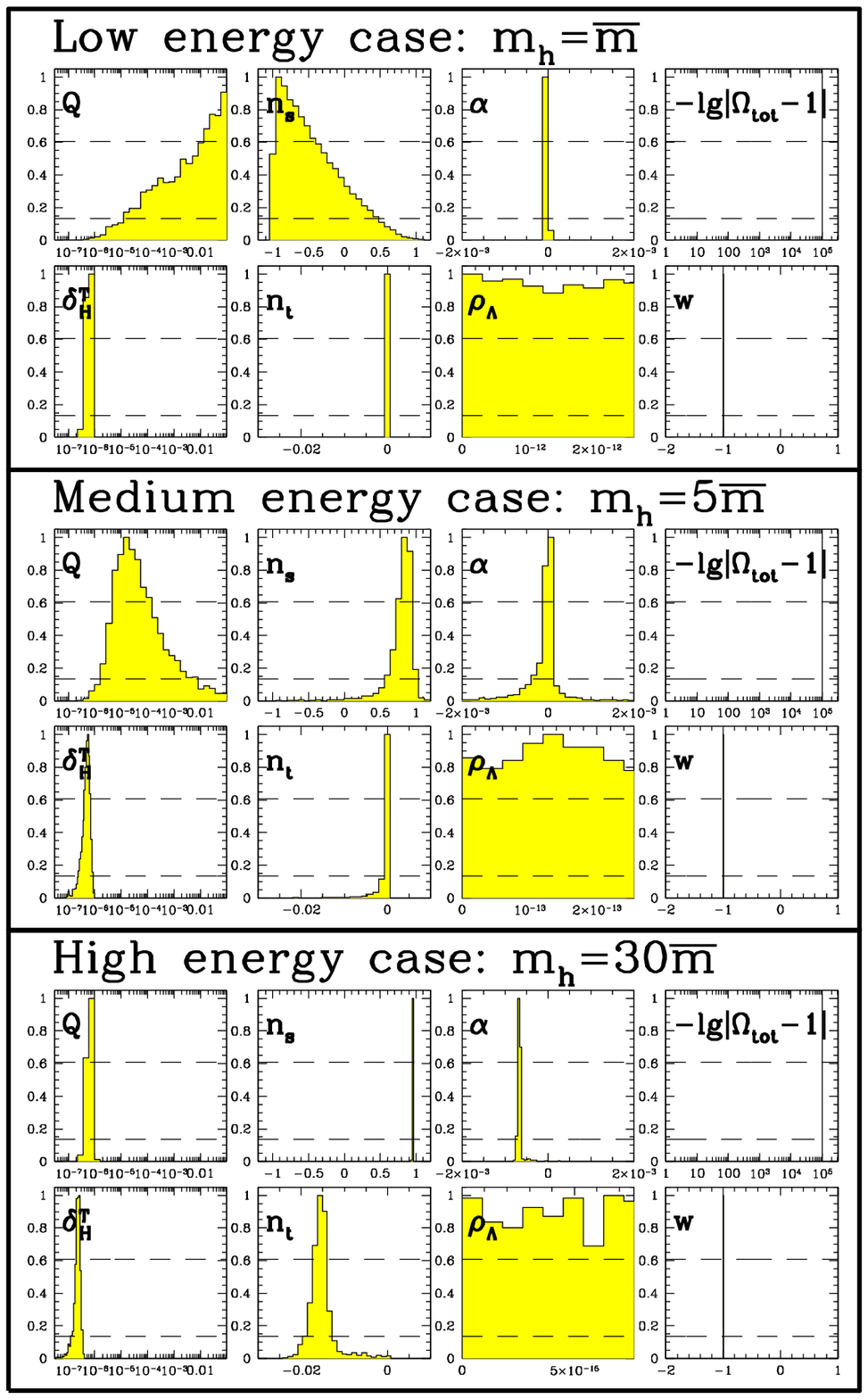}
\vskip-1.3cm
\caption[1]{\label{1DcomboBasinFig}\footnotesize%
The cosmological parameter distributions predicted by three inflation models
with $\mv\sim 0.007\mv$ and Measure A.
}
\end{figure}

\begin{figure} 
%\vskip\smtopskip
\hglue-0.9cm\epsfxsize=15.2cm\epsffile{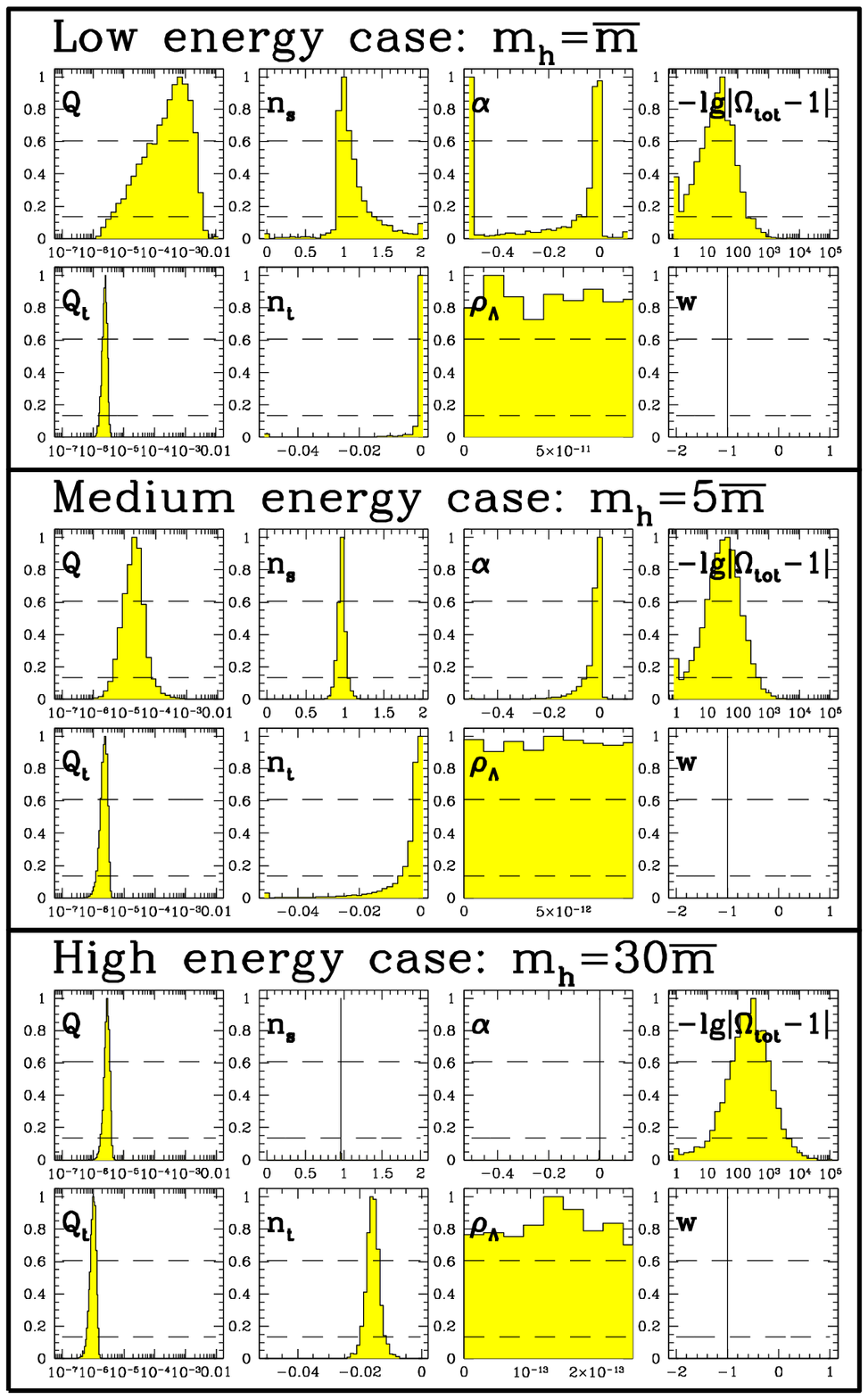}
\vskip-1.3cm
\caption[1]{\label{1DcomboFig}\footnotesize%
The cosmological parameter distributions predicted by three inflation models
with $\mv\sim 0.007\mv$ and Measure B.
}
\end{figure}

% KEEP THESE TWO FIGS TOGETHER ON THE SAME PAGE:

\begin{figure} 
%\vskip\smtopskip
\centerline{\epsfxsize=\figsize\epsffile{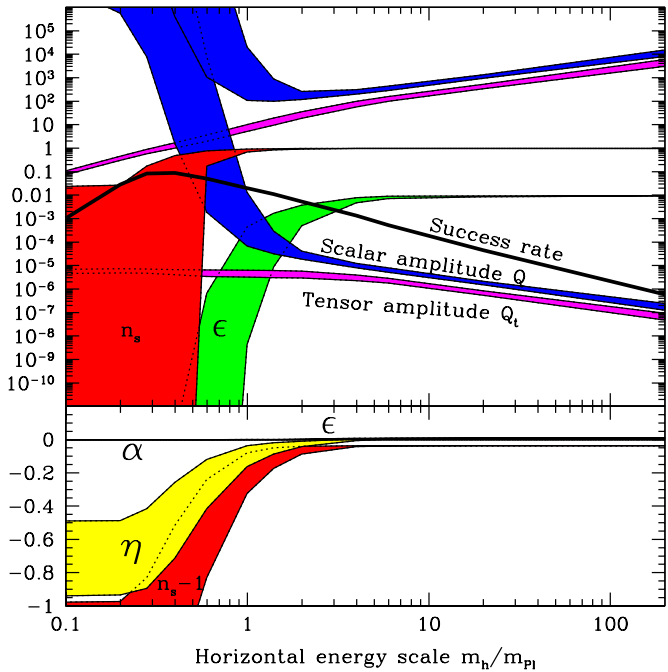}}
%\vskip\smbotskip
\caption[1]{\label{mhBasinFig}\footnotesize%
Each shaded band shows the central $68\%$ of a parameter probability 
distribution (such as those in \fig{1DcomboBasinFig}) 
as a function of horizontal inflation mass scale $\mh$, for Measure A.
All labeled bands have the vertical scale fixed at $\mv=0.007\mbar$.
The second pair of bands for $Q$ and $Q_t$ (unlabeled, above labeled bands with same
colors/shades) have $\mv$ varying so that $\mv=\mh$.
}
\end{figure}

\begin{figure} 
%\vskip\smtopskip
\centerline{\epsfxsize=\figsize\epsffile{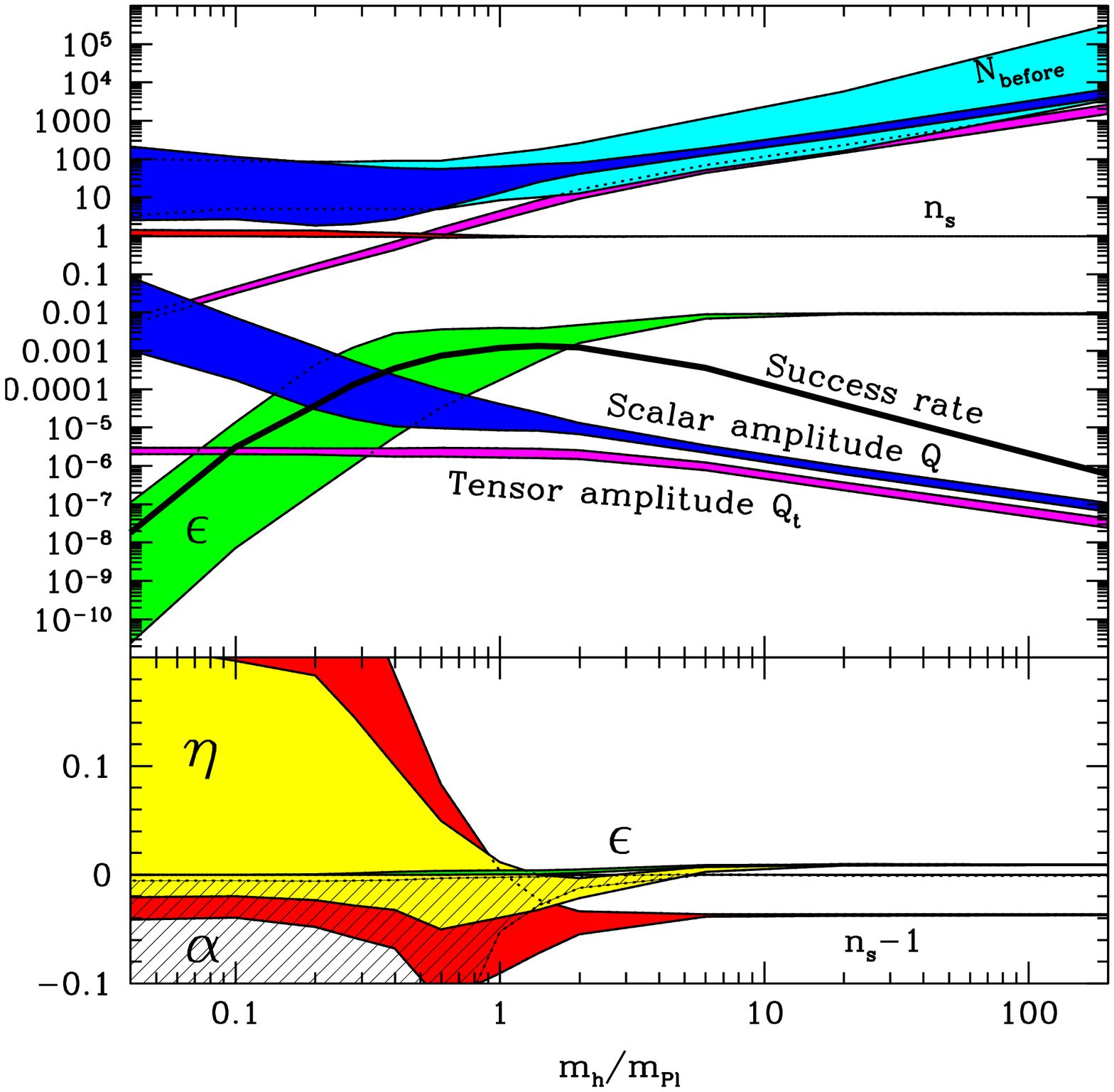}}
%\vskip\smbotskip
\caption[1]{\label{mhBasicFig}\footnotesize%
Same as \fig{mhBasinFig}, but for Measure B. which includes thermalized regions with finite volume.
%the pocket-based measure giving statistical weight only to those thermalized regions that have infinite volume.
}
\end{figure}

\subsection{Basic results}

Figures~\ref{PcomboBasinFig} and~\ref{PcomboFig} 
shows the resulting $\rho(a)$-curves for a small sample of models for the two measures.
From these curves, the cosmological functions $\delt(k)$ and $\deltt(k)$
are readily obtained using equations~(\ref{delttEq}) and~(\ref{deltEq}) --- these two figures show examples.
Our cosmological parameters $\p$ are computed from the $\rho(a)$-curves using 
equations~(\ref{parSummaryEq})-(\ref{ntSummaryEq}), giving histograms such as those shown in 
Figures~\ref{1DcomboBasinFig} and~\ref{1DcomboFig}.
Let us now explore how these results depend on the inflation mass scales $\mh$ and $\mv$ and on the
parameter $\gamma$ controlling the power spectrum of the inflaton potential.

The $\mv$-dependence is readily computed analytically.
Since quantum diffusion is negligible for the cases we consider, 
the vertical inflation scale $\mv$ affects none of the parameters except
the overall amplitudes $Q$ and $Q_t=rQ$, 
which both scale as $\mv^2$ (the $\rhol$-distribution is affected only indirectly via $Q$,
as will become clear in \Sec{AnthroSec}).
The effect of changing the horizontal scale $\mh$ 
(the scale on which the inflaton potential wiggles), however, needs to be
computed numerically, and is complex and interesting for most parameters.
We compute the parameter distribution $f_p(\p)$ for 
the values $m_h/\mbar=$0.2, 0.5, 1, 2, 3, 5, 7, 10, 20, 30, 100 and 1000,
and the results are summarized in Figures~\ref{mhBasinFig} and~\ref{mhBasicFig}.
%The effect of changing $\gamma$ is much less dramatic --- we return to this below.
For our baseline calculations, we take $\gamma=0$.
We repeat the entire grid of calculations for $\gamma=2$ and $\gamma=4$ and find that
this has little effect on the results --- the analytic results below will clarify why.
% The effect of changing $\gamma$ is much less dramatic --- we return to this below.
%Show that altering $\gamma$ doesn't do much, and is partially offset by rescaling $\mh$.
%Perhaps mention random Taylor expansions too?

We find that there are three regimes of horizontal mass scale giving 
qualitatively different behavior: 
low energy $\mh\ll\mbar$, intermediate energy $\mh\sim\mbar$ 
and high energy $\mh\gg\mbar$. We will now discuss these three regimes in turn.

\subsection{The high energy limit $\mh\gg\mbar$}

Although we lack a theory of quantum gravity, it is commonly assumed that
quantum gravity corrections will make contributions to the inflaton potential 
in the form of a power series in $(\phi/\mpl)$ with coefficients of order unity,
thus causing $V(\phi)$ to vary substantially when on a scale $\Delta\phi\sim\mpl$.
In other words, one assumes that quantum gravity predicts $\mh\simlt\mpl\sim 5\mbar$,
rendering the high-energy regime $\mh\gg\mbar$ physically unnatural and poorly motivated.
This classic argument \cite{LindeBook,DodelKinneyKolb97,LythRiotto99,LiddleLythBook,Peiris03,Kinney03,LiddleSmith03,Wands03} 
is also the standard objection to
large-field inflation models like 
$V(\phi)\propto\phi^2$,
%$V(\phi)=\mh^4(\phi/\mh)^2$,
which require this sort of extreme flatness to agree
with observation --- see \cite{Linde02} for a counterarguments and discussion of these issues.

These issues aside, what are the testable predictions in the $\mh\gg\mbar$ limit?
Figures~\ref{mhBasinFig} and~\ref{mhBasicFig}
reveal an interesting limiting behavior: when $\mh$ increases far above the Planck scale,
the predictions for $\ns$, $r$ and $\alpha$ become very sharp, with almost no random scatter, independently of
whether Measure A or Measure B is used.
This behavior can be traced back to 
Figures~\ref{PcomboBasinFig} and~\ref{PcomboFig}, illustrating that as $\mh\to\infty$, all the 
$\rho(a)$-curves have the same shape, differing merely in amplitude.
The slow-roll parameters $\epsilon$, $\eta$ and $\xi_2$, which depend only on the shape of 
$\rho(a)$, therefore approach constants and the corresponding
probability distributions for $r$, $\nt$, $\ns$ and $\alpha$ in 
Figures~\ref{1DcomboBasinFig} and~\ref{1DcomboFig}
approach Dirac $\delta$-functions as $\mh\to\infty$.

The reason for this limiting behavior is easy to understand.
Increasing $\mh$ makes it easier to satisfy 
the slow-roll conditions $\epsilon\ll 1$ and $|\eta|\ll 1$, since 
\eq{VunitEq} implies that 
\beqa{fSRAeq}
\epsilon&=&{1\over 2}\left({\mbar V'\over V}\right)^2
         ={1\over 2}\left({\mbar\over\mh}\right)^2\left({f'\over f}\right)^2  \sim \left({\mbar\over\mh}\right)^2,\\
\eta	&=&{\mbar^2 V''\over V} 
         =\left({\mbar\over\mh}\right)^2 {f''\over f}  
	 \sim\left({\mbar\over\mh}\right)^2\label{fSRAeq2}
%$\epsilon=(\mbar V'/V)^2/2=(f'/f)^2 (\mbar/\mh)^2/2 \sim (\mbar/\mh)^2$.
%$\eta=\mbar^2 V''/V=(\mbar/\mh)^2 f''/f  \sim (\mbar/\mh)^2$.
\eeqa
at generic points $\phi$, so that as $\mh\to\infty$, inflation occurs at almost all $\phi$-values where $V(\phi)>0$. 
If inflation ever ends, it does so because $f$ drops very close to zero, either 
because of $\epsilon$ when $f=(\mh/\mbar)|f'|/2$ 
or because of $\eta$ when $f=(\mh/\mbar)^2|f''|$.
Recall that the dimensionless inflaton potential $f$ is a function whose values and derivatives are generically
of order unity.
In either of the two cases, our $\mh\gg\mbar$ limit therefore implies that $0<f\ll 1$ when inflation ends, 
so that the observational consequences depend only on the behavior of $f$ over a tiny range
very close to zero. Since this range is much smaller than the natural scale on which $f$ varies,
the only aspects of $f$ that have any observational consequences are the first terms
of its Taylor expansion in this range.

The most likely behavior in the region where 
$f\approx 0$ is that $f$ is roughly a straight line with slope of order unity.
For most of our Monte Carlo simulations, inflation is therefore ended by $\epsilon=1$ 
when $f\sim\mbar/\mh$, after which the inflaton fast-rolls \cite{LindeFastroll01} 
down to negative values of the potential and space 
promptly recollapses without producing any galaxies or other reference objects. 

Although these are the most common cases, they are given zero statistical weight by the $w(\p)$-factor in \eq{fpEq}
and hence do not contribute to the predicted parameter probability distribution $f_p(\p)$.
The only cases that produce galaxies are those where $f'$ is unusually small in this region, so
that $f$ flattens out and takes a local minimum where $\rhol=V(\phi)=\mv^4 f(\phi/\mh)$
is close enough to zero to produce galaxies. Since this constraint is roughly 
$|\rhol|\gg 10^{-112}Q^3$ (see \Sec{AnthroSec}), the only segments of a 
potential $V(\phi)$ that contribute to the
parameter probability distribution are those whose minimum $V(\phistop)$ is 
much closer to zero than the $V$-value where inflation ends, in other words those segments that
are well-approximated by a parabola $V(\phi)\propto\phi^2$.
For this well-studied case
\cite{LindeDiffusion86,LiddleLythBook}, 
we have $\epsilon=\eta=2\mbar^2/(\phi-\phistop)^2$,
which gives $\phiend=\phistop\pm\sqrt{2}\mbar$ and the following cosmological functions and parameters 
(see Appendix B and, \eg, \cite{LiddleLythBook}):

\beqa{quadratic_rhoEq}
\rho(a)	&=&\rho(\aend)(1+2N),\\
Q	&=&{2N+1\over\sqrt{75}\pi}\Qstar\approx 4\Qstar\label{quadratic_dHeq},\\
Q_t	&=&{4\over\sqrt{75}\pi}\sqrt{2N+1}\Qstar\approx 1.5\Qstar\label{quadratic_dHTeq}\\
\ns	&=&1-{4\over 2N+1}\approx 1-{2\over N}\approx 0.963,\label{quadratic_nsEq}\\
\al	&=&-{8\over (2N+1)^2}\approx -{2\over N^2}\approx -0.0006,\label{quadratic_alEq}\\
r	&=&{16\over 2N+1}\approx {8\over N}\approx 0.15,\label{quadratic_rEq}\\
\nt	&=&-{2\over 2N+1}\approx -{1\over N}\approx 0.02,\label{quadratic_ntEq}
\eeqa
where $N\equiv\aend/a$ as usual, 
$\Qstar\equiv{\mv^2\over 2\mbar\mh}\sqrt{\fppmin}$
and $\fppmin\sim 1$ is the second derivative of the dimensionless inflaton potential $f$ at the local minimum
into which it rolls down.
The number of e-foldings $N$ corresponding to our current horizon volume is given by \cite{LiddleLeach03} 
\beq{quadratic_Neq2}
N \approx 63.3 - {1\over 12}\ln {\rhoend\over\rhoreh},
\eeq
%\beq{quadratic_epsrreEq}
%{\rho\over\rhoend}\epsilon=1
%\eeq
where the last term is 0, 4, 11 and 15 for reheating instantaneously, at 
$\simlt 10^{11}$ GeV 
(to avoid overproduction of gravitinos in SUSY models), at the electroweak scale 100 GeV and
at the BBN scale 1 MeV (at lower reheat energies, no protons are produced --- but see \cite{Hannestad04}).
We have used $N=55$ in all our plots.
Finally, we expect the observed $\Otot\sim 10^{-5}$ as $\mh\to\infty$, since the decreased slow-roll speed
increases the total number of e-foldings needed to roll down from a given starting point $\phi$,
giving $\Nbefore\equiv\ln(\astart/\aexit)\to\infty$ as $\mh\to\infty$.

The above analytic arguments are confirmed by our numerical calculations. 
Figures~\ref{PcomboBasinFig} and~\ref{PcomboFig} (bottom) 
show that the density histories $\rho(a)$ have the shape predicted
by \eq{quadratic_rhoEq} --- the scatter in amplitudes is due to the fact that
$\rhoend\propto\Qstar^2\propto\fppmin$, and this second derivative of order unity 
differs from one minimum to another.
Figures~\ref{mhBasinFig} and~\ref{mhBasicFig} show that the dispersions in
$\ns$, $r$, $\nt$, $\alpha$, $\epsilon$, $\eta$ and $\xi_2$ drop 
towards zero as $\mh$ increases,
and that the constant values they approach are precisely those 
given by equations~(\ref{quadratic_nsEq})-(\ref{quadratic_ntEq}),
\ie, $(\ns,r,\nt,\alpha,\epsilon,\eta,\xi_2)\approx (0.963,0.15,-0.02,-0.0006,0.009,0.009,0)$.

Also, \fig{mhBasicFig} shows that the success rate drops $\propto\mh^{-2}$ as 
$\mh\to\infty$, as predicted by our above discussion:  inflation ends with $\eta\sim 1$ when
$f=f_{\rm end}\sim(\mh/\mbar)^2f''\sim(\mh/\mbar)^2\ll 1$ according to \eq{fSRAeq2}, and the probability 
that the next minimum lies between this
tiny positive value and zero is thus proportional to $f_{\rm end}\simpropto \mh^{-2}$.
% eps=eta for parabola with rhol=0.
% Empirically, mh=100 and mh1000 always ended by eta while epsilon took rather random 
% values between 0 and 1. Perhaps if I allowed rhol<0 too, I'd occasionally end by epsilon as well?
% summary.dat suggests slope $-2$ or slightly steeper 

\begin{figure} 
%\vskip\smtopskip
\centerline{\epsfxsize=\figsize\epsffile{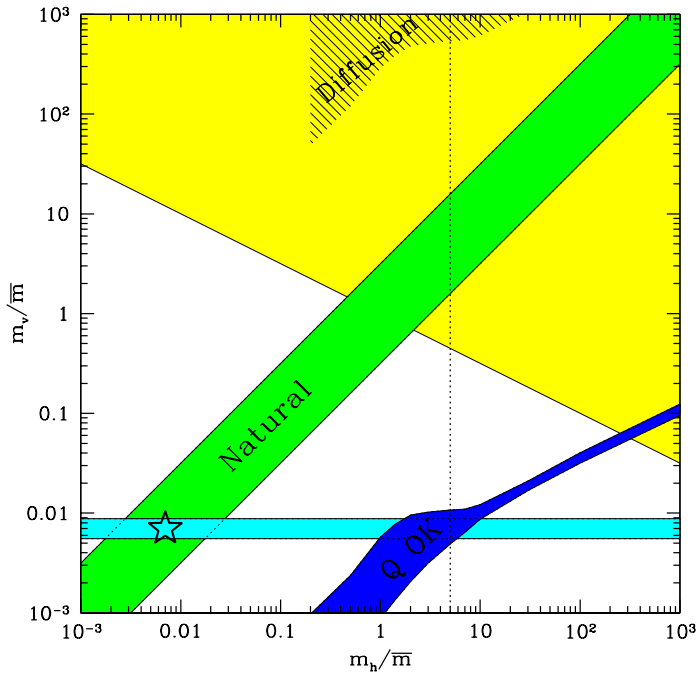}}
%\vskip\smbotskip
\caption[1]{\label{mhmvFig}\footnotesize%
The inflationary model space spanned by the energy scales 
$\mh$ and $\mv$ giving the horizontal and vertical units of the inflaton potential.
Quantum gravity corrections may disfavor models with $\mh\simgt\mpl\approx 5\mbar$ (to the right of
the vertical line).
The physically natural expectation is $\mv\sim\mh$ (green diagonal band), so one may hope 
to find a working model at the star, in the intersection of this band with the 
cyan horizontal band that gives the correct fluctuation amplitude $Q\sim 2\times 10^{-5}$ when
$\mh\sim\mpl$. We find that such hopes are dashed by the fact 
that $\mv$ must be lowered when $\mh$ is decreased
(blue/dark grey band for Measure B), to offset the $Q$-increase caused by the preference for tiny $\epsilon$.
One might naively expect quantum diffusion to be important in the yellow/light grey region,
but the drop in $\expec{\epsilon}$ as $\mh\to 0$ and the drop in $\expec{V(\phiend)}$ 
as $\mh\to\infty$ make diffusion important in the hatched region instead.
}
\end{figure}

\Eq{quadratic_dHeq} shows that reproducing the observed fluctuation level $Q\approx 2\times 10^{-5}$
requires $\mh\sim 0.003\sqrt{\mbar\mv}$, % 0.0027a
so in addition to the possible quantum gravity problems with having $\mh\gg 1$, these models also 
require tuning in the sense that $\mv\ll\mh$ (they are far from the band marked ``natural'' in \fig{mhmvFig}).

In summary, the parabolic inflaton potential $V(\phi)\propto\phi^2$ 
has been widely studied in the literature, and is a classic example giving eternal inflation with an unbounded potential
\cite{LindeDiffusion86}.
It is also the only integer power law potential that is still consistent with observational data now
that the $\phi^4$ model is firmly ruled out \cite{sdsslyaf}.
In this subsection, we have seen that the predictions of this particular model hold also for a much broader class 
of models. Although our Monte Carlo calculations were only for the case of Gaussian random field potentials, our analytic
arguments above clearly hold for {\it any} generic dimensionless inflaton potential $f$, and also independently 
of the initial conditions.
Specifically, in the high-energy limit $\mh\gg\mbar$, {\it any} messy and complicated inflaton potential predicts
the same values of $(\ns,r,\nt,\alpha,\epsilon,\eta,\xi_2)$ as the simple $V(\phi)\propto\phi^2$ model
as long as it is not fine-tuned to have its second derivative vanishing in a large fraction of its minima
that $\phi$ can roll into, \ie, as long as generic minima look locally like parabolas.
Of the eight cosmological parameters from \eq{pEq}, we thus obtain sharp predictions for 
$(\Otot,w,\ns,\al,r,\nt)$. In contrast, the probability distributions retain finite widths for $\rhol$ and $Q$,
which depend on $V(\phi)$ and $V''(\phi)$ at the minimum rolled down into.
This means that in the $\mh\gg\mbar$ limit, the only aspect of the potential $V(\phi)$ 
that affects the cosmological parameter predictions
$f_p(\p)$ is the two-dimensional distribution of $(V,V'')$ at its minima.

\subsection{The low energy limit $\mh\ll\mbar$}

\subsubsection{Motivation}

Although the above-mentioned limit $\mh\gg\mbar$ can match observational data, it has two theoretical blemishes
as illustrated in \fig{mhmvFig}: 
quantum gravity corrections may render $\mh\gg\mbar$ unnatural, and 
the discrepancy $\mh\gg\mv$  between the two mass scales appears contrived. 
A natural expectation would be $\mv\sim\mh\sim m$ for some energy scale $m$ 
corresponding to the physics responsible for
inflation, say the GUT scale or the string scale.\footnote{In
the context of some supersymmetric models, the ``natural'' expectation in \fig{mhmvFig} is different
and perhaps more compatible with what is observed.
The reason is that in a supersymmetric theory, one naturally has two different scales: a high
scale (say $\mbar$) in the ultraviolet and the SUSY
breaking scale $M$.  In many supersymmetric theories, the natural scales for
the potential \eq{VunitEq} of so-called moduli fields
%(fields coupled to the
%supersymmetry breaking, and all other fields, only through gravity and
%irrelevant couplings), takes the form
are $\mv=M$ and $\mh=\mbar$, typically with $\mv\ll\mh$ \cite{Kachru04}.
}

We find that for $\mh\sim\mpl$, the observed fluctuation amplitude 
$Q\sim 2\times 10^{-5}$ is reproduced if $\mh\sim 0.007\mbar$, which suggests exploring whether 
inflation with $\mh=\mv\sim 0.007\mbar$ (five-pointed star in \fig{mhmvFig}) 
makes predictions in agreement with observation.
We will see that this idea, which was one of the original 
motivations for the calculations in this paper, fails.
However, we will find that it clears one of the most challenging hurdles 
and also gives predictions largely independent of the details of $V(\phi)$,
so it will be interesting to see whether future work can resuscitate it in some modified form. 

\begin{figure} 
%\vskip\smtopskip
\centerline{\epsfxsize=\figsize\epsffile{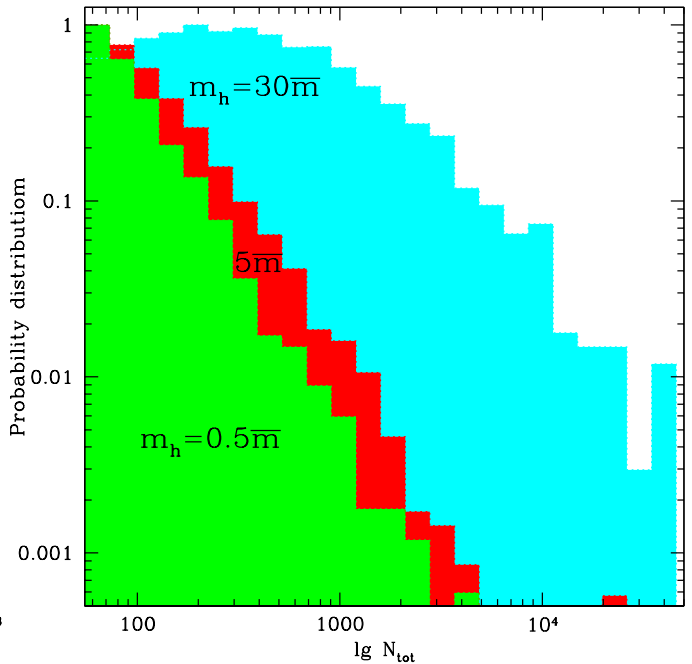}}
%\vskip\smbotskip
\caption[1]{\label{NhistFig}\footnotesize%
The three histograms show the distributions of the total number of e=foldings $\Ntot$ for
high-energy ($\mh-30\mbar$, top)
intermediate energy ($\mh=5\mbar$, middle)
and low-energy ($\mh=0.5\mbar$, bottom)
examples with Measure B.
}
\end{figure}

\subsubsection{The flatness constraint}

Figures~\ref{mhBasinFig} and~\ref{mhBasicFig}
show that as we reduce $\mh$ from the Planck scale to lower values, 
the success rate drops. The reason for this is obvious from equations\eqn{fSRAeq}
and\eqn{fSRAeq2}: $\epsilon$ and $|\eta|$ exceed unity at most $\phi$-values
when $\mh\ll\mbar$, so that we obtain 
substantial inflation only when getting lucky and starting in a patch where 
$\epsilon$ and $|\eta|$ are anomalously small. The smaller $\mh$ is, the luckier we need to get
and the lower the success rate will be.
This well-known fact that very wiggly potentials are less likely to support inflation was
identified early on \cite{Guth81,Starobinsky1980,Linde82,AlbrechtSteinhardt82,Linde83}.

Given that inflation is intrinsically unlikely in this low-energy limit, the observed flatness
constraint $|\Otot-1|\simlt 0.02$ \cite{Spergel03,sdsspars} becomes a challenging hurdle 
that can potentially rule out models with $\mh\ll\mbar$.
If galaxy formation requires only $\Ntot\simgt 55$ (see \Sec{AnthroSec}) and inflation is unlikely, 
is it not highly unlikely to get the additional e-foldings required to 
satisfy the observational flatness constraint, $\Ntot\simgt 57$?

For Measure A, the answer is automatically no, since it automatically gives $\Ntot=\infty$, so let us
limit our discussion in this subsection to Measure B.
Perhaps surprisingly, \fig{mhBasicFig} shows that the answer is no even for Measure B, with low energy 
inflation generally passing this observational test. 
In \fig{mhBasicFig}, the number of additional non-required e-foldings $(\Nbefore)$ is indeed seen to drop
from $\sim 10^5$ to $\sim 10^1-10^2$ as $\mh$ is reduced from $10^3\mbar$ down to the 
Planck scale, but then more or less stops dropping as $\mh$ is further reduced. 
\Fig{1DcomboFig} illustrates that the histogram for $|\Otot-1|\sim e^{-2\Nbefore}$
is almost identical for the $\mh=5\mbar\approx\mpl$ case and that with 5 times smaller $\mh$, 
typically giving values like $\Otot=1\pm 10^{-30}$.
Why is this? The basic reason is that, as seen in 
\fig{NhistFig}, the $\Ntot$-distribution for $\mh\ll\mpl$ is roughly 
a power law $f_N(\Ntot)\sim\Ntot^p$ with a rather modest slope $p\sim -3$
that steepens only slightly when $\mh$ is further decreased (the plotted distribution of 
$\lg\Ntot$ thus has slope $\Ntot^{p+1}$).
The $\mh=0.2\mbar$ case is not plotted in \fig{NhistFig}, 
since it is virtually identical to the $\mh=0.5\mbar$ case shown.
This means that the probability that the total number of e-foldings $\Ntot$ exceeds some
given value $N_*$ satisfies $P(\Ntot<N_*)\simpropto N_*^{p+1}\sim 1/N_*^2$,
so that the probability of satisfying the observed flatness constraint given that
we are observing from a galaxy is 
\beq{NprobEq}
P(\Ntot>57|\Ntot>55)\sim\left({57\over 55}\right)^{p+1}\sim 0.9.
\eeq

In contrast, the flatness constraint would rule out $\mh\ll\mbar$ 
if the $\Ntot$-distribution were a much steeper power law or had an exponential cutoff, so
let us understand why the $\Ntot$-distribution is not steeper.
If we rewrite the slow-roll \eq{nD_SRA_phiprimeEq} as
\beq{SRdNeq}
dN\approx {|d\phi|\over\mbar^2(\ln V)'} = {|d\phi|\over\mbar\sqrt{2\epsilon}},
\eeq
we see that there are two different ways in which we can get lucky 
and obtain large $\Ntot$ when faced with tiny $\mh$: by rolling far (large $\Delta\phi$) or
by being in a region where the potential is extremely flat $(\epsilon\ll 1)$.
As we will now see, our results follow from the fact that the latter way of getting lucky is much more likely
than the former. 
%For our Gaussian random field potentials, the derivative $V'(\phi)$ at some $\phi$-value
%is a Gaussian random variable with zero mean and standard deviation of order
%$\mv^4/\mh$, so $\ln V'=V'/V$ has zero mean and standard deviation of order $\mh^{-1}$.
%To get any inflation, we need $\epsilon<1$, which according to \eq{fSRAeq}
To get any inflation, we need $\epsilon<1$ and $|\eta|<1$ at our starting point, 
which according to Equations\eqn{fSRAeq} and\eqn{fSRAeq} requires
$|(\ln f)'|\simlt\mh/\mbar\ll 1$ and
$|f''|/f\simlt (\mh/\mbar)^2\ll 1$.
Since $f$ and its derivatives are by definition of order unity, the $\eta$-requirement is more stringent
and will be satisfied for only about a fraction  $(\mh/\mbar)^2$ of our simulations, showing that
the success rate must drop at least as fast as $(\mh/\mbar)^2$ as $\mh\to 0$ 
(indeed, \fig{mhBasicFig} shows it dropping roughly as  $\mh^6$).
This argument also predicts that almost all $\mh\ll\mbar$ models will have inflation terminated 
by $|\eta|$ exceeding unity while we still have $\epsilon\ll 1$, which is borne out by our simulations
(for instance, out of our 
1410065405 
%1{,}410{,}065{,}405 
$\mh=0.5\mbar$ simulations, every one of the 4260 that gave $\Ntot>55$ were terminated
by the $\eta$-constraint and none by the $\epsilon$-constraint). This is why the density histories
look so extremely flat in \fig{PcomboFig} (top panel).
There is, however, an even stronger pressure towards tiny $\epsilon$. Since the slow-roll constraints
must keep holding at all $\phi$-values that we roll through, we have to keep getting lucky 
with $\epsilon$ and $\eta$ over and over again as we roll along, and
the probability of this happening drops faster than any power law in 
the rolled distance $\Delta\phi/\mh$ as higher and higher derivatives in the initial 
Taylor expansion become important. 
\Fig{mhBasicFig} shows that empirically, by far the most likely way to get large numbers of e-foldings
us therefore to roll very little, by having an anomalously small $\epsilon$ and hence a tiny
roll distance $\Delta\phi$. 
\Fig{mhBasicFig} shows that $\epsilon\sim (\mh/m)^6$ at the lowest energies probed, \ie,
$(\ln V)'\sim (\mh/m)^3$,
to be contrasted with the expectation $(\ln V)'\sim (\mh/m)^{-1}$ from \eq{fSRAeq}.
In summary, success requires getting lucky 
both with $\epsilon$ (which happens a fraction $\sim (\mh/m)^4$ of the time)
and with $\eta$ (which happens a fraction $\sim (\mh/m)^2$ of the time),
which together explains the empirically observed success rate $\epsilon\sim (\mh/m)^6$.

In conclusion, we can obtain large numbers of $e$-foldings with low-energy inflation
by simply by getting lucky with tiny
$|V''(\phi)|$ and minuscule $V'(\phi)$ at the starting point $\phi$. Since zero is not a special
value in the probability distributions for $V'$ and $V''$ (as opposed to very large values, say, 
which may be exponentially unlikely), this is only polynomially unlikely. 
%Moreover, once we have
%gotten lucky enough to obtain say $N=50$, we can roughly double $N$ by cutting $V'$ in half,
%which is only twice as unlikely. WHAY'S WRONG HERE?
% This explains why the $N$-distrition falls as $1/N$ in the $\mh\to 0$ limit.

% Inspection of 
% ~/trym/t3/inflation/NOLAMBDASCREEENING/SLOWROLL_mh0.2_mv0.004_n100/qaz_slowrollphi38434631.dat et al
% shows that most runs start with |eta\sim 1| but drop to
%\eta ~ 0.002866 (changed sign)

\subsubsection{Predictions for the other 7 parameters}

Above we saw that despite their low success rate,
$\mh\ll\mbar$ models are not ruled out by their predictions for the curvature parameter $\Otot$.
Let us now explore their predictions for the remaining 7 parameters.
We will first focus on the case of Measure A, which is arguably 
the better motivated of the two since it weights by
thermalized volume. We will then turn to Measure B, and find that 
its predictions and problems are qualitatively similar to those of Measure A.

Above we saw that the $\mh\gg\mbar$ limit simplified because the SRA was so easy to satisfy:  
the only properties of $V(\phi)$ that mattered were its minima, since the observable fluctuations 
where produced very near them.
For the $\mh\gg\mbar$ limit, there is an analogous simplification: the SRA is now so hard to satisfy
that the only properties of $V(\phi)$ that really matter are are the maxima, since the observable fluctuations are produced near 
them. For Measure A, where the thermalized pockets produced are infinite, we know that we began by rolling off a peak 
where $-1<\eta_0\le 0$. Generic peaks have non-vanishing second derivative, \ie, $\eta_0<0$, and thus look locally 
like an upside-down parabola --- for this well-know case, reviewed in Appendix B.2, the distance of $\phi$ from the peak
grows $\propto a^{\eta_0}$, \ie, exponentially with the number of e-foldings. 
As shown in Appendix B.2, inflation around more general peaks is usually terminated when higher derivatives 
make $\eta$ exceed unity, and for the observable regime $N\sim 55$ gives
the cosmological parameter predictions (see Appendix B)
\beqa{hill_epseq2}
\epsilon&\sim&{\eta_0^2\mh^6\over\mbar^6} e^{2\eta_0 N}\approx 0,\\
\rho(a)	&\approx&V(\phi_0),\label{hill_rhoEq}\\
%Q	&=&{Q_t\over\sqrt{\epsilon}}\sim V(\phi_0)^{1/2}{\mbar^3\over\mh^3\eta_0} e^{-\eta_0 N},\label{hill_Qeq}\\
Q	&=&{Q_t\over 4\sqrt{\epsilon}}\sim {\mv^2\mbar\over\mh^3\eta_0} e^{-\eta_0 N},\label{hill_Qeq}\\
Q_t	&\approx&\sqrt{8V(\phi_0)\over 75\pi^2\mbar^4}\sim{\mv^2\over\mbar^2} ,\label{hill_QtEq}\\
\ns	&\approx&1+2\eta_0,\label{hill_nsEq}\\
\al	&\approx&-2\xi_2 \sim \eta_0 e^{\eta_0 N}\approx 0,\label{hill_alEq}\\
r	&=&16\epsilon\approx 0,\label{hill_rEq}\\
\nt	&=&-2\epsilon\approx 0,\label{hill_ntEq}.
\eeqa
Here the $\sim$ symbols indicate scatter of order unity in the relations stemming from $f$ and its derivatives.
The most important scatter therefore comes not from this, but from the strong dependence on $\eta_0$
which, by the infinite-volume requirement, is a random variable in the range $(-1,0)$.
\Eq{hill_epseq2} shows that for $N\sim 55$, 
the four parameters $\epsilon$, $\alpha$, $r$ and $\nt$ will be exponentially small unless $\eta_0 N\simlt 1$,
\ie, unless $-0.02\simlt\eta_0<0$. Since \eq{fSRAeq2} shows that the curvature $\eta_0$ is much larger at generic
extrema ($|\eta|\gg 1$), we thus expect this exponential suppression of order 98\% of 
the time\footnote{Specifically, this suggests that the $\eta_0$-distribution cannot have a characteristic
scale on this comparatively tiny sub-interval $(-1,0)$, and therefore can be well approximated by a power law.
The $\ln Q$-distribution is the convolution of the broad and featureless power-law distribution for $N\eta_0$ with
a much narrower distribution of width $\Delta\ln Q\sim 1$ reflecting when exactly the $f$-factors end inflation,
and will therefore have essentially the same power-law distribution as $N\eta_0$ over most of the range, 
\ie, be for all practical purposes determined by the distribution of peak curvatures $\eta_0$ alone.
}.
Even for the remaining $\sim 2\%$ of the models, these four parameters will be tiny, since \eq{hill_epseq2}
gives $\epsilon\simlt N^{-2}(\mh/\mbar)^6\simlt 10^{-9}$ for $\mh<0.1\mbar$.

Figures~\ref{PcomboBasinFig}, \ref{1DcomboBasinFig} and~\ref{mhBasinFig} show that 
these analytic predictions are borne out by our numerical calculations:
for $\mh\ll\mbar$, the parameters $\epsilon$, $\alpha$, $r$ and $\nt$ are all
near zero and $\ns$ asymptotes to a broad distribution 
with range $-1<\ns<1$. As suggested by equations\eqn{hill_Qeq} and\eqn{hill_nsEq},
the distributions for $\ln Q$ and $\ns$ are seen to have the same shape up 
to a sign reversal, this common shape being simply that of the $\eta_0$-distribution 
since up to additive factors, $-\ln Q$ and $\ns$ equal
$2\eta_0$ and $N\eta_0$, respectively.

Whereas the predictions when $\mh\gg\mbar$ agreed well will observational data for a suitably tuned $\mh$-value,
the predictions when $\mh\ll\mbar$ agree quite poorly with the observed spectral index $\ns\approx 0.98\pm 0.02$, 
mostly being way too red.
In addition, \eq{hill_Qeq} shows that generic $\eta_0$-values predict $Q$-values that are exponentially large compared to 
the vertical energy scale $\mv$:
for the extreme case $\eta_0=1$, 
matching the observed value $Q\approx 2\times 10^{-5}$ would require
\beq{mvExtremeEq}
\mv\sim\sqrt{\mh^3Q\over\mbar}e^{-N/2}\sim 60\>{\rm TeV}\>\times\left({\mh\over\mbar}\right)^3.
\eeq
% calc sqrt(2e-5*exp(-55))*1.22e19
These two potential problems are clearly linked, since they both correspond to $\eta_0\ll 0$.
We will return to these interesting issues in \Sec{AnthroSec}, since they require a careful treatment of selection effects related to
the $Q$-parameter.

Although our Monte Carlo calculations were only for the case of Gaussian random field potentials, our analytic
arguments above clearly hold for {\it any} generic dimensionless inflaton potential $f$, and also independently 
of the initial conditions.
Specifically, in the high-energy limit $\mh\gg\mbar$, {\it any} messy and complicated inflaton potential predicts
negligible values of $(r,\nt,\alpha,\epsilon,\xi_2)$
as long as it is not fine-tuned to have its second derivative vanishing in a large fraction of the
maxima with $\eta>-1$, \ie, as long as generic maxima look locally like upside-down parabolas
(we cover the case of unbounded potentials in \Sec{DiffusionSec}).
Of the eight cosmological parameters from \eq{pEq}, we thus obtain sharp predictions for 
$(\Otot,w,\al,r,\nt)$. 
In contrast, the probability distributions retain finite widths for $\ns$ and $Q$ (which are highly correlated since
they both depend on the peak curvature $\eta_0$) and for $\rhol$ which depends on $V(\phi)$ at the subsequent minimum.
In \Sec{AnthroSec}, we will see how the $\rhol$-distribution can be computed analytically from the distribution of the other
parameters, so this means that in the $\mh\gg\mbar$ limit, the only aspect of the potential $V(\phi)$ 
that significantly affects the cosmological parameter predictions $f_p(\p)$ is the distribution of curvatures 
$\eta$ at the peaks.

The situation for Measure B is somewhat more complicated since, as seen in \Fig{PcomboFig}, 
it includes cases where the only substantial period of
inflation occurred at a near-inflection point rather than close to maximum, and cases where most of the inflation
occurred {\it after} our current horizon scale fluctuations had been generated. 
\Fig{1DcomboFig} shows that this favors $\ns$-values exceeding unity and gives a broad $\al$-distribution in poor
agreement with the observed constraint $|\al|\simlt 0.01$ \cite{sdsslyaf}. However, Measure B is seen to share a 
problematic broad distributions for $\ln Q$. The reason is the same for both
measures: Long-lasting inflation requires a tiny $\epsilon$ which can give huge $Q\propto Q_t/\sqrt{\epsilon}$, 
and the value of the $\epsilon$-parameter 55 $e$-foldings before the end of inflation depends sensitively on the second 
(and for Measure B also higher) derivatives of the potential.
% Explain that a non-negligible fraction of the models do their main inflating {\it after} we observe.

\subsection{The intermediate case $\mh\sim\mbar$: Planck-scale inflation}

\begin{figure} 
%\vskip\smtopskip
\centerline{\epsfxsize=\figsize\epsffile{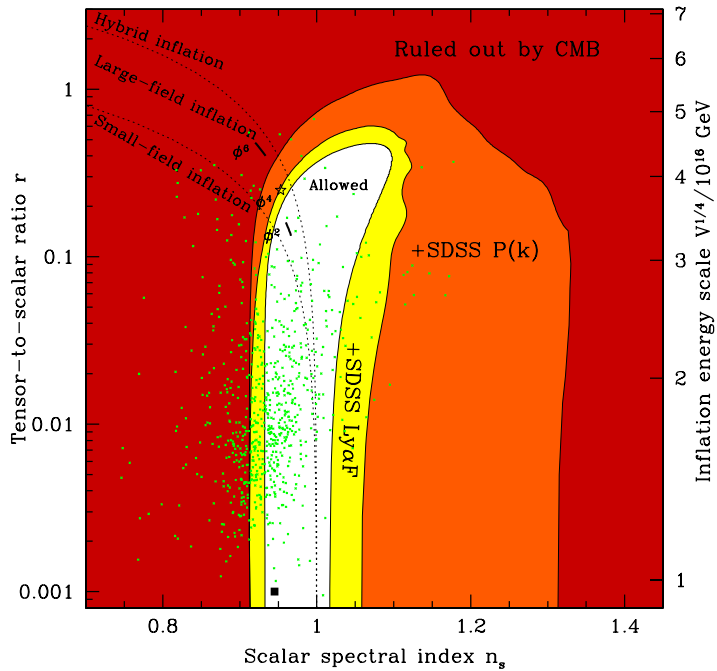}}
%\centerline{\epsfxsize=\figsize\epsffile{2d_nsr_7parr_5.ps}}
%\vskip\smbotskip
\caption[1]{\label{nsrFig}\footnotesize%
Constraints and predictions in the $(\ns,r)$-plane. 
The nested shaded regions are ruled out at 95\% confidence from WMAP alone \cite{Spergel03},
when adding SDSS galaxy clustering information \cite{sdsspower,sdsspars} and
when also adding SDSS Lyman $\alpha$ Forest information \cite{sdsslyaf}.
The green/grey points are the predictions from our simulations 
with $\mh=\mpl$ and Measure A. 
The two dotted curves delimit the three classes of inflation models known 
as small-field, large-field and hybrid models.
Some single-field inflation models make highly specific predictions in this plane as indicated.
From top to bottom, the figure shows the predictions for
$V(\phi)\propto\phi^6$ (line segment; ruled out), 
$V(\phi)\propto\phi^4$ (star; ruled out),
$V(\phi)\propto\phi^2$ (line segment, $\mh\gg\mpl$; still allowed), and
$V(\phi)\propto 1-(\phi/\phi_*)^4$ (horizontal line segment with $r\sim 10^{-3}$; still allowed),
assuming $N=64$ for the $\phi^4$ model and $50<N<60$ for the others
as per \cite{LiddleLeach03}.
}
\end{figure}

Above we saw that models with both $\mh\gg\mbar$ and $\mh\ll\mbar$ face
severe problems, the former theoretically and the latter observationally. 
Let us therefore briefly discuss the third option, $\mh\sim\mbar$, \ie, that 
the horizontal scale of the inflaton potential is of order the Planck scale.

This is the most complicated of the three cases: the predictions $f_p(\p)$ will depend
on the full details $V(\phi)$, not merely on the peaks or troughs, and none of
the above-mentioned analytic approximations hold. The only relation between the observables
that still holds is therefore the familiar slow-roll consistency relationship $\nt=-r/8$.
To make interesting predictions for this case, it is thus crucial to have
a physically motivated model for $V(\phi)$, which we currently lack.
We will therefore limit this section to one interesting feature which should be
rather generic. 

\Fig{nsrFig} compares the predictions the spectral index $\ns$ and the tensor-to-scalar ratio $r$
for about $10^3$ of our $\mh=\mpl$
simulations with Measure A. The observational constraints will continue to improve 
dramatically in coming years  --- for instance, the completed SDSS survey combined with 
the Planck CMB measurements have been forecast \cite{parameters2} to measure both 
of these two parameters to an accuracy of about $0.01$.
For $\ns$, this means that {\it all three} of the cases we have considered
($\mh\ll\mbar$, $\mh\sim\mbar$ and $\mh\gg\mbar$) generically predict departures
from scale-invariance ($\ns=1$) that should soon be detectable.
As regards gravitational waves, the $r$-predictions 
are seen to scatter up to $r$-values greatly exceeding $0.01$, 
with the mean being at the readily detectable level $r\approx 0.03$.
This bodes well for upcoming CMB polarization missions aiming to detect 
$B$-polarization.

Both of these qualitative predictions for the $\mh\sim\mpl$ case
(good prospects for detecting gravitational waves and lack of scale invariance)
are rather generic, since we have seen that they with a greater margin
($\ns=0.96$ and $r\approx 0.15$) for essentially {\it any} potential when $\mh\gg\mpl$.

%So perhaps this is the best bet case?
% Also highest acceptance rate. 
% Worth showing any of my $3\times 3$ 2D plots? FORGET IT!

%SRA requires $|\dot\vphi|\ll\sqrt{2V(\phi)}$.
%Find that almost all inflation that I get occurs during slow-roll, either
%SRA is valid from the beginning or it becomes valid later because 
%I roll up a hill and slow down just to the point that $|\dot\vphi|\ll\sqrt{2V(\phi)}$.
%So for any $\phi$, there's some magic range of $\dot\vphi$ that will 
%get me to a slow-roll region with negligible kinetic energy, $|\dot\vphi|\ll\sqrt{2V(\phi)}$.
%It's therefore pretty much equivalent to simply start with 
%$\dot\phi=0$ and reject the cases where SRA isn't valid from the outset. 
%That's what I do.

\subsection{The effect of quantum diffusion and unbounded potentials}
\label{DiffusionSec}

Our calculations above ignored the effect of quantum diffusion \cite{LindeDiffusion86}
and did not include unbounded potentials where $V\to\infty$ --- let us briefly comment on these 
two issues.

For these calculations, diffusion makes essentially no difference for Measure B and only a minor 
difference for Measure A, changing the half-basin weights by factors of order unity.
Specifically, quantum diffusion is negligible on parts of the inflaton potential producing fluctuations 
$Q\ll 1$ \cite{LindeDiffusion86}. 
For the calculations in this section, 
diffusion is therefore important only right near the top of peaks in $V(\phi)$ 
(where we start with Measure A), 
and will then have the effect of giving roughly equal odds for rolling down from this peak to the left and to the right. 
In other words, including it numerically would merely readjust the weights of each pair of half-basins
separated by a peak by a factor of order unity, making the two weights 
equal rather than proportional to their length.
% Including diffusion would have essentially no effect for measure B.
% We will return to the effect of quantum diffusion for more general cases is discussed in \Sec{AnalyticSec}.

Having the potential blow up ($V\to\infty$) in places is either irrelevant or observationally ruled out, depending on 
the horizontal mass scale $\mh$. If $\mh\gg\mbar$, then as we saw above, 
the observable cosmological parameters depend only the behavior of $V$ near its minima and are unaffected by how 
the field rolled there (except in so far as it affects the inter-basin weighting).
If $\mh\ll\mbar$, then 55-e-foldings before the end of inflation, $\phi$ was most likely 
either exponentially close to 
a maximum of $V$ or extremely high up a $V\to\infty$ hill. Although our lack of a theory of quantum gravity precludes 
us from calculating predictions when $V(\phi)\gg\mbar^4$, even the less extreme case $V(\phi)\sim\mbar^4$ gives
$Q_t\sim 1$, \ie, tensor modes producing large-scale unpolarized CMB fluctuations about 5 orders or magnitude
larger than the current observational limits.

\section{Conditioning on reference objects}
\label{AnthroSec}

Above we computed probability distributions for the cosmological parameter vector $\p$ for 
various potentials and measures. However, as discussed in \Sec{MeasureSec}, 
these distributions (\fig{1DcomboBasinFig} and 
\fig{1DcomboFig}) are not the final theoretical predictions that we are entitled to confront with 
cosmological observation, since they do not include selection effects and hence 
lack the appropriate weight function $w(\p)$ in \eq{fpEq}.
They implicitly used protons or thermalized physical volume as reference objects, 
but the vantage points from which we make our observations are not a random protons but highly unusual
protons (on planets, \etc).
Instead, we can think of distributions like those in \fig{1DcomboBasinFig} and 
\fig{1DcomboFig} as the raw probability distributions that come out of inflation, 
to be modulated by conditioning on whatever is appropriate to condition on \cite{conditionalization,BostromBook}.
To make this more explicit, let us rewrite \eq{fpEq} as
\beq{fpEq2}
f_p(\p) = \finf_p(\p)w(\p),
\eeq
defining $\finf_p(\p)\equiv\expec{\expec{f_p(\p;\vPhi,V)}_\vPhi}_V$ as the
raw parameter distribution emerging from inflation.
These two parts of the problem (computing $\finf_p$ and computing $w$) thus decouple completely, 
simply being multiplied at the end.
Likewise, the resulting distribution $f_p$ for some rudimentary reference objects (protons, say) 
simply gets multiplied by an additional weight function if we compute the distribution $f_p$ for more complex
reference objects made of the rudimentary ones (galaxies, say).
We will therefore explore only a limited set of reference objects in this section, making 
various simplifying approximations,
leaving a more thorough exploration of weight functions for future work.
Specifically, we will explore various combinations of the following weight factors:
\beqa{wEq}
\wvol(\p)	&\propto&e^{3\Ntot},\label{wvolEq}\\
\whalo(\p)	&\approx&\erfc\left[{0.1\rhol^{1/3}\over\xi^{4/3}Q}\right],\label{whaloEq}\\
\wN(\p)		&\approx&\theta[\Ntot-\Ngal(Q)],\label{wNeq}\\
\wQ(\p)		&=&\hbox{broad function around $Q\sim 10^{-5}$},\label{wQeq}
%\wgal(\p) 	&\propto&\whalo(Q,\rhol)\wQ(Q)\wN(Q)
%\whalo(Q,\rhol)&=	&\fgal(Q,\rhol)\approx\erfc\left[{1.17\rhol^{1/3}\over\xi^{4/3}Q}\right]\\
%\Ngal(Q)	&\approx&\ln\left({\rinf^{1/4}\over\xi Q^{1/2}}\right)
\eeqa
where $\xi\equiv\xib+\xic+\xin\approx 3.3\times 10^{-28}$ (see Table 1) is the matter-to-photon ratio in Planck units.
$\wvol(\p)$ is simply the factor by which inflation expands the volume.
$\whalo(\p)$ is the fraction of all
protons that end up gravitationally bound in nonlinear objects (in dark matter halos),
so including the $\whalo$-factor in $w(\p)$ corresponds to narrowing the class of reference 
objects to gravitationally bound protons.
If the reference objects are galaxies, then 
the $\wN$-factor and the lower cutoff in the $\wQ$-factor are arguably required as well. As described below, 
the $\wN$-factor reflects the fact that long-lived galaxies are only formed if the number of 
e-foldings exceeds 
\beq{NgalEq}
\Ngal(Q)\approx\ln\left({\rinf^{1/6}\rhoreheat^{1/12}\over\xi Q^{1/2}}\right).
\eeq
--- otherwise, pre-inflationary fluctuations cause dark matter halos to be engulfed by black holes.
%--- otherwise, pre-inflationary fluctuations place the protons in black holes or deep voids before
%they have time to form galaxies. 
The lower cutoff in the $\wQ(Q)$-factor reflects gas
cooling physics related to galaxy formation and the upper cutoff incorporates effects that are relevant if
the reference objects are stable planets or observers \cite{Q}.

Since the inflationary parameters $\Ntot$, $Q$ and $\rhol$ upon which equations\eqn{wEq}-(\ref{wQeq}) depend
are correlated with our other five parameters, applying these selection effects modifies the observational 
predictions for all eight inflationary parameters in Table 1 (and also for $\xi$ if this can vary from basin to basin).
Arguably our most striking result in this section will be that 
when taking the selection effects into account, 
broad classes of inflation models are ruled out by the low observed CMB fluctuation amplitude 
$Q\sim 2\times 10^{-5}$ --- we will refer to this as the {\it smoothness problem} below.

In the next four subsections, we will derive and discuss the weight factors in equations\eqn{wEq}-(\ref{wQeq}).
We will then discuss the implications in sections~\ref{HorizonSec} and~\ref{SmoothnessSec}.
\Fig{1DconditionedFig} shows an example of the final outcome of these calculations: 
inflationary predictions with these weight factors (selection effects) included.

\subsection{Volume weighting and conditioning on protons}

Early on, the number density of quarks (later to end up bound into protons and neutrons) 
is comparable and proportional to the number density of photons $\ng$. Then matter and antimatter annihilate, leaving only a 
small residual proton density  
$n_p=\eta\ng$, where the baryon-to-photon ratio $\eta\approx 6\times 10^{-10}$ (in terms of the parameters in Table 1, $\eta=\xib/m_p$).
As long as the physics of baryogenesis that determines $\eta$ has nothing to do with inflation, 
the number of protons $N_p$ in a given comoving region of space that had volume $\Vreheat$, temperature $\Treheat$
and density $\rreheat$
at the end of reheating is thus given by 
\beq{NpEq}
N_p = \Vreheat n_p\propto \Vreheat\ng \propto \Vreheat\Treheat^3 \propto \Vreheat\rreheat^{4/3}.
\eeq
If the reheat density is independent of the basin into which we roll down (hence constant across our ensemble)
and $\rho\propto a^{-3}$ during reheating as usual \cite{LiddleLeach03}, then
the number of protons is thus proportional to the physical volume at the end of inflation.
The key question is whether this justifies including the volume factor of \eq{wEq} 
in the weight function $w(\p)$ of \eq{fpEq}.

When comparing finite volumes, the answer appears to be a clear yes, since 
there will be only a finite number of protons and hence no ambiguities associate with
their ordering in \eq{OrderingEq}.
This implies that when comparing finite and infinite pockets, the finite ones get zero statistical
weight since the infinite ones contain almost all reference objects.
In terms of the previous section, it argues for Measure A over Measure B.
% all but a measure zero.
% So using this for finite volumes is enough to prove that volume produced is infinite
% (only need to show that for any finite number of protons made, there's always one more.
When comparing two different infinite volumes, the answer is less clear, and is equivalent to
specifying the inter-pocket weighting.
Are all infinite volumes equal, or are some more infinite than others?
We will return to this important question in \Sec{AnalyticSec}.

\subsection{$\whalo$: conditioning on halos}
\label{whaloSec}

This particular selection effect (that we are observing from a region of space where 
gravitationally bound object have been able to form) was the one used to obtain the classic 
anthropic upper bounds on the dark energy density $\rhol$ 
\cite{BarrowTipler,LindeLambda,Weinberg87,Efstathiou95,Vilenkin95,Martel98,GarrigaVilenkin03}.
Its inclusion is clearly not optional when testing theories where $Q$, $\rhol$ or $\xi$ vary,
and it is more elegant than many other selection effects
in that it is physically clean, involving only well-understood gravitational physics and no uncertainties
related to the formation of galaxies, planets of observers. Any future definition of an observer is likely
to require the inclusion of the $\whalo$-factor, since no complex objects whatsoever (let alone observers) ever 
form in the linear regime.

The first mass scales $M$ (if any) to go nonlinear are the smallest, which enter the horizon well before matter-radiation 
equality with a fluctuation amplitude $\sigma_*(M)\sim Q$ and then grow over time 
due to gravitational instability as (see Appendix A of \cite{anthroneutrino})
\beq{sigmaMeq}
\sigma(M)\approx \left[1+{3\over 2}\left({\rhomeq\over\rhol}\right)^{1/3}\Gl\left({\rhol\over\rhom}\right)\right]\sigma_*(M),
\eeq
where the dimensionless function 
\beq{GlambdaFitEq}
\Gl(x)\approx x^{1/3}\left[1+\left({x\over G_\infty^3}\right)^{0.795}\right]^{-{1\over 3\times 0.795}}
\eeq
describes how fluctuations grow as the cosmic scale factor $a$ as long as dark energy is negligible 
($\Gl(x)\approx x^{1/3}=(\rhol/\rhom)^{1/3}\propto\rhom^{-1/3}\propto a$ for $x\ll 1$) and then asymptote to 
a constant value as $a\to\infty$ and dark energy dominates: $\Gl(x)\to G_\infty$ as $x\to\infty$,
where 
\beq{GmaxEq}
G_\infty\equiv {5\Gamma\left({2\over 3}\right)\Gamma\left({5\over 6}\right)\over 3\sqrt{\pi}}\approx 1.43728.
\eeq
\Eq{GlambdaFitEq} assumes negligibly small neutrino masses, $\rhol\ge 0$ and 
a flat Universe --- we will discuss the $\rhol<0$ case below and include the effect of spatial curvature in \Sec{BlackHoleSec}.
Neutrino masses lower the $\rhol$-predictions marginally \cite{anthroneutrino,anthrolambdanu}, but within the stringent
current observational limits \cite{sdsslyaf}, they have negligible effect on the qualitative results of this paper.

In Planck units, the matter density at matter-radiation equality is 
\beq{rhomeqEq}
\rhomeq = \ng^{\rm eq}\xi = {2\zeta(3)\over\pi^2}\Teq^3\xi,	
\eeq
where the Riemann zeta function $\zeta(3)\approx 1.20206$ and the matter-radiation equality temperature is 
\beq{TeqEq}
\Teq = {30\zeta(3)\over\pi^4}{\left[1+{21\over 8}\left({4\over 11}\right)^{4/3}\right]^{-1}}\xi\approx 0.220189\xi,
\eeq
where the second term in square brackets reflects the contributions from neutrinos to the radiation density.
\Eq{sigmaMeq} thus shows that the net fluctuation growth is controlled by the factor
\beq{AvacEq}\left({\rhomeq\over\rhol}\right)^{1/3}\approx 0.13751{\xi^{4/3}\over \rhol^{1/3}},
\eeq
observed to be $3215\pm 639$ \cite{sdsspars,anthroneutrino},
which is the factor by which the universe expanded between matter domination 
(when growth effectively begins) and dark energy domination (when growth effectively ends).

We approximate the fraction of matter collapsed into dark matter halos by
the standard
Press-Schechter formalism \cite{PressSchechter}, which gives
\beq{whaloEq2}
\whalo(\p)=\erfc\left[{\delta_c\over\sqrt{2}\sigma(M)}\right]
%\approx \erfc\left[{\delta_c\over{3\over 2}\sqrt{2}\left({\rhomeq\over\rhol}\right)^{1/3}G_\infty\sigma_*(M)}\right],
%\approx \erfc\left[{1.133928895\rhol^{1/3}\over{3\over 2}\sqrt{2}1.43728*0.13751\xi^{4/3} \sigma_*(M)}\right],
\approx \erfc\left[{2.7046\rhol^{1/3}\over\xi^{4/3} \sigma_*(M)}\right],
% calc 1.133928895/(sqrt(2)*1.5*1.43728*0.13751)
% ~ 2.704606056
\eeq
where $\delta_c=(9/5) 2^{-2/3}\approx 1.63$ (not 1.69) is the fluctuation threshold corresponding
to the infinite future (see \cite{Weinberg87} and appendix D.4 of \cite{anthroneutrino}).
In the second step of \eq{whaloEq2}, we used equations\eqn{GlambdaFitEq},\eqn{GmaxEq} and\eqn{AvacEq}
and took the limit $t\to\infty$, \ie, $\rhom\to 0$.
To derive \eq{whaloEq} from \eq{whaloEq2}, all that remains is to relate the initial fluctuation amplitude $\sigma(M)$ to $Q$.
For the small scales relevant to galaxy halos, 
the ratio $\sigma(M)/Q$ is a constant of order unity which depends very weakly (roughly logarithmically) 
on the mass scale $M$. To be specific, we use the observed ratio $\sigma(M)/Q\approx 29$
corresponding to a galactic halo scale $M=10^{12}M_{\odot}$ \cite{sdsspars,anthroneutrino}.
% (0.000579\pm 0.000064)/0.00002 ~ 28.95000305.
% $\sigma(M) =\Delta(k) = (ck/H)^2 Q T(k)$
% $\sigma(M)/Q = (ck/H)^2 T(k)$
% calc 1.133928895/(sqrt(2)*1.5*1.43728*0.13751*29)       ~ 0.09326227941 ~ 0.1
% calc 1/(1.133928895/(sqrt(2)*1.5*1.43728*0.13751*29))   ~ 10.7
% set eta = 3e-28
% set rhol = 
% set eta  = 6.4e-10
% set rhol = 9.3e-124
% set xi  = 3.3e-28 
% set Q   = 2e-5
% set z = rhol**(1/3)/(xi**(4/3)*Q)/sqrt(2)
% set Qstar = 0.1*rhol**(1/3)/xi**(4/3)
This particular choice of scale is irrelevant to the qualitative conclusions we will draw in this paper, \
since slight variations in the numerical prefactor $0.1$ in \eq{whaloEq} are dwarfed 
by the order-of-magnitude changes in $\rhol$ and $Q$.

\Eq{whaloEq} applies to the case $\rhol\ge 0$ and gives $\whalo\approx 0$ if $\rhol\gg\xi^4 Q^3$.
%We will return to the case $\rhol<0$ below. 
If $\rhol\ll -\xi^4 Q^3$, then space will recollapse
before any reference objects have had time to 
form \cite{BarrowTipler,LindeLambda,Weinberg87,Efstathiou95,Vilenkin95,Martel98,GarrigaVilenkin03}.
Although the physics is more complicated than for the $\rhol>0$ case (since fluctuations {\it do} have time to go nonlinear
before the Big Crunch), 
there is no indication that the constraints on negative values of $\rhol$ are much weaker than 
those on positive values.
This implies that 
\beq{whaloEq3}
\whalo(\p)\approx\left\{\begin{tabular}{lll}
%0\quad\hbox{if}\quad |\rhol|\gg\xi^4 Q^3,
$1$	&if	&$|\rhol|\ll\xi^4 Q^3$,\\
$0$	&if	&$|\rhol|\gg\xi^4 Q^3$,
%$(\rhol+\Lmin)^{1+k\over 3}$ if $k\le -1$,\\
%$(\rhol^{-1}+\Lmax^{-1})^{-(1+k)\over 3}$ if $k\ge -1$,
\end{tabular}\right.
\eeq
in Planck units,
which is all we need for the qualitative conclusions of the present paper.
The origin of this result is that the 
parameter combination $\rhol/\xi^4 Q^3$ is simply the 
ratio of the dark energy density to the matter density at the epoch 
when halos would form if $\rhol=0$. Regardless of the sign of $\rhol$,
dark energy domination is when time starts running out for the formation of reference objects:
either quite literally for the $\rhol<0$ case (because space recollapses) or for the $\rhol>0$ case
(because fluctuation growth stops).
% typically form either before dark energy domination or not at all.

\subsection{$\wN$ and $\wQ$: conditioning on galaxies}

Above we computed the weight factor $\whalo$ that entered when the reference objects were
dark matter halos.
Let us now compute the additional weight factors that enter when 
the reference objects are galaxies. We will see that they depend primarily on $\Ntot$ and $Q$.

\subsubsection{When galaxy-scale fluctuations were generated}

%Making the standard assumption that that $\rho\propto a^{-3}$ during reheating
%and $\rho\propto a^{-4}$ during radiation domination (\fig{arhoFig}), one readily
%obtains \cite{LiddleLeach} 
%\beq{AendEq}
%{\aend\over\aeq}=\left({\rhoend\over\rhoeq}\right)^{-1/4}\left({\rhoend\over\rhoreheat}\right)^{-1/12}.
%\eeq

Consider a comoving Hubble volume (either ours or a smaller one) 
that left the horizon at $\aexit$ during inflation (open triangle in \fig{arhoFig}, say)
and reentered at $a_0$ during matter domination (filled triangle, say).
As discussed in \Sec{Vsec} and Appendix A,
these two points are connected by $a_0^2\rho(a_0)=\aexit^2\rho(\aexit)$.
By making the standard assumptions that 
$\rho\simpropto a^{-3}$ during matter domination, 
$\rho\simpropto a^{-4}$ during radiation domination,
$\rho\simpropto a^{-3}$ during reheating and
$\rho\simpropto a^0$ during inflation (the five above relations 
are illustrated in \fig{arhoFig} by lines of slope 
$-3$, $-4$, $-3$, $0$ and $-2$), one readily derives the following expression for
the number of e-foldings $N_0$ between horizon exit and the end of inflation \cite{LiddleLeach03}:
\beq{N0eq}
N_0=\ln{\aend\over\aexit}\approx
\ln\left[{\rinf^{1/6}\rhoreheat^{1/12}\over\rhoeq^{1/4}}\left({a_0\over\aeq}\right)^{1/2}\right].
\eeq
Working in Planck units, combining \eq{rhomeqEq} with \eq{TeqEq} shows that $\rhoeq\sim\xi^4$.
We use the symbol $\sim$ to indicate that we are ignoring numerical factors of order unity for simplicity, 
since \eq{N0eq} shows that they affect $N_0$ only logarithmically.
%  the dependence on $\rhoeq$ in \eq{N0eq} is only logarithmic.
The number of e-foldings $\Neq$ between horizon exit of the matter-radiation equality scale 
(open square in \fig{arhoFig}) and the end of inflation (five-pointed star) is thus simply
\beq{NeqEq}
\Neq\approx
\ln\left({\rinf^{1/6}\rhoreheat^{1/12}\over\xi}\right).
%= \ln\left({\rinf^{1/6}\rhoreheat^{1/12}\over\xi}\right)
\eeq
Fluctuations that entered the horizon before matter-radiation equality 
grew during matter domination with amplitude of order $(a/\aeq)Q$, so 
the first nonlinear objects formed at the epoch $a\sim\aeq/Q$.
Since galaxies cannot form before this time, 
the number of e-foldings $\Ngal$ between horizon exit of the galaxy scale 
and the end of inflation is therefore at least 
\beq{NgalEq2}
\Ngal\approx\ln\left({\rinf^{1/4}R^{1/12}\over\xi Q^{1/2}}\right),
%\left({\rinf^{1/6}\rhoreheat^{1/12}\over\rhoeq^{1/4}}\right)\left({a_0\over\aexit}\right)^{1/2}.
%\ln\left({\rinf^{1\over 3}\over\rhoreheat^{1\over 12}\xi Q^{1\over 2}}\right)
%=\ln\left[{\rinf^{1\over 4}\left({\rinf\over\rhoreheat}\right)^{1\over 12}\over\xi Q^{1\over 2}}\right]
\eeq
where $R\equiv\rhoreheat/\rinf$ equals unity if reheating is instantaneous and decreases if the
duration of the reheating epoch grows.
For canonical values $\xi\approx 3.3\times 10^{-28}$, $Q\approx 2\times 10^{-5}$ and
$\rinf\sim (10^{16}\GeV)^4\sim 10^{-12}$, we thus obtain 
$\Ngal\approx 55$ for instant reheating $(R=1)$.
Lowering the reheat energy to the extreme value $1\,\TeV$ 
(arguably about the lowest energy that is still observationally allowed)
gives $\Ngal\approx 65$ and, conversely, lowering the inflation energy scale reduces
$\Ngal$.
% set lgrhoinf = -12
% set lgM = 0		# M = rhoinf/rhoreheat
% set lgxi = lg(3.3e-28)
% set lgQ  = lg(2.2e-5)
% set N = (lgrhoinf/2 + lgM/12 - lgxi - lgQ/2)*ln(10)
% M=0  ==> N~55
% Increasing lgM by 12/ln(10)~5.2 increases N by 1.
% rhoplanck ~ (1.2210e19 GeV)^4
% rhoinfl   ~ rhoplanck/10^12 ~ 1e+16 GeV
% Pushing rhoreheat down to (1 TeV)**4 gives lgM ~ lg(1e16/1e3)**4 ~ 52, boosting N by ~10.
% THIS IS Ngal, NOT THE SAME THING AS THE USUAL N.

\subsubsection{The black hole constraint on galaxy formation}
\label{BlackHoleSec}

In the context of galaxy formation, there are 
two sources of fluctuations, one good and one bad:
\begin{enumerate}
\item Inflation-generated fluctuations that are of order $Q$ when they enter the horizon.
\item Pre-inflationary fluctuations that we assume to be of order unity when they enter the horizon.
\end{enumerate}
Consider what happens in a spatial region that inflated by only a finite number of
e-foldings $\Ntot$ that exceeds $\Neq\sim 50$, say. 
Since inflation diluted away sub-horizon pre-inflationary fluctuations, the first fluctuations to
enter the horizon after matter-radiation equality have amplitude $Q$, then grow $\propto a$ due to 
gravitational instability as long as $\rhol$ is negligible, perhaps forming galaxies if they are able 
to grow enough to go nonlinear. If $\rhol=0$, then the horizon will keep growing until,
eventually, pre-inflationary fluctuations of order unity start entering the horizon.
This is bad news, since once the current horizon scale has gone nonlinear, the horizon volume
lies within its own Schwarzschild radius and will recollapse into a giant black hole on 
the local Hubble timescale, \ie, roughly by the time
the cosmic age has doubled.\footnote{Note that fluctuations of order unity in a 
flat Friedman-Robertson-Walker background only form black holes
if they are on or above the {\it current} horizon scale.
In contrast, fluctuations with $Q\ll 1$ grow to order unity only much later, when the current horizon
scale is vastly larger than the fluctuation scale. 
}
Even if the first order-unity fluctuations to enter the horizon happen to be negative
and void-like, still larger-scale fluctuations will keep entering and ensure that 
our local spatial patch is engulfed by a black hole before long.
In summary, horizon-sized black holes form at the epoch 
\beq{abhEq}
\abh\sim\aeq e^{2(\Ntot-\Neq)}
\eeq
when pre-inflationary fluctuations start entering the horizon.

If our reference objects are merely nonlinear structures, then fluctuations of 
the above-mentioned types 1 and 2 are equally 
useful for producing them, and there is no constraint whatsoever on $\Ntot$, the duration of inflation.
Indeed, no inflation at all is needed, since objects would be nonlinear from the outset.
In contrast, reference objects such as galaxies, planets and observers all require extra time to form.
Specifically, they require a period of peace and quiet after their parent dark matter halo has gone
nonlinear, and will not have time to form if their parent halo is promptly swallowed by a black hole.
Such reference objects will therefore only form if halos form substantially
before the epoch $\abh$, \ie, if $\Ntot\simgt\Ngal(Q)$.
This constraint on the duration of inflation is embodied by \eq{wNeq}.

None of the qualitative conclusions in this paper depend on whether $\Ngal=50$, $60$, or some other number
of order $10^2$.
Since the dependence of $\Ngal$ on $Q$ is only logarithmic, \ie, much weaker than that of
the $\whalo$ factor of \eq{whaloEq}, we therefore made the crude approximation $\Ngal=55$ for 
simplicity in \Sec{MonteSec}.

\subsubsection{$w_Q$: The cooling constraint on galaxy formation}
\label{wQsec}

To end up in a galaxy, a proton must first fall into a dark matter halo (a fraction $\whalo$ factor does this),
then avoid promptly being engulfed by a black whole (a fraction $w_N$ avoids this).
Next, the gas to which it belongs needs to cool efficiently enough to be able to 
contract, become self-gravitating and form stars
(see, \eg, \cite{Binney77,ReesOstriker77,Silk77,WhiteRees78}).
This problem is analyzed in terms of fundamental physical constants in \cite{Q}.
If we decrease $Q$, then dark matter halos form later and have lower escape velocities,
hence heating infalling gas to lower temperatures where cooling is less efficient, causing 
the weight function $w_Q(Q)$ to fall off towards very low $Q$-values.
% set eta  = 6.4e-10
% set xi  = 3.3e-28 
% set fvir = 0.03
% set alphag = 5.90464e-39
% set Ob = 0.05
% set alpha = 1/137.0359895
% set beta = 1/1836.153 
% set Q1 = fvir**(2/3)*alpha**(-1)*ln(alpha**(-2))**(-16/9)*alphag*beta**(4/3)*xi**(-4/3)*Ob**(-2/3)
% set me = sqrt(alphag)*beta
% set mp = sqrt(alphag)
% set Q2 = 0.2*alpha**(-1)*ln(alpha**(-2))**(-16/9)*me**(4/3)*eta**(-2/3)*xi**(-2/3)
% calc Q1/Q2
After changing variables to reflect the notation in this paper, equation (11) in \cite{Q} states
that $w_Q(Q)\approx 0$ unless
\beq{QcoolingEq}
Q\simgt 0.2\alpha^{-1}\ln[\alpha^{-2}]^{-16/9} m_e^{4/3}\eta^{-2/3}\xi^{-2/3}\sim 10^{-6},
%Q\simgt \alpha^{-1}\ln[\alpha^{-2}]^{-16/9} \alpha_g\beta^{4/3}\xi^{-4/3}\tvirfudge^{2/3}\Omega_b^{-2/3}
\eeq
where 
$\alpha\approx 1/137.036$ 
%$\alpha\approx 1/137.0359895$ 
is the fine structure constant and
$m_e\approx 4.2\times 10^{-23}$ is the electron mass in Planck units.

\subsection{$\wQ$: conditioning on stable planets}

If the reference objects are observers who, like us, have required 
billions of years of evolution on a planet, 
we need to include an additional weight factor $w_Q(Q)$ incorporating
the fraction of the protons in galaxies that end up in planets in long-lived
stable orbits. As discussed in \cite{Q}, 
if we increase $Q$, then galaxies form earlier with higher stellar density,
increasing the rate of catastrophic orbit disruptions from near encounters
with other stars.
Equation (19) in \cite{Q} states
that $w_Q(Q)\approx 0$ unless 
\beq{QdistruptionEq}
Q\simlt 10^{-4}.
\eeq
In summary, combining equations\eqn{QcoolingEq} and\eqn{QdistruptionEq} suggests
that $w_Q(Q)\sim 0$ unless $10^{-4}\simlt Q\simlt 10^{-6}$.
However, as detailed in \cite{Q}, the physical assumptions underlying the calculations
of $w_Q$ are not nearly as clean and clear-cut as those underlying, say, 
the $\whalo$-factor, so this conclusion should be taken with a healthy grain of salt,
and deserves further investigation.

% The only cosmological parameters upon which this depends are $\xi$ and $\eta$.

\subsection{Predictions for the horizon problem}
\label{HorizonSec}

In the four preceding subsections, we derived the weight factors in equations\eqn{wEq}-(\ref{wQeq}).
Let us now explore their implications, starting in this section with \eq{wNeq}.

Inflation solves the horizon problem if $\Ntot\simgt N_0$, so
that all fluctuations that have entered our horizon so far have 
inflationary rather than pre-inflationary origin.
In \Sec{CurvatureSec}, we saw that if inflation lasts for an additional few e-foldings, 
it solves the flatness problem as well: since $\Nbefore=\Ntot-N_0$, 
we found that $\Nbefore\simgt 0$ solves the horizon problem and $\Nbefore\simgt 3$
reproduces the observed spatial flatness.

One of the original motivations for 
the early work on
% inventing 
inflation was that it {\it could} solve the horizon problem and the flatness problem 
\cite{Guth81,Starobinsky1980,Linde82,AlbrechtSteinhardt82,Linde83}
(for certain potentials).
To what extent can we make the prediction that inflation {\it does} solve these problems?

When the total number of e-foldings 
$\Ntot$ varies across the ensemble,
% as for the second measure of \Sec{MonteSec}, 
it will generally exceed
$\Nnow$ in some spatial regions but not in others.
For observations made from galaxies, we automatically have $\Nnow\ge\Ngal$,
but the existence of galaxies guarantees only that $\Ntot>\Ngal$, not that $\Ntot$ also exceeds the
larger quantity $\Nnow$ corresponding to the current observation time.
The probability of observing the horizon problem to be solved is thus
the probability of observing $N>\Nnow$ given that $N>\Ngal$, \ie,
\beq{HorizonProbEq}
\Phorizon = P(\Ntot>\Nnow|\Ntot>\Ngal) = {P(\Ntot>\Nnow)\over P(\Ntot>\Ngal)}.
\eeq
In \Sec{MonteSec}, we found that Measure A predicted 
$\Ntot=\infty$, automatically solving the horizon and flatness problems, whereas our other measure
gave a probability distribution for $N$ falling off roughly as 
$\Ntot^{-\gamma}$ with $\gamma\approx 3$ in the worst-case scenario $\mh\ll\mbar$ (plotted in \fig{NhistFig}).
\Eq{N0eq} shows that $N_0-\Ngal={1\over 2}\ln(1+\zgal)\sim 1$, 
where $\zgal$ is the galaxy formation redshift.
For $\Ngal=55$, \eq{HorizonProbEq} therefore gives
\beq{HorizonProbEq2}
\Phorizon =
\left\{\begin{tabular}{l}
$1$ with volume-weighting,\\
$\left({\Nnow\over\Ngal}\right)^{-\gamma}\approx 1-{\gamma\over 2}{\ln(1+\zgal)\over\Ngal}\sim 95\%$ otherwise.
\end{tabular}\right.
\eeq
% calc 1-(3/2)*ln(1+10)/55
For either measure, inflation thus {\it predicts} that the horizon problem is solved from the vantage point of
most galaxies. If we were to retrodict whether our own horizon-scale fluctuations were inflationary rather than
pre-inflationary, varying $\zgal$ over the observationally rather extreme range $1<\zgal<100$ would 
give $87\%<\Phorizon<98\%$, so our qualitative conclusion that $\Phorizon\sim 1$ is independent of
uncertainties regarding what to take as the present epoch.
We can also make this argument in a less anthropocentric way, noting that since the $\whalo$-factor tends to postpone
dark energy domination tends to only slightly 
after galaxy formation, the horizon stops growing and precludes observers from observing more
than a few e-foldings beyond $\Ngal$ no matter how long they wait.

In conclusion, none of the inflation models we have considered are ruled out by the horizon problem.
In \Sec{MonteSec}, we saw that they were not ruled out by the flatness problem either.

\subsection{The smoothness problem and other implications}
\label{SmoothnessSec}

If a physics problem involves two vastly different scales, this fact can often be used to simplify it.
We will now see that precisely this happens in our inflationary prediction problem, 
because the inflation density vastly exceeds the dark energy density $\xi^4 Q^3\mpl^4$
allowed by \eq{whaloEq3}. Conveniently, this will allow us to deal analytically with the
$\rhol$-parameter and dispense with it once and for all. 

In \Sec{MonteSec}, we saw that inflation typically ended at a density $\rhoend\sim\mv^4$ for $\mh\simlt\mbar$ and 
$\rhoend\sim(\mh/\mbar)^2\mv^4$ for $\mh\simgt\mbar$. For any relevant values of $Q$, $\mh$ and $\mv$, both of these two 
densities vastly exceed the density scale $\xi^4 Q^3\mpl^4\sim 10^{-110}Q^3\mpl^4$ above which 
% set xi  = 3.3e-28 
% set Q   = 2e-5
% calc xi**4 * Q**3 ~ 1.185921e-110
\eq{whaloEq} cuts off the statistical weight (no halos form). The lower limit on negative $\rhol$-values 
from requiring reference objects to form before space recollapses is of comparable magnitude.
This means that the only part of the probability distribution for $\rhol=V(\phistop)=\mv^4 f(\phistop/\mh)$ that matters is 
that in a region around zero which is way smaller in magnitude than any relevant physical scales. Since the dimensionless inflaton potential 
$f$ by definition varies of order unity and zero is not in any way a special value for a random minimum of a complicated function,
it follows that we can for all practical purposes take $\rhol$ to have a uniform distribution, uncorrelated with all the other parameters. 
This prediction is confirmed by our numerical results and illustrated by the $\rhol$-histograms in 
figures (\ref{1DcomboBasinFig}) and~(\ref{1DcomboFig}).

\subsubsection{Marginalizing over $\rhol$}

Since $\whalo$ is the only one of the weight functions from equations\eqn{wEq}-(\ref{wQeq})
that depends on $\rhol$, this result means that we can 
once and for all marginalize over $\rhol$ analytically and eliminate it from the discussion of our other cosmological parameters.
The resulting parameter distribution observed from a random halo is 
\beqa{fpEq3}
\fhalo_p(\p)	&\propto&\int_{-\infty}^\infty\finf_p(\p)\whalo(\p)d\rhol\nonumber\\
		&\approx&\int_{-\infty}^\infty \finf_p(\p)_{|\rhol=0}\> g\left({\rhol\over\xi^4 Q^3}\right) d\rhol\nonumber\\
		&\propto&\xi^4 Q^3  \finf_p(\p)_{|\rhol=0}.
\eeqa
On the second line, we used the above uniform distribution result to
approximate the first factor $\finf_p(\p)$ by its value at $\rhol=0$ because it is $\rhol$-independent 
whenever the second factor is appreciably nonzero.\footnote{
Because our simulations predict the $\rhol$-distribution to be uniform over a much broader range ($|\rhol|\ll\rhoend$) than the range forming 
halos, it would be numerically inconvenient to apply \eq{whaloEq} to our simulations: it would give a 
success rate $\ll 10^{-100}$ and require many Hubble times worth of computer simulating to obtain even a single successful model.
Requiring $\rhol=0$ as in \eq{fpEq3} is of course even less convenient numerically.
In practice, we therefore use the result of \eq{fpEq3} with
the constraint $\rhol=0$ replaced by $0\le \rhol<\rhol^*$, where 
$\rhol^*\ll\rhoend$. We choose the cutoff $\rhol^*$ sufficiently  small that reducing it further has
no effect on our predicted distributions, but large enough that we retain a statistically
large sample of models.
}
Here we also reexpressed $\whalo(\p)$ as $g(\rhol/\xi^4 Q^3)$ to emphasize that 
the final result in \eq{fpEq3} applies to much more general weight functions $\whalo(\p)$ than
of that of \eq{whaloEq} --- which corresponds to the particular case $g(x)=\erfc(0.1 x^{1/3})$.
Replacing the Press-Schechter formula by the more accurate approximation of Sheth \& Tormen \cite{ShethTormen99} would merely
modify the functional form of $g(x)$, not the final result in \eq{fpEq3}.
More importantly, modifying $\whalo(\p)$ to explicitly give zero weight to models where space recollapses 
before reference objects form (which is necessary to prevent the integral in \eq{fpEq3} from diverging as $\rhol\to -\infty$)
would again simply modify the functional form of $g(x)$, this time to ensure that the condition $g(x)\approx 0$  
for $|x|\gg 1$ held also for negative $x$, as per \eq{whaloEq3}.
%The reason for this is that the 
%parameter combination $\rhol/\xi^4 Q^3$ is simply the 
%ratio of the dark energy density to the matter density at the epoch 
%when halos would form if $\rhol=0$ --- regardless of the sign of $\rhol$, 
%halos typically form either before dark energy domination or not at all.

\begin{figure} 
%\vskip\smtopskip
\centerline{\epsfxsize=\figsize\epsffile{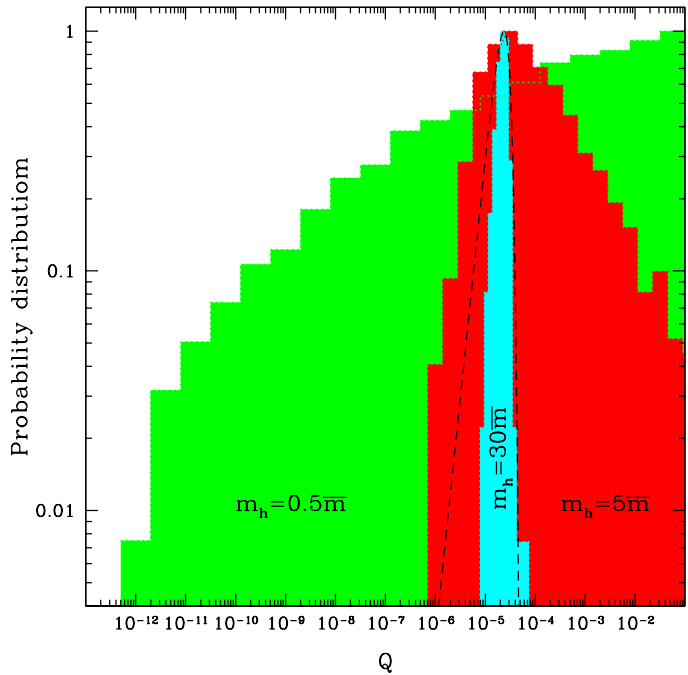}}
%\vskip\smbotskip
\caption[1]{\label{QhistFig}\footnotesize%
The smoothness problem is seen to afflict 
quantum-gravitationally ``natural'' inflation models with
$\mh\ll\mpl\approx 5\mbar$, predicting $Q$ way above the observed value 
$Q\sim 2\times 10^{-5}$ since $\rhol$-marginalization multiplies the above 
distributions by an extra $Q^3$-factor.
The three histograms are for Measure A and 
show the distribution of $\lg Q$ 
for $\rhol\approx 0$, marginalized over all other cosmological parameters.
The $\mh=0.5\mbar$ distribution is seen to approach 
the analytic $\mh\to 0$ prediction of a uniform distribution in the high $\lg Q$ tail, 
and the $\mh=30\mbar$ curve is seen to be approximated by the 
analytic $\mh\to\infty$ prediction $\propto Q^2 e^{-(Q/Q_*)^4}$ for $Q_*\approx 5\mv^2/\mh$
near the peak (dashed line).
}
\end{figure}

The $Q^3$-factor on the last line of \eq{fpEq3} poses great difficulties for 
many inflation models, since it can easily overpower a slight inflationary 
preference for low $Q$-values in $\finf(\p)$ (\fig{QhistFig}) and give predictions far exceeding the
observed value $Q\sim 2\times 10^{-5}$.
As a concrete illustration of this {\it smoothness problem}, consider 
the models with $\mh\simlt\mbar$ from \Sec{MonteSec}. These were the only models that were 
natural, in the sense that quantum gravity corrections may preclude larger $\mh$-values.
We found that all these models predicted a $Q$-distribution with a high tail 
falling off roughly as $f_Q(Q)\propto Q^{-1}$, corresponding to a near-uniform distribution for $\ln Q$,
as confirmed numerically in \fig{QhistFig}.
The basic reason for this was that 
$Q\sim V^{3/2}/|V'|$, and that $V'$ was often exponentially small.
% $\ln Q\sim (3/2)\ln V - \ln|V'|$, and that $V'$ had a uniform distribution near zero.
\Eq{fpEq3} now shows that these models predict a {\it rising} tail, 
with a random halo seeing a distribution $f_Q(\p)\propto Q^2$ for large $Q$.
Quantum diffusion cuts off this distribution at $Q\sim 1$, so 
$Q$ observed from a random halo would be of order unity.
\Eq{mvExtremeEq} shows that this result is independent of the choice of the vertical energy scale $\mv$ unless
it is down in the TeV range --- for that extreme case, equations\eqn{hill_Qeq} and\eqn{hill_nsEq})
show that $\eta_0\approx -1$ and $\ns\approx -1$.

All these $\mh\ll\mbar$ models are thus firmly ruled out by the observed measurement $Q\sim 2\times 10^{-5}$
unless they are rescued by the weight function $\wQ(Q)$.\footnote{The
$\wN$-factor of \eq{wNeq} provides essentially no help --- although it does depend on $Q$
for one of the two measures we studied, the
dependence is very weak because of the logarithm and the broad intrinsic width of the $N$-distribution.
}
However, as discussed in \cite{Q} and \Sec{wQsec}, the anthropic upper limits on $Q$ from planetary orbit
disruptions and other effects appear rather weak, both in terms of physical uncertainties
and in terms of the actual numbers. In particular, there is currently no compelling 
argument for why there would be much fewer observers if $Q$ were say 3 times larger than observed,
yet these inflation models predict the $(\ln Q)$-distribution $\propto Q^3$ so that
this larger $\ln Q$-value would be a priori about 30 times more likely. Having
$\ln Q$ up by an order of magnitude (bumping up against the disruption limit of \cite{Q})
would be a priori about 1000 times more likely.
It would be much more satisfying if this smoothness problem could be solved by 
a well-motivated inflation model firmly predicting smaller $Q$-values.

Analogously, one can argue that \eq{fpEq3} rules out any particle physics model 
for the matter density parameter parameter $\xi$ if it predicts an ensemble with a high-$\xi$ tail 
shallower than about $\xi^{-5}$, unless galaxy formation can be shown to be strongly inhibited by 
small increases in $\xi$ --- for a discussion of such 
dark matter issues, see \cite{conditionalization,Wilczek04}.

\begin{figure} 
%\vskip\smtopskip
\centerline{\epsfxsize=\figsize\epsffile{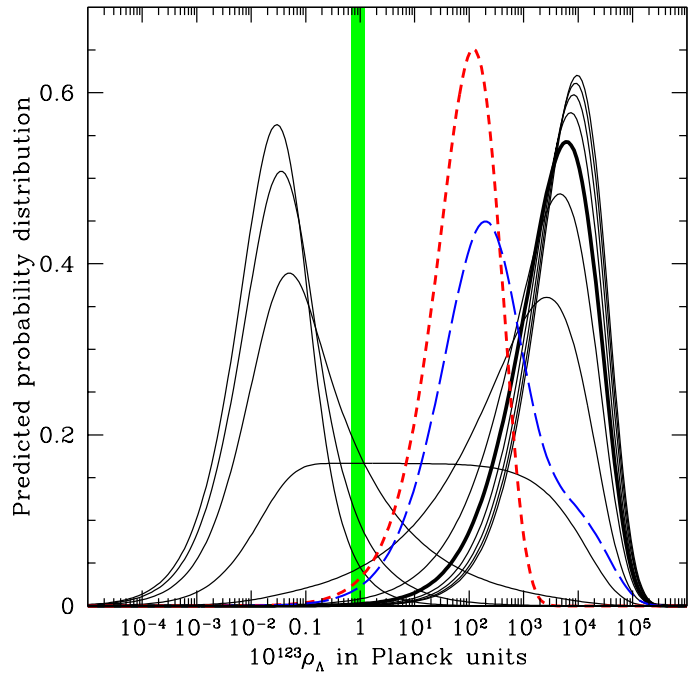}}
%\vskip\smbotskip
\caption[1]{\label{rholmargFig}\footnotesize%
Predictions for the dark energy density $\rhol$ for the toy model where
reference objects get produced if $10^{-6}<Q<10^{-4}$ and 
inflation predicts a power law $f_Q(Q)\propto Q^n$ across this range.
Peaking from left to right, the solid curves correspond to 
$n=-8$, $-7$, $-6$, $-5$, $-4$, $-3$, $-2$, $-1$ (heavy), $0$, $1$, $2$ and $3$,
respectively. The short-dashed curve shows the case where inflation predicts 
$Q=2\times 10^{-5}$ without any scatter, and the long-dashed curve is for the actual 
$Q$-distribution for $\mh=10\mbar$, $\mv=0.01\mbar$ from \fig{1DconditionedFig}. 
% 0.006741645746*sqrt(2) ~ 0.009534125216
The shaded band shows the WMAP+SDSS measurement from Table 1.
}
\end{figure}

\subsubsection{Predictions for $\rhol$}

Above we marginalized over $\rhol$ to obtain predictions for the other parameters.
Let us now do the opposite, obtaining inflationary predictions for $\rhol$.
The inflationary prediction for the dark energy probability distribution $f_\Lambda(\rhol)$ 
is obtained by marginalizing $f_p(\p)$ from \eq{fpEq2} 
over the other seven parameters in the $\p$-vector, \ie,
\beq{LambdaPosteriorEq}
f_\Lambda(\rhol')\propto\int\finf(\p)w(\p)\delta(\rhol-\rhol') d^8 p.
\eeq
Above we found that $\finf(\p)$ was $\rhol$-independent over the narrow range of $\p$-values 
where $w(\p)$ was nonzero. %  so we can factor the $\finf$-term out of the integral as a mere constant - NO!
The only $\rhol$-dependence in \eq{LambdaPosteriorEq} therefore comes from the weight function $w(\p)$,
where $\rhol$ enters in the halo term $\whalo(\p)$ of \eq{whaloEq} and gets coupled with $Q$.
We can thus integrate trivially over all parameters except $Q$ in \eq{LambdaPosteriorEq}, obtaining
\beq{LambdaPosteriorEq2}
f_\Lambda(\rhol)\simpropto\int_0^\infty\finf(Q)\wQ(Q)\>\erfc\left[{0.1\rhol^{1/3}\over\xi^{4/3}Q}\right] dQ,
\eeq
where $\finf(Q)$ is the raw inflation prediction for $Q$ for $\rhol=0$
(\ie, $\finf(\p)|_{\rhol=0}$ marginalized over all parameters except $Q$), as plotted in \fig{QhistFig} and in 
the $Q$-panels of figures (\ref{1DcomboFig}) and~(\ref{1DcomboBasinFig}).
Since the values $Q_i$ in our simulations with $\rhol\approx 0$ are drawn from the distribution $\finf(Q)$,
we can evaluate this expression in practice as
\beq{LambdaPosteriorEq3a}
f_\Lambda(\rhol)\simpropto\sum_i\wQ(Q_i)\>\erfc\left[{0.1\rhol^{1/3}\over\xi^{4/3}Q_i}\right],
\eeq
where the sum is over those simulations for which $\rhol\approx 0$.
If inflation were to predict that $Q$ took a specific value $\Qstar$-value (\ie, if $\finf(Q)=\delta(Q-\Qstar)$,
as for a one-dimensional inflaton potential with only a single minimum that either was symmetric or
could only be slow-rolled to from one side, 
then \eq{LambdaPosteriorEq2} would give the classic prediction $\rhol\simlt 10^3\xi^4\Qstar^3$ 
\cite{BarrowTipler,LindeLambda,Weinberg87,Efstathiou95,Vilenkin95,Martel98,GarrigaVilenkin03,Q}
(more specifically, the dashed curve in \fig{rholmargFig}).
However, as stressed in \cite{Aguirre01}, degeneracies with other parameters can invalidate such arguments. 
Sure enough, in the cases we have considered in this paper, the distribution $\finf(Q)$ has a nonzero width,
and \eq{LambdaPosteriorEq2} shows that a tail towards large $Q$-values will give more statistical weight
to higher $\rhol$-values. 
This important coupling between $\rhol$ and $Q$ was ignored in the early work predicting $\rhol$-distributions.
It was emphasized in \cite{Q} and recently explored numerically in \cite{Graesser04} for some instructive
toy models, with the conclusion that it can invalidate anthropic explanations for $\rhol$ for a broad class of
primordial $Q$-distributions. \cite{Huang04} explores how $w\ne -1$ affects this link.
Using our results from \Sec{MonteSec}, we can for the first time 
explore this issue with $Q$-distributions from actual inflation calculations.

Before doing this, however, let us first build some 
intuition for \eq{LambdaPosteriorEq2}. Consider the toy model where
$\finf(\p)\propto Q^n$ for some power law index $n$ and the weight function $\wQ(Q)=1$ if 
$\Qmin\le Q\le\Qmax$, vanishing otherwise. Limiting our attention to the case $\rhol\ge 0$ and 
performing the integral in \eq{LambdaPosteriorEq2} numerically
now gives the curves shown in \fig{rholmargFig}, which are easy to understand intuitively if we
define the quantities $\Lmin\equiv\xi^4\QQQmin$ and $\Lmax\equiv\xi^4\QQQmax$.
Since $\erfc x\approx 1$ for $x\ll 1$ and  $\erfc x\approx 0$ for $x\gg 1$,
\eq{LambdaPosteriorEq2} shows that 
$f_\Lambda(\rhol)$ is constant if $\rhol\ll\Lmin$, vanishes if $\rhol\gg\Lmax$
and is otherwise approximated by
\beq{LambdaPosteriorEq3}
f_\Lambda(\rhol)\simpropto\int_{0.1\rhol^{1/3}\over\xi^{4/3}}^\Qmax Q^n dQ
\simpropto\left|1-\left({\rhol\over\Lmax}\right)^{n+1\over 3}\right|.
\eeq
\Fig{rholmargFig} shows the distribution of $\lg\rhol$ rather than $\rhol$, which introduces
an extra $\rhol$-factor since $d\rhol = \rhol d\ln\rhol$, making the exponent of the 
second term of \eq{LambdaPosteriorEq3} equal to
$(n+4)/3$. Whether $n$ is smaller or larger than $-4$ therefore determines the qualitative
behavior of the resulting distributions: \fig{rholmargFig} shows that 
for $n=4$, all values of $\lg\rhol$ between $\Lmin$ and $\Lmax$ are roughly equally 
likely, whereas smaller or larger $n$ cause the distributions to be peaked at 
$\Lmin$ and $\Lmax$, respectively. 
There are thus only three classes of theories that are not ruled out by the
observed $\rhol$-value (shaded band):
\begin{enumerate}
\item Theories where $n\sim -4$.
\item Theories where inflation predicts a very narrow distribution $\finf(Q)$ centered 
around or slightly below the observed $Q$-value.
\item Theories where the weight function $w(Q)$ is sufficiently sharply peaked
around the observed $Q$-value to overpower the pull towards $\Lmin$ or $\Lmax$.
\end{enumerate}
None of these classes are problem free.
Class 1 involves inflationary fine-tuning in the sense that it requires the sharp
inflationary prediction for $\lg Q$ to 
coincidentally fall within the range around $-5$ where $\wQ(Q)$ is substantial.
We argued in \Sec{wQsec} that $\wQ(Q)$ may be rather broad, suggesting that models in class 3 will 
only work if  $n$ is not too far from $-4$.
We saw in \Sec{MonteSec} that inflation models with a ``natural'' horizontal energy scale $\mh\simlt\mbar$ do
not fall into class 1, but rather that  
their raw predicted $Q$-values display a high $Q$ tail with $n\approx -1$ (heavy curve in \fig{rholmargFig}). 
Such models thus tend to overpredict $\rhol$ by orders of 
magnitude, which is yet another manifestation of the smoothness problem discussed in 
\Sec{SmoothnessSec}.

\begin{figure} 
%\vskip\smtopskip
\centerline{\epsfxsize=\figsize\epsffile{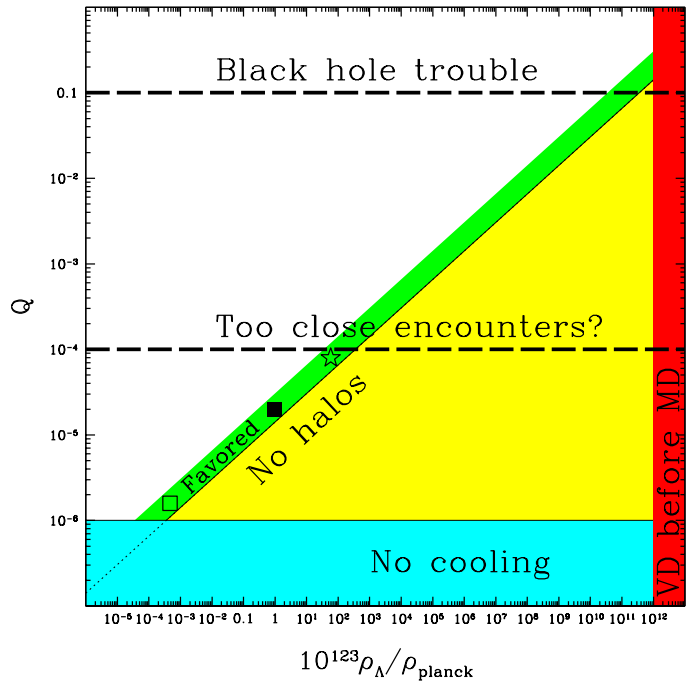}}
%\vskip\smbotskip
\caption[1]{\label{lqFig}\footnotesize%
The weight function $w(\p)$ (the number of reference objects produced per unit thermalized volume)
causes important modulations of the raw inflationary predictions
in the $(\rhol,Q)$-plane, with $\whalo(\p)$ giving essentially no weight to 
the region marked ``no halos'' and $w(Q)$ strongly downweighting regions 
more than about an order of magnitude above or below $Q=10^{-5}$.
The fact that inflation typically predicts $\rhol$ to have a uniform distribution in this range,
uncorrelated with $Q$, gives probability $\propto\rhol$ in this plane, so nearly all probability
lies within the diagonal green/grey favored band near the halo limit.
The logarithmic slope of the inflationary Q-distribution determines the most likely outcome,
with $d\ln\finf_Q/d\ln Q\ll -4$ giving the open square and $d\ln\finf_Q/d\ln Q\gg -4$
giving the star. The filled square shows the measurements from Table 1.
}
\end{figure}

\Fig{lqFig} is an attempt to illustrate the smoothness problem and this intimate relation between
$\rhol$ and $Q$. In summary, we have found no compelling solution to the smoothness problem.
High energy inflation models with $\mh\gg\mpl$ can have $\mh$ fine-tuned to fall into class 1, but 
may flounder on quantum gravity corrections. 
Low energy inflation models with $\mh\simlt\mpl$, on the other hand, predict $(\rhol,Q)$ near 
the five-pointed star in \fig{lqFig}, quite far from the observed value.
If a future candidate inflation model provides a compelling solution to the smoothness problem,
this will thus be a non-trivial and noteworthy achievement.

\subsection{Calculation summary}

Let us summarize the recipe that we have derived for computing the cosmological 
parameter distribution $f_p(\p)$. 
\begin{enumerate}
\item Use one of the two Monte Carlo approaches of \Sec{MonteSec} 
(depending on whether you wish to exclude finite pockets or not)
to generate a large number of parameter vectors $\p_i$ drawn from the raw inflationary 
distribution $\finf_p(\p)$, keeping only those with $\Ntot>\Ngal\approx 55$.
\item To approximate the $\rhol\to 0$ limit of \eq{fpEq3}, 
discard all vectors with $\rhol<0$ or $\rhol>\rhol^*$, reducing $\rhol^*$ until the results stop
changing.
\item Compute $f_\Lambda(\rhol)$ using \eq{LambdaPosteriorEq3a}.
\item Compute the distributions for all other parameters as
$f_p(p)\propto w_Q(Q) Q^3\finf_p(\p)_{|\rhol=0}$ by weighting the points
from Step 2 by $w_Q(Q_i) Q_i^3$ and then making histograms.
\end{enumerate}
In practice, Step 4 can be done either by unequal point weights $w_Q(Q_i) Q_i^3$ 
in the histograms or by
resampling, drawing a large number of points from Step 2 at random with probability $\propto w_Q(Q_i) Q_i^3$.
We use the latter approach.
\Fig{1DconditionedFig} shows a complete worked example for the case of 
Measure A, $\mh=10\mpl$ and $\mv=0.002\mpl$.

\section{Analytic Results}
\label{AnalyticSec}

In the two preceding sections, we computed cosmological parameter predictions for specific potentials and measures,
finding that unless $\mh\sim\mpl$, the dependence on the inflaton potential could be understood analytically
in terms of its peak and trough statistics.
We did this with pocket-based orderings, so to better understand how the choice of measure affects the
predictions, let us now revisit this computation using the other class of orderings from 
\Sec{OrderingSolutionSec}: global orderings.
%finding that it was quite difficult to come up with compelling inflation energy scales that matched
%observational constraints, particularly for the $Q$-parameter.
We will find that simple analytic calculations allow us to rule out a broad class of global orderings, notably 
many
% those 
that give attractor behavior and make the predictions independent of pre-inflationary initial conditions.

In \Sec{OrderingSolutionSec}, we saw that a global ordering is uniquely defined by specifying a ``time'' variable to order the
reference objects by. We found that all {\it observable} quantities (\eg, $\rho$, $T$ or $H$) are inadequate as 
such global time variables, leading instead to pocket-based orderings, so let us now explore time variables that are
{\it not observable} (except in differences or ratios, like $t$ and $a$). 

We will focus most of our discussion on the case of ordering by time $t$\footnote{
Specifically, we define $t$ as the time variable in synchronous coordinates where the metric is 
\beq{SynchronousMetricEq}
ds^2 = dt^2-a(\x,t)^2 d\x^2.
\eeq
The lines of constant $\x$ in this metric are timelike geodesics corresponding to the 
worldlines of {\it Gedanken} co-moving observers, and $t$ is the proper time of these observers.
This coordinate system remains well-defined until these
geodesics start to cross, which happens only long 
after thermalization when fluctuations go non-linear --- the coolness problem discussed
below becomes severe long before then, so it cannot be circumvented by a creative choice of gauge at late times.
When we discuss $t$ for a particle in the present epoch, the rigorously inclined 
reader can simply take this to mean its proper time, since this provides a well-defined ordering 
even after geodesic crossing. 
% By a synchronous gauge condition, I mean that each equal-time hypersurface is
% obtained by propagating every point on the previous hypersurface by a fixed infinitesimal time interval
% in the direction normal to the hypersurface.
}, then discuss how these conclusions can be generalized to other global orderings.
This natural-sounding choice is one of the first to have been explored in detail, 
and was used in some of the key early work on eternal inflation
\cite{LindeBook,Vilenkin83,Starobinsky84,Starobinsky86,Goncharov86,SalopekBond91,LindeLindeMezhlumian94}.

\subsection{The origin of attractor behavior}

One of the most appealing features of $t$-ordering is that it generically gives attractor behavior, 
allowing a unique parameter probability distribution $f_p(\p)$ to be computed independently of the initial 
conditions.
This is most elegantly seen using the 
Fokker-Planck equation formalism,
giving the time-evolution of $\Vphi(\vphi)$, the physical volume associated with different $\vphi$-values.
Thorough and detailed treatments of this are given in, \eg, 
\cite{LindeBook,Vilenkin83,Starobinsky84,Starobinsky86,Goncharov86,SalopekBond91,LindeLindeMezhlumian94}, so we will only 
summarize the results here and attempt to provide physical intuition for them.
The Fokker-Planck equation takes the form 
\beq{FokkerPlanckEq}
\dot\Vphi(\vphi) = L\Vphi(\vphi),
\eeq
where $L$ is a linear operator involving $0^{th}$, $1^{st}$ and $2^{nd}$ derivatives with respect to 
the inflaton field $\vphi$ to incorporate the effects of expansion, slow-roll and quantum diffusion, 
respectively.
The total volume at a given time is thus
\beq{FP_Veq}
V(t)\equiv \int\Vphi(\vphi,t)d^d\phi,
\eeq
which is conveniently partitioned as the sum of two contributions 
$\Vinf(t)$ and $\Vtherm(t)$ corresponding to integrating 
over the parts of $\vphi$-space where space is inflating and has thermalized, respectively.
(The standard prescription is to reclassify a volume of space from inflating to thermalized
as soon as its $\vphi$-value leaves the part of $\vphi$-space where the slow-roll approximation is valid 
and to not evolve it further \cite{LindeBook,Vilenkin83,Starobinsky84,Starobinsky86,Goncharov86,SalopekBond91,LindeLindeMezhlumian94},
so that $\Vtherm(t)$ is simply the volume of the U-shaped hypersurfaces
in \fig{OrderingFig} that lie in the part of spacetime with time coordinate $\le t$.)

If the total volume $V(t)$ is finite at some initial time $t$, then both $\Vinf(t)$ and $\Vtherm(t)$ 
will remain finite for all $t$ since the expansion rate $H$ cannot be infinite. 
Inflation is usually said to be eternal\footnote{
Throughout this paper, we use eternal to mean eternal to the future.
For recent discussions of whether inflation can be eternal to the past, see 
\cite{AguirreGratton02,AguirreGratton03,Guth00,Guth00b,Guth04,BordeGuthVilenkin03} and references therein. 
For an up-to-date discussion of initial conditions and whether they are needed, see \cite{Albrecht04}.
}
if $\Vinf(t)\to\infty$ as $t\to\infty$.
If the number of reference objects is proportional to the thermalized volume produced, \ie, $\Vtherm(t)$
(as is the case for points, protons, planets and all other reference objects we have discussed),
then inflation produces an infinite number of reference objects provided 
that $\Vtherm(t)\to\infty$ as $t\to\infty$.
Such eternal inflation with infinitely many reference objects occurs
quite generically, whenever the inflaton potential $V(\vphi)$ contains an ``eternal inflation region''.
It is well known that there are three types of eternal inflation regions \cite{Guth00,Guth00b,Guth04,Vilenkin04}:
\begin{enumerate}
\item A local maximum in $V(\vphi)$ where the slow-roll approximation is valid \cite{Vilenkin83,Steinhardt83}.
\item A $\vphi$-range where $V(\vphi)$ is large enough for quantum diffusion to dominate the dynamics \cite{LindeDiffusion86}.
\item A local minimum in $V(\vphi)$ from which $\vphi$ can eventually tunnel out (``old inflation'').
\end{enumerate}
In the first case, the volume rolling off the peak in various directions is more than replenished by
the expansion of the volume that has not yet rolled off.
In the second case, quantum diffusion up the hill causes faster 
expansion that more than compensates for the fact that it would have been more likely to roll downward
--- this happens in $\vphi$-regions where the slow-roll approximation predicts $Q\simgt 1$.

If there are multiple eternal inflation regions, the fastest expanding one (the one where $V(\vphi)$ is largest) always 
wins,
completely dominating the volume distribution $\Vphi(t)$ as $t\to\infty$.
\cite{LindeBook,Vilenkin83,Starobinsky84,Starobinsky86,Goncharov86,SalopekBond91,LindeLindeMezhlumian94}.
The reason is that if two volumes are eternally expanding with rates $H_1$ and $H_2$, and $H_1>H_2$, then
the lower second peak will asymptotically contribute a volume fraction 
%$V_2(t)/V_1(t)=V_2(0) e^{3H_2 t}/V_1(0)e^{3H_1 t} = {V_2(0)\over V_1(0)} e^{-3(H_1-H_2)t}\to 0$ 
\beq{AsymptoticsEq}
{V_2(t)\over V_1(t)}={V_2(0) e^{3H_2 t}\over V_1(0)e^{3H_1 t}} = 
{V_2(0)\over V_1(0)} e^{-3(H_1-H_2)t}\to 0 
\eeq
as $t\to\infty$ whatever the 
initial volume ratio $V_2(0)/V_1(0)$ was.\footnote{Measure ambiguities are thus linked to the fact that
we cannot reverse the order of the two limits $n\to\infty$ in \eq{OrderingEq} and $t\to\infty$.}

Moreover, the fastest expanding eternal inflation region of all (around $\vphi_*$, say) 
will completely dominate the asymptotic 
$t\to\infty$ volume distribution regardless of the 
initial conditions, \ie, for any initial distribution $\Vphi(\vphi)$.
This is because quantum diffusion will populate all $\vphi$-values 
and even an exponentially tiny contribution $\Vphi(\phi_*)$ created, say, by
exponentially suppressed uphill diffusion/tunneling at low density, will eventually overpower
all other slower expanding regions as $t\to\infty$ just as described above.
As a result, the asymptotic distribution $\Vphi(\phi)/V(t)$ will be determined simply by 
how long time it takes to roll/diffuse down from $\vphi_*$ to $\vphi$.
Since we are evaluating the distribution at fixed time, faster is better.
If a volume of space has the inflaton descending from $\phi_*$ along a trajectory $\phi(t)$ during
the time interval $t_1\le t\le t_2$, then 
it will expand by $\int_{t_1}^{t_2} H(\phi(t))dt$ e-foldings while the eternal region expanded more, by
$(t_2-t_1)H(\phi_*)$ e-foldings ($H(\phi)=\mbar^{-1}\sqrt{V(\phi)/3}$), 
so the asymptotic probability distribution is 
\cite{LindeBook,Vilenkin83,Starobinsky84,Starobinsky86,Goncharov86,SalopekBond91,LindeLindeMezhlumian94}
\beq{FPequilibEq}
\Vphi(\phi) = \Vphi(\phi_*)\exp\left[-3\int_{t_1}^{t_2}[H(\phi_*)-H(\phi(t))]dt\right].
\eeq
Since this asymptotic distribution satisfies ${\dot V}_\phi(\phi)=3H_*\Vphi(\phi)$,
it can be computed directly from the 
Fokker-Planck \eq{FokkerPlanckEq} as the solution to the eigenvalue problem 
$L\Vphi=3H_*\Vphi$
that has the largest eigenvalue $3H_*$. (If this largest eigenvalue is negative, inflation is not eternal.)
This eigenvalue problem can be transformed into a manifestly Hermitean form,
and has been shown to reduce to a Schr\"odinger equation \cite{WinitzkiVilenkin96}.

What we care about empirically is not the probability distribution for $\vphi$, but the 
parameter probability distribution $f_p(\p)$. The latter is readily extracted
from the former by keeping track of how inflation ends, \ie, by how the thermalized volume 
$\Vtherm(t)$ is distributed among the different basins of attraction and, for each basin, what
the probability distribution is for the direction from which it rolled down towards the minimum.
A mathematically equivalent alternative to solving the Fokker-Planck equation (see below) is to evolve the inflaton $\vphi$
using a stochastic ordinary differential equation 
that incorporates 
% both the causal effect of slow-roll and
the effects of quantum diffusion as a random noise term.

\subsection{What $t$-ordering predicts}

A strength of the $t$-foliation solution to the ordering problem is that it makes 
a specific and well-defined prediction for the parameter probability distribution $f_p(\p)$ which,
because of the attractor dynamics, is independent on initial conditions and unknown pre-inflationary
physics. Let us now discuss what this prediction is.

We saw above that in the $t\to\infty$ limit, all reference objects except for a set of measure zero 
are in comoving regions of space where $\vphi$ came down from the globally highest peak in the potential, 
$V(\vphi_*)$. 
There is ambiguity as to what happens if more than one peak has $V$ rising to above the Planck scale,
because we lack a theory of quantum gravity, 
but we will not worry about this issue here in light of the more pressing problems described below.
Note that the attractor behavior makes not only the initial conditions irrelevant, but also 
most of the properties of the inflaton potential $V(\vphi)$ --- all that matters is the shape 
of the highest peak and its immediate surroundings.
%* STRESS THAT FOR ATTRACTOR CASE, DETAILS OF V DON'T MATTER EITHER!
%  AS LONG AS "GENERIC"! 
%  Almost all functions V (in mathematical sense) predict the same thing!

Because of \eq{FPequilibEq}, the fastest way to come down from this peak is the most favored,
and it has been shown \cite{LindeLindeMezhlumian} that the exponential preference for speed is so strong that
the dominant descent mechanism can in some cases be 
quantum diffusion/tunneling rather than slow roll, giving the strange appearance that we are living in 
the center of a spherical void (at least for Planck-scale inflation). 
This preference for rapid descent can cause $\vphi$ to be quite near the top of the 
peak $55$ e-foldings before inflation, so that the tensor amplitude 
$Q_t\sim V(\vphi_*)^{1/2}/\mbar^2$. 
% If $\vphi$ is kept near the top by a very flat potential 
% with a sharp cliff where inflation ends, then $\epsilon\ll 1$ and the scalar amplitude
% $Q\sim [V(\vphi_*)/\mbar^4\epsilon]^{1/2} \gg V(\vphi_*)^{1/2}$.
This suggests that a generic inflaton potential with many random-looking peaks will predict 
$Q_t\gg 10^{-5}$ and hence be ruled out by observation unless 
$V(\vphi)\ll 10^{-10}\mbar^4 \sim 10^9$ GeV$^4$ for all $\vphi$.
For instance, the Gaussian random field potentials we explored in \Sec{MonteSec} 
are guaranteed to contain arbitrarily high peaks (exponentially rarely, but this is irrelevant),
and are hence all ruled out if $t$-ordering is used.
It should be emphasized, however, that there are infinitely many potentials that do not overpredict 
the fluctuation level with time ordering, most obviously popular single-basin ones like 
$V(\phi)=m^2\phi^2$ with $m\approx 5\times 10^{-6}\mbar$.
% Q ~ 4(m/mbar)
% calc mbar*2e-5/4
%
%most likely way to get elsewhere is to have come down from that peak (as long as it takes only a finite time, which is 
%always arrangeable by quantum diffusion/tunneling).
%This equilibrium distribution will have $L\fphi(\vphi) = H_* \fphi(\vphi)$ for some 
%$H_*$ corresponding to the highest peak in $V(\vphi)$.

\subsection{The coolness problem}

We will now present an argument for why the $t$-foliation solution to the ordering 
problem is observationally ruled out. This is closely related to arguments given in 
\cite{LindeLindeMezhlumian,Guth00,Guth00b,Guth04} as described below.

Consider any reference objects the number of which is proportional to physical thermalized volume, say
protons, planets or people. Let us compute the probability distribution $f_T(T)$ for the CMB temperature $T$ observed
at such reference objects at a given time $t$. We will find that our universe is much cooler than predicted.

After a given volume thermalizes at time $\ttherm$, it keeps expanding and its photon temperature $\Tcmb$ drops.
We can therefore write $\Tcmb=f(t-\ttherm)$ for a known and monotonically decreasing function $f$.
In Planck units ($\mpl=1$; used throughout this section), we have 
\beq{fEq}
f(\Delta t)\sim 
\left\{\begin{tabular}{ll}
$\Delta t^{-1/2}$		&during radiation domination,\\
$\xi^{-1/3}\Delta t^{-2/3}$ 	&during matter domination,
\end{tabular}\right.
\nonumber
\eeq
%$f(t)\sim t^{-1/2}$ during radiation domination and 
%$f(t)\sim \xi^{-1/3}  t^{-2/3}$ during matter domination, 
where $\xi\approx 3.3\times 10^{-28}$ is the  
matter-to-photon ratio from Table 1. 
($\Tcmb\propto a^{-1}$ at all times.)
Since the number of refence objects is proportional to the thermalized volume, 
the fraction of them in domains thermalized by time $t$ is 
\beq{PthermDistrEq}
P(\ttherm<t)\propto\Vtherm(t)\propto e^{3H_* t}.
\eeq
The fraction of the reference objects where $\Tcmb$ is below some given value $T$ is therefore
\beqa{PcalcEq}
P[\Tcmb<T] 	&=& P[f(t-\ttherm)<T]\nonumber\\
		&=& P[\ttherm<t-f^{-1}(T)]\nonumber\\
		&\propto&  e^{3H_*[t-f^{-1}(T)]}\nonumber\\
		&\propto& e^{-3H_* f^{-1}(T)},
\eeqa
where the inverse function $\Delta t= f^{-1}(T)$ is approximately
\beq{finvEq}
f^{-1}(T)\sim 
\left\{\begin{tabular}{ll}
$T^{-2}$			&during radiation domination,\\
$\xi^{-1/2}T^{-3/2}$ 		&during matter domination,
\end{tabular}\right.
\nonumber
\eeq
Since the expression $e^{-3H_* f^{-1}(T)}$ in \eq{PcalcEq}
is a perfectly regular and well-behaved function, 
approaching zero as $T\to 0$ $(t\to\infty)$ and approaching unity as 
$T\to T_{\rm therm}$, 
eternal inflation with $t$-ordering thus makes the firm testable
prediction that the temperature probability distribution 
is given by $P[\Tcmb<T]=e^{-3H_* f^{-1}(T)}$.
During matter domination, we thus have the prediction
\beq{TpredictionEq}
P(\Tcmb<T) \approx e^{-\left({T\over T_*}\right)^{-3/2}}, 		
\eeq
where $T_*\equiv \xi^{-1/3} (3H)_*^{2/3} % xi ~ 3.3e-28
\approx 10^9 H_*^{2/3}$. % 1.447090304e9
For definiteness, let us take the eternal inflation expansion rate
to be a typical inflationary rate $H_* \sim 10^{-5}/10^{-38}$s, % 6.167877928e-06/tplanck
%(0.004\mbar)^2$, 
% V(phi) = m^4 (phi/mbar)^2 gives delta_H~4(m/mbar) ~ 0.00002 for m=0.002 mbar,
% which gives phi ~ sqrt(2*2N)mbar ~ 15 mbar, V~(0.002 mbar)^4 * 15**2,
% H =sqrt(V/3)/mbar  ~ (0.002**2 * 15/sqrt(3)) mbar ~ 3.5e-05 mbar
% V ~ 4N m^4, H ~ sqrt(4N/3) (m/mbar)^2 ~ sqrt(4*55/3)*0.002**2 mbar = sqrt(4*55/(3*8*pi))*0.002**2 mpl
% H^{-1} ~ tpl/(sqrt(4*55/(3*8*pi))*0.002**2) ~ 5.39056e-44/(sqrt(4*55/(3*8*pi))*0.002**2) ~ 7.793338225e-39 seconds.
% Qt ~ H/mpl, plain and simple, and for phi^2-potential, Qt ~ sqrt(8/N)Q~0.4Q~ 0.8e-5.
% T* ~ \xi^{-1/3} H_*^{2/3} ~ (3.3e-28)**(-1/3) * (sqrt(4*55/(3*8*pi))*0.002**2)**(2/3) ~ 521061.9375 Tplanck
%  ~ 521061.9* 1.41696e32 K ~ 7.383238745e+37 K
which gives 
$T_*\sim 10^{38}$K % 7.383238745e+37 K
--- the conclusions we will draw are completely unaffected by varying $H_*$ within the observationally
allowed range.
This gives the spectacularly small probability
\beq{T3Eq}
P(\Tcmb<3\K)\approx 10^{-10^{56}}, 	%  calc (3/7.383238745e+37)**(-3/2)  
% calc -(3/2)*lg(3/7.383238745e+37) ~ 56.08
\eeq
and the probability of finding $\Tcmb$ as low as observed by COBE/FIRAS, \ie, 
$\Tcmb=2.728\pm 0.004$, is of course even smaller.
Even if we take as a given that $\Tcmb<4\K$,  
this probability remains minuscule:
%\beq{T4Eq}
$P(\Tcmb<3\K|\Tcmb<4\K)\approx 10^{-10^{55.6}}.$ %  calc (3/Tstar)**(-3/2) - (4/Tstar)**(-3/2) 
%\eeq
% set xi = 3.3e-28 
% set T = 2.728/1.41696e32
% set t = 13.7e9*(3600*24*365.25)/5.39056e-44
% set T = xi**(-1/3)*t**(-2/3) # T
% set dTdt = -(2/3)*xi**(-1/3)*t**(-5/3) # dT/dt 
%%% set dtdT = -1.5*xi**(-0.5)*T**(-5/2) # dt/dT 
Indeed, the probability is small even of having the CMB temperature a mere $10^{-17}$K
warmer, corresponding to a one second younger (relative to the thermalization time) universe:
about $e^{-\hbox{(1 sec})H_*}\approx 10^{-10^{38}}$. % Calc (1 sec)*Hstar; this exponent 38 is essentially 56-17.
Although it is easy to envision an anthropic weight factor $w(\p)$ that may slightly 
skew this distribution towards lower temperatures (by requiring sufficient time for evolution, say),
such corrections will be clearly be way too week to overpower this double exponential
preference for high temperature --- indeed, before CMB temperature was measured in 1965,
scientists saw no fundamental problem with it being say $5\K$ as predicted by 
Alpher \& Herman \cite{AlpherHerman50}.
The FIRAS measurement of the CMB temperature therefore rules out eternal inflation 
with time ordering at a significance level of about
$99.999...9\%$, where the dots denote $10^{56}$ nines.
Few scientific theories have had the honor of being 
ruled out with greater statistical significance.

%\beq{TpredictionEq}
%f_T(T) = {d\over dT} P[\Tcmb<T] 
%\propto \left({T\over T_*}\right)^{-5/3} e^{-\left({T\over T_*}\right)^{-3/2}}, 		
%\eeq
%eternal inflation with $t$-foliation thus makes a firm 
%prediction for the CMB temperature probability distribution 
%$f_T(T)$ which is simply the derivative of $P[\Tcmb<T]$ with respect to $T$.

The root of the failure is clearly that this ordering rewards rapid expansion, so that 
at a fixed time, the vast majority of all reference objects formed only very recently.
Although we focused on $\Tcmb$ above to be specific, 
the coolness problem clearly afflicts many other observable parameters as well. 
Our universe is not only cooler than predicted, but also older and less dense than $t$-ordering predicts.
These paradoxical implications have been discussed in the literature, although in contexts 
leading to somewhat less negative conclusions. 
This issue is at the core of the surprising findings of Linde \& Mezhluminan \cite{LindeLindeMezhlumian}
that we might expect to find ourselves in the center of a rare spherical void, 
and the authors mention this as a perhaps a possible way to
obtaining an open $\Otot<1$ universe from inflation.
Guth \cite{Guth00,Guth00b,Guth04} discusses the problem in detail, terming it the ``youngness paradox''.
He speculates that this could explain why the SETI project has failed to discover extraterrestrial
intelligence (since the preference for low cosmic age implies that we are overwhelmingly likely 
to be the first observers in our Hubble volume), but finds it 
more plausible that it is merely a symptom of the probability calculation technique
not being the right one. A measure with $t$-ordering implies 
not only this, but also more 
%Byzantine 
extreme consequences such as that you personally 
are the first observer in your Hubble volume since thermalization, and that
it is way more likely that you evolved on the first habitable planet to form,
under circumstances where many apparent flukes accelerated your arrival.
We conclude that our living in a cool, leisurely universe means nothing other than that 
$t$-ordering is the wrong ordering.

%\subsubsection{Implications of the failure, why attractor behavior is ruled out, sense in which inflation is eternal}
\subsection{Why does not rule out eternal inflation but limits attractor behavior}
%\subsection{Why this rules out attractor behavior but not eternal inflation}
\label{CoolnessImplicationsSec}

%We have found that $t$-ordering is {\it not} the solution to the ordering problem.
%What does this mean?
There is presumably some correct solution to the ordering problem, and
we have found that $t$-ordering is not the one. It is tempting to speculate that this is because it flaunts some important 
yet-to-be-established underlying physical principle, in which case the coolness problem is a helpful clue
for identifying this principle.
Here is a guess as to such a principle: {\it Physical questions must be expressible in terms of observables.}
This implies that it is a big no-no to condition on quantities that are not physically observable, 
like $t$ or $a$. $t$-ordering violates this principle since, as described above, it corresponds
to defining reference objects such as, say, ``protons at time $t$".
The global time $t$ is a completely unobservable quantity, because all we can measure in our 
thermalized region is time intervals, \eg, the time interval $(t-\ttherm)$ since inflation ended here.
Moreover, there are infinitely many alternative choices of time variable, and between causally disconnected regions of
spacetime (such as two different thermalized pockets), gauge ambiguities imply that there is no objective
way of defining simultaneity.

Let us now discuss the implications of the coolness problem for eternal inflation and attractor behavior.
Since the same Fokker-Planck probability approach that elegantly predicts eternal inflation is ruled out
by the coolness problem, is inflation really eternal in any meaningful sense?
A standard definition of eternal inflation (\eg,\cite{WinitzkiVilenkin96}) is that 
{\it ``at any given time, part of space is inflating, and the inflating volume increases over time"},
yet this definition uses precisely the infamous time foliation by considering the situation ``at any given time''. 
% Or, for short, ``the physical volume that is inflating increases over time''.
%Although perhaps true in some moral sense, this statement is undefined and hence ``not even wrong''.

Inflation can nonetheless be eternal, in the following well-defined sense:
{\it A finite comoving volume can produce an infinite physical volume and an infinite number of particles} (and other 
reference objects).
Equivalently, inflation can be eternal in the sense of producing infinite Level I multiverses according to the
classification of \cite{multiverse}.
This is readily seen in the following simple example, which is
illustrated in \fig{OrderingFig}.
Consider a single-field inflaton potential $V(\phi)$ with a peak (local maximum) at $\phi=0$, 
and a comoving spatial region where the inflaton potential varies extremely slowly in the $x$-direction:
\beq{SmoothPhiEq}
\phi(x,y,z)\approx \varepsilon x,
\eeq
where the constant $\varepsilon$ is so small that spatial gradients in $\phi$ are negligible.
If quantum diffusion is negligible, then inflation will eventually end everywhere in our comoving region 
except at the one point $\phi=0$, but 
it clearly takes longer for $\phi$ to roll off the
peak and end inflation the closer to $\phi=0$ it starts.
In the slow-roll approximation and considering the very top of the peak where it can be approximated by an upside down 
parabola, one readily finds that the number of $e$-foldings required to roll
down from $\phi$ is given by (Appendix B.2) 
\beq{PeakNeq}
e^{N(\phi)}\propto|\phi|^{\eta^{-1}},
\eeq
where $\eta\equiv\mbar^2 V''(0)/V(0)$ is the usual slow-roll parameter at the origin.
The total thermalized volume produced by our comoving region is thus
\beq{VtotEq}
V\propto \int e^{3N(\phi(\r))}dV \propto \int e^{3N(|\varepsilon x|)}dx \propto \int |x|^{3\eta^{-1}}dx,
\eeq
which will diverge near the origin and give an infinite volume as long as $\eta>-3$.
This simply reexpresses the well-known result that slow-roll peaks are eternal inflation regions, but without 
involving the Fokker-Planck equation or $t$-foliation. 
An alternative way of seeing this is to compute the probability distributions in \eq{fpEq2}.
If the initial conditions give a uniform distribution for $\phi$ around the peak,
then since $\eta<0$, \eq{PeakNeq} readily gives well-behaved probability distributions $\finf$
for $\Ntot$, $V\propto e^{3\Ntot}$, $a\propto e^\Ntot$ and $t\approx \Ntot/H_0$ at thermalization, 
where $H_0\equiv\mbar^{-1}\sqrt{V(0)/3}$:
\beqa{finfExampleEq}
P(\Ntot>N_*)&=& e^{\eta N_*},\label{NprobEq1}\\
P(V>V_*)&\propto&V_*^{\eta/3},\\
P(a>a_*)&\propto&a_*^\eta,\\
P(t>t_*)&\approx&e^{\eta H_0 t_*}.
\eeqa
Multiplying the corresponding differential probability distributions $\finf$ by the volume factor $\wvol\propto e^{3\Ntot}$ in
\eq{fpEq2} gives the predicted probability distribution $f_p$ for the parameters of 
a random thermalized volume element:
\beqa{fExampleEq}
P(\Ntot<N_*)&\propto& e^{(3+\eta) N_*},\\
P(V<V_*)&\propto&V_*^{1+\eta/3},\\
P(a<a_*)&\propto&a_*^{3+\eta},\label{aProbEq2}\\
P(t<t_*)&\simpropto&e^{(3+\eta)H_0 t_*}.\label{tProbEq2}
\eeqa
Since $\eta\ge -1$, these distributions are all unnormalizable, so the probability is zero that
$\Ntot$, $V$, $a$ or $T$ is finite.

The second type of eternal inflation region (the kind dominated by quantum diffusion) also survives even though 
the standard justification for it becomes invalid.
The standard argument \cite{LindeDiffusion86} 
is that quantum diffusion up the hill causes faster 
expansion that more than compensates for the fact that it would have been more likely to roll downward,
but it is of course precisely such measures favoring faster expansion that lead to the coolness problem.
An alternative justification is to again consider the evolution of a single comoving spatial region, without
applying any volume weighting until after thermalization. 
When quantum diffusion is included, $\vphi(t)$ will be not the deterministic slow-roll trajectory, but 
a random walk. This is guaranteed to end after some finite time $t_{\rm end}$ when 
$\vphi$ leaves the slow-roll region of $\vphi$-space, after which the comoving region thermalizes
with a volume
\beq{VtotEq2}
V\propto e^{3\Ntot},\quad\Ntot\equiv \int_{0}^{t_{\rm end}} H(\phi(t)) dt.
\eeq
Although both $N$ and $V$ are clearly finite for any trajectory $\phi(t)$,
the volume expectation value $\expec{V}$ can easily be infinite.
As a simple example, if the random variable $\Ntot$ has an exponential distribution 
$f_N(\Ntot)\propto e^{-\Ntot/\bar N}$ for some constant $\bar N$ 
(\eq{NprobEq1} is such an exponential distribution with $\Nbar=-\eta^{-1}\ge 1$), 
then just as in the slow-roll peak case above,
%then $\expec{V}=\expec{e^{3N}}=\infty$ if $\Nbar>1/3$ and 
the probability that $\Ntot$ is less than any finite value vanishes if $\Nbar>1/3$.
Such an ensemble averaging is equivalent to what happens when 
averaging (integrating) over a finite comoving volume as in \eq{VtotEq}, since the quantum fluctuations
are effectively uncorrelated in causally disconnected patches, and infinitely many such patches develop.
In other words, both of the ``new inflation'' mechanisms that can generate eternal inflation as per the
$t$-foliated Fokker-Planck formalism can do so also in the sense of producing infinitely many reference 
objects in a finite comoving region. 

A different way of seeing this is to note that 
the thermalization surfaces are perfectly well-defined physical hypersurfaces in spacetime, so there are no ambiguities
regarding the question of whether they have finite or infinite volume. 
Therefore the Fokker-Planck approach to computing this volume must give the same answer as the 
differential equation approach 
described above --- indeed, 
\eq{PthermDistrEq} agrees with \eq{tProbEq2} with $H_*=(1+\eta/3)H_0$.
Ambiguities only arise if we ask whether one infinite volume is larger than another infinite volume.

%The former will happen if the field $\phi(\r)$ in the volume is draped over a maximum of the inflaton potential
%where the slow-roll approximation is valid ($epsilon\ll 1$, $|\eta|\ll 1$) and also if you're in the quantum diffusion regime.
%The latter will also happen as long $\phi$ rolls down to a place where inflation ends and reheating occurs 
%(as opposed to getting stuck in a local minimum of $V$ where inflation continues forever).

Although eternal inflation thus survives the coolness problem unscathed, 
the attractor dynamics giving testable prediction $f_p(\p)$ independent of the initial conditions
is severely challenged. 
There is of course nothing wrong with the Fokker-Planck equation {\it per se}:
it can be consistently reexpressed in terms of any time variable one choses
\cite{LindeBook,Vilenkin83,Starobinsky84,Starobinsky86,Goncharov86,SalopekBond91,LindeLindeMezhlumian94},
so the problem lies entirely with the limiting procedure used to extracting probabilities from it.
However, modulo a tunneling caveat discussed below in \Sec{TunnelingSec}, 
the Fokker-Planck equation only exhibits this attractor dynamics when expressed in such a way that
it rewards rapid expansion, since as described above, this was the mechanism by which the 
highest peak in the potential was able to dominate the asymptotic solution.
To circumvent the coolness problem, one must therefore avoid rewarding rapid expansion.
There are two ways of doing this:
\begin{enumerate}
\item Changing the reference objects to be comoving
rather than physical volume elements, \ie, replacing the volume weight factor of \eq{wvolEq} by $\wvol(\p)=1$.
\item Changing the time variable to something not rewarding faster expansion, like $a$ or $\rho$.
\end{enumerate}
Either way, the attractor dynamics described above goes away, and the asymptotic distribution (if it exists) will depend
on initial conditions (modulo tunneling issues to which we return in \Sec{TunnelingSec}).
The former approach 
has been studied in, \eg, 
\cite{LindeBook,Vilenkin83,Starobinsky84,Starobinsky86,Goncharov86,SalopekBond91,LindeLindeMezhlumian94}, and 
can be argued to be unphysical
since all reference objects that may be related to observers (particles, planets, \etc) are 
proportional to {\it physical} rather than comoving volume. 
Let us now consider the latter.
Although changing the time variable to $\rho$ or other observable quantities solves the coolness problem,
we have seen that this leads to pocket-based orderings which we have already discussed, so let us turn instead
to unobservable time variables such as $a$.

\subsection{$a$-ordering and others not favoring rapid expansion} 

There are infinitely many possible choices of time variable, 
and different choices can give different predictions.
\cite{Vilenkin95,WinitzkiVilenkin96,LindeMezhlumian96}.
As an interesting example giving qualitatively different predictions from $t$-ordering, let us consider
ordering by the cosmic scale factor $a$.

First of all, $a$-ordering is not afflicted by the coolness problem.
This is readily seem from comparing \eq{aProbEq2} with \eq{tProbEq2} and noting that
the exponential divergence is replaced by much milder one that is between quadratic and cubic.
Since $\Tcmb=\Ttherm\atherm/a$, repeating the derivation of \Sec{SmoothnessSec} 
shows that \eq{PcalcEq} gets replaced by
\beqa{PcalcEq2}
P[\Tcmb<T] 	&=&P[\Ttherm\atherm/a<T]\nonumber\\
		&=&P[\atherm<aT/\Ttherm]\nonumber\\
		&\propto&(aT/\Ttherm)^{3+\eta}\nonumber\\
		&\propto&T^{3+\eta}.
\eeqa
The probability of observing $\Tcmb<3$K given than $\Tcmb<4$K therefore has the 
acceptable value   
$(3/4)^{3+\eta}\approx 0.5$ rather than $10^{-10^{56}}$ as with $t$-ordering.

As elaborated in, \eg, 
\cite{Vilenkin95,WinitzkiVilenkin96,LindeMezhlumian96},
replacing $t$ by $a$ as a time-variable reverses some qualitative conclusions. 
With the initial conditions of \eq{SmoothPhiEq}, the amount of thermalized volume produced up to a given time 
is readily found to be 
\beq{VproducedEq}
%V\propto  \left(1-3\eta^{-1}\right)^{-1}e^{(3+\eta)H_0 t} 
%\approx   \left(1-3\eta^{-1}\right)^{-1}a^{3+\eta}.
\Vtherm\propto  {e^{(3+\eta)H_0 t}\over 1-3\eta^{-1}}
\approx   {a^{3+\eta}\over 1-3\eta^{-1}}.
\eeq
Since $\eta\ge -1$ at a the peak, the amount of thermalized volume produced thus grows steadily. Over time, the freshly thermalized volume
comes from an increasingly tiny initial volume very close to the peak (reflected by the $\eta$ in the exponent), 
but this negative effect is more than offset by the
huge inflationary expansion (giving the $3$ in the exponent).

\Eq{VproducedEq} shows that when the inflaton potential has multiple peaks, $t$-ordering and $a$-ordering give radically different 
predictions. 
With $a$-ordering, taking $a\to\infty$ shows that the flattest peak (which has the largest $[3+\eta]$)
will completely dominate the thermalized volume, and that peak height $H_0$ is irrelevant.
With $t$-ordering, on the other hand, taking $t\to\infty$ shows that the peak with the largest value of $(3+\eta)H_0$
will completely dominate the thermalized volume, \ie, that the most important factor is peak height $H_0$, modulated by
a slight preference for flatness. By constructing new time variables that are functions of 
$t$, $a$ and observables like $H$, one can readily construct a range of other predictions.

In conclusion, global orderings based on a non-observable time variable are not all ruled out by observation,
but lack compelling physical motivation and, as a class, fail to make unambiguous predictions. 
Another argument against global ordering with ``time'' variables such as $a$, $T$, $H$ and $\rho$ is that
they can change direction (when a spatial region starts to contract) and hence fail to smoothly cover the entire spacetime manifold.
Most importantly, if we accept the observability principle of \Sec{CoolnessImplicationsSec} that physical questions must be expressible
in terms of observables, then this entire class of orderings is ruled out.

\subsection{Ordering by an observable and the pothole paradox}
\label{potholeSec}

\Eq{NpEq} shows that the number of reference objects is proportional to the volume at thermalization, 
so if two initially equally large volumes roll off the same peak in $V(\phi)$ in opposite directions,
the one that requires more $e$-foldings to thermalize will produce more reference objects in the end.
As we will now see, this has an important implication for pocket-based orderings.

Consider the number of reference objects that will ultimately form in a given comoving volume. 
\Eq{NpEq} shows that it is proportional to the physical volume at thermalization, \ie, 
that it grows during inflation and then stays constant.
This implies that whatever time variable we use, most reference objects at a 
given ``time'' are in regions that stopped inflating as recently as allowed by their existence.
Above we saw that this effect was rather mild with the time variable $a$ but gave the 
% egregious 
catastrophic coolness problem with the time variable $t$. 
%Despite this difference, these cases share a key
%feature: there is an upper limit on the total amount of inflation, since the ``time'' spent inflating and the
%time spent subsequently must add up to the present time.
When conditioning not on a monotonically increasing time variable but
on an observable quantity such as $\rho$, $T$ or $H$ (implying a pocket-based ordering), the effect
is that the longest possible inflation scenario will completely dominate the statistics, {\it no matter how long it takes}.
This means that a flatter peak dominates over a shallower one. However, it also means that 
both will be outperformed if the inflaton potential has
a pothole in the roll path where $\phi$ will get stuck for an exponentially long 
time until it tunnels out \cite{Guth81}. The deeper the pothole, the better,
so one could argue that almost all observers can trace the origin of their pocket back to an exponentially rare tunneling 
event through a high peak in the potential. This event in turn would be overwhelmingly likely to have 
been preceded by a many other freak tunneling events. This is a disturbing scenario, since for any extremely long 
inflationary history there exists
an even longer one, and and it is far from clear that it gives well-defined and observationally allowed
cosmological parameter predictions $f_p(\p)$. Note that the exponentially small probability per unit time to tunnel out of
a deep inflating minimum is irrelevant, since $\vphi$ is guaranteed to tunnel out eventually with probability unity. 

The ``pothole paradox''
% seemingly paradoxical conclusion 
above follows from strict adherence to weighting pockets by their number of reference objects,
even though each pocket contains infinitely many\footnote{Note that global orderings lack this ambiguity, since there are
only finitely many reference objects at any finite ``time'' (say $t$ or $a$) as compared to an infinite
number at finite ``time'' $\rho$, $T$, \etc}.
In other words, it follows from assuming that some infinities are greater than others. 
This argues against inter-pocket weightings 3, 4 and 5 from \Sec{InterPocketSec}, 
suggesting that that all countable infinities receive equal weight.

%\begin{itemize}
%\item $t$-ordering: favors inflation ending as recently as possible; fatal coolness problem
%\item $a$-ordering: favors inflation ending as recently as possible; mild coolness problem
%\item $\rho$-ordering: favors slowest possible end to inflation
%\end{itemize}

%zzz \subsection{The multidimensional case}

\subsection{Quantum tunneling and attractor behavior}
\label{TunnelingSec}

One area in need of further work is the effect of quantum tunneling in inflation. 
In some cases, for example ones involving the above-mentioned pothole paradox,
this tunneling process may hold the key to understanding the qualitative predictions.

As a specific example, consider the question of whether $a$-ordering gives attractor behavior or 
gives predictions that depend on initial conditions.
As discussed above, $a$-ordering is equivalent to weighting by comoving volume $\Vcom$, which is a conserved quantity. 
This means that, ignoring quantum diffusion and tunneling, the distribution $\Vcom(\vphi,a)$ simply flows down into the valleys,
never moving from one basin of attraction to another. In other words, initial conditions matter, 
since the probability of ending up in a given basin of attraction is simply the probability of starting out in that basin of attraction.
Quantum tunneling alters this conclusion, allowing transitions between basins either during inflation or after reheating, 
in the second inflation phase corresponding to a cosmological constant $\rhol>0$ \cite{GarrigaVilenkin98}.
Modeling this as a Markov process with a matrix of jump probabilities per unit ``time'' $a$ between the different basins, 
one can imagine convergence to an equilibrium state as $a\to\infty$ where the basin weighting is independent of initial conditions.
Further work should clarify whether such attractor behavior actually occurs. For instance, 
although ``downward'' tunneling to lower energy through the 
Coleman-de Lucia bubble nucleation mechanism is fairly well-understood, the situation regarding ``upward'' tunneling is less clear.
Also, there will be no tunneling out of a minimum with $V(\vphi)=0$ to first order, thereby spoiling attractor behavior if
there is more than one such minimum, but it is not clear that there is no tunneling out to second order.

\vskip4cm % FOR LESS DISASTROUS PAGE BREAKS FURTHER ON

\section{Conclusions}
\label{ConcSec}

If string theory or some other fundamental theory 
predicts an effective inflaton potential $V(\vphi)$, then it will clearly be worthwhile to
compute the predicted cosmological parameter probability distribution $f_p(\p)$ to confront it with
observation.
In this paper, we have attempted to strengthen the groundwork necessary for such a calculation
in the context of classic slow-roll inflation, exploring 
how the predictions for the 8 parameters $\p=(\Otot,\rhol,w,\dH,\ns,\al,r,\nt)$ 
depend on both the inflaton potential and the measure. 
We found that the results depend on a complex and interesting interplay between the two, 
except for the very simplest potentials with a single minimum that can only be rolled to in one way.
It is a persistent myth that models where ensembles and anthropic constraints enter are 
untestable and hence unscientific \cite{Smolin04}. %  {\frenchspacing\it c.f.} \cite{Susskind04}. 
We have seen that, quite to the contrary,
most such models that we have explored in the present paper are predictive enough to be eminently testable
--- indeed, so much so that they have already been ruled out by observational data!

Since we found that the measure problem is currently the weakest link in the 
calculation of predictions, we will 
classify models below primarily by their potentials and only secondarily by their measure.
This is intended not only to extend the shelf life of our discussion, but also to place the 
seemingly daunting measure problems in context by clarifying what they do and do not affect.

\subsection{How the inflaton potential affects the cosmological parameters}

The effective cosmological constant $\rhol$ depends only on which basin of attraction $\vphi$ rolls into, 
since it is determined by the minimum in which $\vphi$ ends up.
The five power spectral parameters $(Q,\ns,\al,r,\nt)$ vary within the basin, since they depend on the direction from which 
$\vphi$ rolls to the minimum --- for one-dimensional potentials, there are only two possibilities (left and right), so
they are determined uniquely by the half-basin (region between a potential maximum and minimum).
The curvature parameter $\Otot$ and total pocket volume $\propto e^{3\Ntot}$ depend on the starting point and hence 
crucially on the measure.
Finally, the dark energy equation of state $w=-1$ as elaborated below in \Sec{DarkEnergyImplSec}.

% It's all extremum statistics

One of our key findings is a useful simplification: in all cases except $\mh\sim\mpl$, 
the cosmological parameter predictions are determined by extremum statistics alone.
In other words, unless the characteristic scale $\Delta\phi$ on which the inflaton potential $V$ varies happens to be
close to the Planck scale, the only aspect of $V$ that matters observationally is the statistical
distribution of peaks and/or troughs.

The $\mh\gg\mpl$ case predicts that
\beqa{bigmhEq}
\ns&=&1-{r\over 4},\\
\al&=&-{r^2\over 32},\\
\nt&=&-{r\over 8},\\
r&\approx&0.15\pm 0.02
\eeqa
% r = 8/N => N = 8/r
% 1-ns=2/N=r/4 
% al = -2/N^2 = -2*(r/8)**2 = -r**2/32
and that the parameters $(Q,\rhol)$ are determined by trough statistics, specifically by 
the joint distribution of $V$ and its curvature at minima.
Assuming that $V=0$ is not special in the fundamental theory, 
all that matters is the curvature probability distribution at minima with $V\approx 0$, 
given for the one-dimensional case by the
distribution for $V''(\phi)$ where $V'(\phi)=0$ and $V(\phi)\approx 0$.

The $\mh\ll\mpl$ case predicts that
\beqa{smallmhEq}
\ns&\sim&0\pm 1,\\
\al&\approx&0,\\
\nt&\approx&0,\\
r&\approx&0,
\eeqa
and that the parameters $(Q,\rhol,\ns)$ are determined by peak statistics ---
for the one-dimensional case and $V=0$ not special, 
they depend mainly on the  
the joint probability distribution for $(V,\eta)$ at the peaks.

This means that aside from $w$ and $\Otot$, the number of precise and testable quantitative predictions 
is 3 when $\mh\ll\mpl$ and 4 when $\mh\gg\mpl$, dropping to only one ($\nt=-r/8$) when $\mh\sim\mpl$.
In other words, in the 5-dimensional parameter space $(Q,\ns,\al,r,\nt)$, the predictions
populate a 2-dimensional hypersurface when $\mh\ll\mpl$, a 4-dimensional region when $\mh\sim\mpl$ and 
a 1-dimensional curve when $\mh\gg\mpl$. 
In essence, the $\mh\ll\mpl$ limit simplifies since $\epsilon\to 0$ and 
the $\mh\gg\mbar$ limit simplifies since $(\ns,\alpha,r)$ are all determined by $N$, which
is in turn known to about 10\%.

\subsection{How the measure affects the cosmological parameters}

Above we saw that except when $\mh\sim\mpl$, the predictions for the parameters 
$(Q,\ns,\al,r,\nt,w,\rhol)$ could all be calculated from extremum statistics alone.
In other cosmological examples where extremum statistics enter, such as the 
BBKS Gaussian-peaks formalism for galaxy clustering and the subsequent 
peak-patch formalism \cite{BondWeb96}, the relevant extremum statistics can be
extracted unambiguously from the statistical properties of the function in question (in these examples, the 
cosmic density field). 
In our case, however, these extremum statistics are not necessarily those one would obtain by
simply locating the appropriate extrema in $V(\vphi)$ and tallying up the distribution.
Rather, this ``raw'' distribution is modulated by the measure to give  
the above-mentioned extremum statistics that determine the parameter prediction.
For example, for a homogeneous Gaussian random field $V(\phi)$, the 4-dimensional vector 
$[V(\phi),V'(\phi),V''(\phi),V'''(\phi)]$ has a simple
multivariate Gaussian distribution that can analytically be conditioned on $V(\phi')=0$ and whatever else is
appropriate, but the measure (both 
initial conditions, ordering and the choice of reference objects) can multiply the resulting  
distribution of extremum properties by a factor that radically modifies it.
Conversely, it is reassuring to note that modifying the extremum distribution is the {\it only} way in which 
the pesky measure problems we have discussed affect the cosmological parameter distribution unless 
$\mh\sim\mpl$. 

Our simulated Measure A case essentially weighted all peaks/troughs equally 
(strictly speaking, troughs by the lengths of their basins of attraction
and peaks by the lengths of the regions to which they flow), 
except for the rather straightforward conditioning on reference objects in \Sec{AnthroSec}.
%Our simulated Measure A essentially left the raw extremum distribution alone, except for the 
%rather straightforward conditioning on reference objects in \Sec{AnthroSec}.
% a pretty regular probability distribution, derivably from Gaussian statistics. 
% $(V,V',V'',V''')$ has a 4D Gaussian distribution, and you condition on $V=0$ as in BBKS.
%
For other measures, the extremum statistics can get highly weighted towards certain peaks/troughs over
others, either because of their intrinsic properties
(like $t$-ordering favoring height and $a$-ordering favoring flatness) or because
of good neighbors (like a deep valley to tunnel from in the $a$-ordering case). 
We saw that $t$-ordering is an extreme case where 
only the highest slow-roll peak matters, so that the  
probability distribution is a $\delta$-function corresponding to the highest peak (if $\mh\ll\mbar$) or
the trough that $\phi$ can reach the fastest from the highest peak (if $\mh\gg\mbar$).
%In both cases, readily extractable once you know $V(\phi)$.
%If $V\to\infty$ somewhere in a slow-roll region, that will completely dominate. The steeper, the better, as
%long as SRA is valid to make it eternal.

This author considers the inflationary measure problem wide-open, so let us merely summarize our 
limited progress with it.
We have argued that global orderings such as $t$-ordering that reward 
rapid expansion are observationally ruled out by the coolness problem (overpredicting the CMB temperature).
If we accept the notion that we may only condition on physically observable quantities (\eg, $\rho$ or $T$), then
we are lead to pocket-based orderings that are specified by an intra-pocket ordering and an inter-pocket weighting. 
Radial intra-pocket ordering provides a specific working recipe that may and may not be the correct one. 
Weighting all spatially infinite pockets equally 
likewise provides a specific prescription that may and may not be right. It avoids the pothole paradox,
but is computationally difficult to apply because of the challenge of calculating the relative
frequencies with which different 
basins of attraction are rolled down into when quantum diffusion is taken into account.
This computation becomes still more complicated in recycling universe \cite{GarrigaVilenkin98} and string landscape
scenarios 
\cite{Bousso00,Feng00,KKLT03,Susskind03,AshikDouglas04,KKLMMT03,WKV04,Denef04,GK04,Freivogel04,Conlon04,DeWolfe04,Nima05,Acharya05}
where each thermalized region can trigger new inflation leading effectively to a 
Markov process of jumps between basins of attraction.

Since there appears to be only a small-number of well-motivated measures and they make quite distinct parameter predictions,
it may be possible to settle the measure problem empirically by ruling out all but one of them observationally.

%What to use for say the peak probability distribution depends strongly on both
%$V(\phi)$ and the measure. 

\subsection{Implications for gravitational waves and spectral indices}

Great worldwide efforts are currently being devoted to measuring CMB polarization in the hope 
of measuring the primordial gravitational wave amplitude $Q_t$ via large-scale $B$-modes
\cite{bb04,Carlstrom03}.
A key theoretical question is therefore what amplitude $Q_t$ to expect,
since this determines the prospects for upcoming experiments to discover what they are looking for
and influences their funding and design.
Our results are very encouraging in this regard.
Models with $\mh\gg\mpl$ predict $r\approx 0.15$, which (\fig{nsrFig}) is close to current sensitivities
and should be detectable already within five years by Planck $+$ SDSS \cite{parameters2}.
Models with $\mh\ll\mpl$ predict unobservably small $r$-values, but appear to agree poorly with 
existing observations: they fail to explain the near scale invariance $\ns=0.98\pm 0.02$
that is observed \cite{sdsspower,sdsslyaf} by giving instead a wide distribution with $-1\le\ns\le 1$
and they flounder on the smoothness problem by generally overpredicting $Q$.
Finally, models with $\mh\sim\mpl$ typically predict $r$-values that are 
clearly within reach of a next generation CMB polarization satellite; \fig{nsrFig} shows typical values $r\sim 0.03$.

The most imminent observational breakthrough may well be sharpening the current limits spectral index limits
$\ns=0.98\pm 0.02$ to exclude the scale-invariant Harrison-Zel'dovich case $\ns=1$, and this will severely 
reduce the number of viable inflation models. We have seen that models with $\mh\gg\mpl$ predict $\ns\approx 0.96$
whereas models with $\mh\ll\mpl$ predict a broad distribution in the range $-1\le\ns\le 1$ --- \ie, they
fail to explain why $\ns\approx 1$ and cannot give blue ($\ns>1$) spectra, 
the only exception being arguably unphysical measures that do not volume weight, like Measure B.
%, where inflation near inflection points can give $\ns>1$ (\fig{1DcomboFig}).

Our $\mh\simgt\mpl$ predictions agree well with other arguments suggesting that classic single-field inflation
produces both gravitational waves and departures from scale invariance that will soon be observable
\cite{LiddleLythBook,KhourySteinhardt0302012,Babich04}.

\subsection{Implications for dark energy}
\label{DarkEnergyImplSec}

% 
% Basics:
% 
We saw that in order for the standard inflationary calculation to be self-consistent 
(give the correct expansion rate $H$),
the inflaton potential $V(\phi)$ must be defined to include any $\phi$-independent 
vacuum energy contributions from other sectors of physics, including any bare cosmological constant
term that may be present in the Einstein field equations. The effective cosmological constant 
that we observe in our Hubble volume is therefore simply the 
height $\rhol\equiv V(\vphi_0)$ of the potential at the minimum that the inflaton has rolled
down to here. Any messy inflaton potential $V(\vphi)$ thus naturally predicts $\rhol\ne 0$ unless
all its minima happen to have exactly the same height and this common height happens to be exactly zero.

\subsubsection{Vanilla or not?}
%
% w:
%
If no new physics is added beyond this inflaton potential, a generic prediction is that
this dark energy will exhibit ``vanilla'' properties indistinguishable from a
cosmological constant, \ie, $w=-1$, a time-independent 
density at late times (including the present) and no spatial fluctuations.
The reason for this is that the effective inflaton mass scale $m$ is huge by current 
standards\footnote{If we Taylor expand a 1-dimensional potential around a $V\approx 0$ 
minimum as $V(\phi)\approx m^2\phi^2/2$, then \eq{VunitEq} gives
$m=\sqrt{V''}\sim\mh^2/\mv$.
For the observationally favored case $\mh\simgt\mpl$, \eq{quadratic_dHeq} 
thus shows that $m\sim Q\mpl\sim 10^{-5}\mpl\sim 10^{14}$ GeV.
%For the observationally disfavored case $\mh\ll\mpl$, \eq{hill_Qeq} gives
%$m\sim e^{\eta N} Q\mh^2/\mpl$, which is typically exponentially small ---  but even 1 TeV is huge by present-day standards.
}.
% V(phi) ~ V'' phi^2 = m^2 phi^2,
% m^2 ~ V'' ~ mh^4/mh^2
%
This means that on a very rapid reheating timescale, $\phi$ will settle down into its minimum and
stay there, either promptly recollapsing space if the minimum is substantially below zero
or giving $\rhol(t)=V[\phi(t)]$ constant at late times.
The huge mass scale $m$ will not only ensure that the minimum is rapidly attained, 
but also that there are no observable dark energy fluctuations, since excitations
in $\phi$ are too massive to be excited by present-day energies.
It is noteworthy that the substantial improvements (most recently \cite{Riess04})
in dark energy measurements since the first discovery \cite{Riess98,Perl99} 
have not established any departures from ``vanilla'' \cite{supernovae,supernovae2}.
% no departure whatsoever from ``vanilla'' \cite{supernovae,supernovae2}.
However, improved dark energy observations are clearly crucial, since the fact that  
messy inflation may already explain $\rhol\ne 0$ and its rough magnitude does not in any way preclude
the existence of additional physics producing non-vanilla dark energy at late times, and
there are interesting hints at modest statistical significance (\eg, \cite{GordonHu04}).
% A 2D V(\vphi) would do this if you have a perfectly separable minimum with 
% one large positive eigenvalue (m) and one tiny negative one corresponding to unstable 
% dark energy. You simply have to be very careful not to trigger the instability
% of the latter during reheating.
%
%  So $w\ne -1$ would require additional physics for which we lack observational evidence.
As shown in \fig{arhoFig}, improved dark energy observations also complement CMB polarization observations by 
measuring the exact same curve $\rho(a)$ at a vastly lower density.

\subsubsection{Continuum or discretuum?} 

%
% discretuum:
%
\def\ntot{n_{\rm tot}}
\def\nsm{n_{\rm sm}}

Potentials $V(\vphi)$ with infinitely many minima 
(such as Gaussian random fields as explored in \Sec{MonteSec}) 
can give a continuous probability distribution for $\rhol$. 
In contrast, the landscape picture emerging from string theory suggests 
a very large but yet finite number of minima, say $\ntot\sim 10^{300}$ 
\cite{Bousso00,Feng00,KKLT03,Susskind03,AshikDouglas04,KKLMMT03,WKV04,Denef04,GK04,Freivogel04,Conlon04,DeWolfe04,Nima05,Acharya05}\footnote{There may indeed be
a hierarchy of minima in some sort of generalized potential:
\begin{enumerate}
\item Different ways in which extra dimensions can be compactified
\item Different discrete fluxes that stabilize the extra dimensions (this sublevel appears to be where the largest 
number of choices enter, perhaps $10^{300}$).
\item Once these two choices have been made, there may be a handful of different minima in the effective supergravity potential.
\end{enumerate}
}
Suppose we find a fundamental theory that has $\nsm$ minima which all reproduce our 
$SU(3)\times SU(2)\times U(1)$ standard model of particle physics at low energies
and differ only in their predictions for $\rhol$. 
It has been argued that $\nsm\ll\ntot$, \ie, that only a tiny fraction of 
the minima reproduce our observed low-energy physics \cite{Ovrut04}.
Our results show that the key for testable predictions is whether $\nsm\gg \xi^4 Q^3\sim 10^{123}$ or not.
If $\nsm\gg 10^{123}$, then there may be so many minima able to produce galaxies  
that $\rhol$ will for all practical purposes  have a continuous probability distribution 
such as those plotted in this paper, \ie, the ``discretuum'' \cite{Bousso00,KKLT03,Susskind03} of $\rhol$-values may be 
indistinguishable from a continuum.
If $\nsm\simlt 10^{123}$, on the other hand, then there may be only one or a handful of 
minima producing galaxies, raising the exciting possibility that a precision measurement 
of $\rhol$ would allow us to determine precisely which minimum we are in, predict the remaining
decimal places of $\rhol$ and bring closure to our quest for a fundamental theory of physics.

\subsubsection{The smoothness problem}
%
% Q:
%

It is striking that even though many people start foaming at the mouth when encountering anthropic 
$\rhol$-arguments, feeling that they render theories untestable,
we have found that requiring inflation models to predict a $\rhol$-value consistent with observation 
is in fact a very strong test that rules out broad classes of models. 
The key reason for this is that, as emphasized in \cite{Q,Graesser04}, 
requiring nonlinear structures to form strongly constrains the quantity $\rhol/Q^3$ whereas 
the selection effects acting on the fluctuation amplitude $Q$ alone 
(with $\rhol/Q^3$ held fixed) appear to be substantially weaker.
The suppression of galaxy formation from increasing $\rhol$ can therefore be largely offset by
increasing $Q$, and we found that the inflation models we explored 
typically overpredicted $\rhol$ by a couple of orders of magnitude or more.
%This illustrates the hazard of performing anthropic calculations without an underlying theory.
This is directly linked to the smoothness problem, where the same coupling between $\rhol$ and $Q$ multiplies
the $Q$-distribution by $Q^4$ and overpredicts the CMB fluctuation level for most inflation models. 
In other words, the standard anthropic explanation of the observed $\rhol$-value does not appear 
to work well unless we can solve the smoothness problem.

\subsection{Implications for spatial curvature}

Since almost all reference objects are in pockets of infinite volume,
all measures except arguably unphysical ones (that do not weight by reference object and hence do not 
favor infinite volumes over finite ones)
predict that we should observe $\Otot\approx 1\pm 10^{-5}$. 
A second implication is that the thermalized pocket that we inhabit is
spatially infinite, \ie, that our Hubble volume is but one in an infinite 
``Level I multiverse'' \cite{multiverse} with distant Doppleg\"angers
and other Byzantine implications \cite{multiverse,GarrigaVilenkin01}.
The reader finding these implications disturbing can take heart in the fact that these inflation models
remain quite falsifiable, with Planck $+$ SDSS forecast to shrink the currently
allowed range $\Otot=1.01\pm 0.02$ \cite{Spergel03,sdsspars} by a 
factor of four to $\Delta\Otot\approx 0.005$ \cite{parameters2}
and therefore potentially excluding this inflationary prediction.

\subsection{Implications for quantum gravity}

If attractor dynamics makes the cosmological parameter predictions $f_p(\p)$ independent of pre-inflationary initial conditions,
then inflation provides a cosmic censorship that erases all quantitative clues about preceding events
at higher energy scales. 
We have argued that many measures providing such attractor behavior are observationally ruled out by the coolness problem,
which suggests that perhaps pre-inflationary conditions {\it do} matter.
If so, drawing conclusions about these conditions will not be easy, requiring both 
finding $V(\vphi)$ and solving the ordering problem,
but this nonetheless offers a glimmer of hope that
precision measurements of cosmological parameters 
may ultimately teach us something about quantum gravity.

\subsection{Outlook}

In conclusion, we have studied the important but difficult problem of how 
to compute testable cosmological parameter predictions from complicated inflaton potentials. 
Although we have obtained a number of results summarized above, 
% this paper should be viewed as only a humble beginning 
much work remains on multiple fronts --- here are a few examples:
\begin{enumerate}
\item
Although many of our results apply also to the multidimensional case $d>1$,
%zzz (\Sec{MultidimSec}), 
others do not, and a detailed study study of $d>1$ would be of great interest --- particularly 
$d=2$ and very large dimensionalities like $d\simgt 10^2$.
\item
A better understanding of ``upward'' quantum tunneling during inflation should settle the issue of if/when initial conditions matter.
\item 
As concrete potentials $V(\phi)$ emerge from string theory or other fundamental approaches,
it will be of great interest to repeat our calculations of $f_p(\p)$ for these physically motivated potentials.
\item 
Being the Achilles heel of the entire endeavor, the ordering problem is in grave need of further 
work. The urgency of this goes beyond cosmology, afflicting {\it any} ensemble theory with distinguishable 
infinite volumes.\footnote{
For instance, much recent work in the string theory landscape context has involved counting minima and quantifying their 
statistical properties. However, it is far from clear 
with what statistical weights these minima should be counted --- equally 
(as done in most papers), by basin of attraction area or by some other measure related to, \eg, tunneling or inflation.
Solving the measure problem may thus be required for string theory to qualify as a testable physical theory.}
\end{enumerate}
This will be challenging work. However, the avalanche of high-precision cosmological measurements is continuing unabated,
so if the goal is to find an inflation model that predicts rather than postdicts the correct answer, 
then the time to rise to this challenge is now.

\bigskip

%{\bf Acknowledgements:}

Thanks to Andy Albrecht, Anthony Aguirre, Edmund Bertschinger, Ted Bunn, Dick Bond, Alan Guth, Shamit Kachru, 
Andrei Linde, Burt Ovrut, Martin Rees, Paul Steinhardt, Wati Taylor, Neil Turok,
Alex Vilenkin and Matias Zaldarriaga for interesting discussions  and helpful comments.
Thanks to Jean Levitties for providing the park bench 
upon which most of this manuscript was written.
This work was supported by NASA grant NAG5-11099,
NSF CAREER grant AST-0134999, and fellowships from the David and Lucile
Packard Foundation and the Research Corporation.  

\vskip4cm

%\clearpage
\appendix

\section{Computing the observables from $\rho(a)$}

In this appendix, we review how the cosmological observables are derived from the cosmic density
history $\rho(a)$ and derive the equations of \Sec{Vsec}. This is mainly review material
\cite{LindeBook,DodelKinneyKolb97,LythRiotto99,LiddleLythBook,Peiris03,Kinney03,LiddleSmith03,Wands03}, 
presented here to 
clarify when and how most observables can be derived from $\rho(a)$ alone without reference to $V(\phi)$. 

Given the cosmic density history $\rho(a)$, the expansion history $a(t)$ is readily obtained by integrating 
the Friedman equation
\beq{FriedmanEq}
H^2 = {8\pi\over 3\mpl^2}\rho = {1\over 3\mbar^2}\rho,
\eeq
where the Hubble parameter $H\equiv d\ln a/dt=\dot a/a$.
%and we have defined the 
%{\it reduced planck mass}
%\beq{mbarDefEq}
%\mbar\equiv{\mpl\over\sqrt{8\pi}}\approx {\mpl\over 5}.
%\eeq
Luminosity distances, angular diameter distances, and all other classical cosmological observables
similarly follow from $\rho(a)$, typically via integrals involving this function. %$\rho(a)$.
We will now summarize the calculation of the observables $\rhol(a)$, $\deltt(k)$, $\deltt(k)$ 
and the eight cosmological parameters from Table 1.

\subsection{Horizons and e-foldings}
\label{Nsec}

Since the horizon radius is $H^{-1}$, the
comoving horizon size $a/H^{-1}=\dot a\propto 1/a^2\rho$
is constant on the dotted lines of slope $-2$ in \fig{arhoFig}, increasing towards the upper right.
Comoving wavelengths leave the horizon whenever the comoving horizon size increases, 
\ie, whenever the curve $\rho(a)$ is shallower
than these dotted lines (when $d\ln\rho/d\ln a>-2$, \ie, when $\ddot a>0$ so that the expansion is accelerating), and
enter the horizon when the slope is steeper, $d\ln\rho/d\ln a<-2$.
When two points lie on the same dotted diagonal, it therefore means that that the 
horizon volume at the two epochs is the same comoving spatial region.
For instance, the comoving spatial volume corresponding to our observable universe now at the present epoch
(filled triangle in \fig{arhoFig}) was the horizon volume during the inflationary epoch corresponding to
the open triangle, and the perturbations that left the horizon then are those that we see entering our 
horizon now.

Key epochs in \fig{arhoFig} are 
$\astart$ (when inflation began; $\astart=0$ if there was no beginning)
$\aexit$ (open triangle, when our current Hubble volume left the horizon),
$\ak$ (when a fluctuation of wavenumber $k$ leaves the horizon; open square or circle, say),
$\aend$ (star, when inflation ended),
$\areheat$ (cross, when reheating ended),
$\aeq$ (filled square, matter-radiation equality) and
$\anow$ (filled triangle, now).
The number of e-foldings before the end of inflation is
\beq{NdefEq}
N(a)\equiv\ln\left({\aend\over a}\right).
\eeq
Of particular importance is $\Nexit\equiv N(\aexit)$, often referred to as just $N$,
which is simply $\ln 10$ times horizontal distance between the open circle and the star
in \fig{arhoFig}.
The figure illustrates that $\Nexit$ 
depends not only on the behavior of $\rho(a)$ during inflation, but also
on when reheating ends.\footnote{As discussed in \cite{LiddleLeach03}, 
if $\vphi$ oscillates around a generic
(parabola-shaped) minimum in the potential during reheating, 
the logarithmic slope $d\ln\rho/d\ln a=-3$ just like during matter domination, 
so ending reheating at higher density increases $\Nexit$.
If the second derivative of the potential vanishes in the oscillation 
direction(s), however, as for $V(\phi)\propto\phi^4$, then $d\ln\rho/d\ln a=-4$
like during radiation domination and $\Nexit$ becomes independent of 
when reheating ends.}
The strong bounds on this quantity derived in \cite{LiddleLeach03,HuiDodel03}
can be readily read off geometrically from \fig{arhoFig}.
We must have $\rho(\aexit)\simlt 10^{-12}\mpl$ to avoid overpredicting
gravitational waves and density fluctuations (as detailed below) and presumably need 
$\rho(\aend)\simgt 1\GeV^4$ for reheating to be able to produce protons ({\cf} \cite{DimopoulosHall87,Hannestad0403291}).
% It seems you can go a fair bit below that, saved by the exponential tail of the Boltzmann distribution, 
% since you only need a tiny baryon-to-photon ratio.
% cf.: An abbreviation meaning Òcompare.Ó It is short for the Latin word confer and instructs the reader to compare one thing with another.
This gives the extreme bounds are $35\simlt\Nexit\simlt 85$, whereas most
popular inflation models favor the narrower range $\Nexit\sim 55\pm 5$ \cite{LiddleLeach03,HuiDodel03}.

\subsection{Spatial curvature}
\label{CurvatureSec}

The quantity $|\Otot-1|^{-1/2}$ is simply the ratio of the curvature radius ($\propto a$) to the
horizon radius $H^{-1}$. The spatial curvature parameter $\Otot$ therefore evolves as 
\beq{OtotEq}
\Otot-1\propto {1\over (aH)^2} = {1\over{\dot a}^2} \propto {1\over a^2\rho},
\eeq
\ie, $\Otot$ stays constant on the dotted lines of slope $-2$ in \fig{arhoFig} and increases towards
the upper right. 
This means that $\Otot$ is identical on horizon exit and subsequent horizon entry, so
that our currently observed $\Otot$ equals $\Otot(\aexit)$.
We will make the standard assumption that, if inflation had a beginning, 
things were a mess at that time with fluctuations of order unity on the horizon scale, \ie, 
$\Otot(\astart)\sim 1$. The inflationary prediction is therefore
\beq{OtotEq2}
|\Otot-1|\sim  {(aH)^2_{\rm exit}\over (aH)^2_{\rm start}} =  \left({\rhoexit\over\rhostart}\right)^2 e^{-2\Nbefore},
\eeq
where $\Nbefore=\ln(\astart/\aexit)$ is the number of e-foldings of inflation before our comoving Hubble 
volume exited the horizon (open triangle in \fig{arhoFig}).
In typical slow-roll inflation models like the one plotted in \fig{arhoFig}, 
the $\rho(a)$-curve is nearly horizontal and so 
\eq{OtotEq} is totally dominated by the second term, giving simply
$|\Otot-1|\sim e^{-2\Nbefore}$.
To match the current observational constraint $|\Otot-1|<0.02$ \cite{Spergel03,sdsspars}
therefore requires $\Nbefore\simgt 2.3$.
If the true curvature were exactly zero, density fluctuations on the horizon scale 
would propagate into our measurement and give $|\Otot-1|\sim\dH\sim 10^{-5}$, corresponding to
$\Nbefore \approx 5.8$, so the prediction for the {\it observed} curvature is
\beq{OtotEq3}
|\Otot-1|_{\rm observed}\sim  10^{-5} + e^{-2\Nbefore},
\eeq
and $\Nbefore$-values in the range $6\simlt\Nbefore<\infty$ are observationally indistinguishable.

\subsection{Dark energy}

\begin{figure} 
%\vskip\smtopskip
\centerline{\epsfxsize=\figsize\epsffile{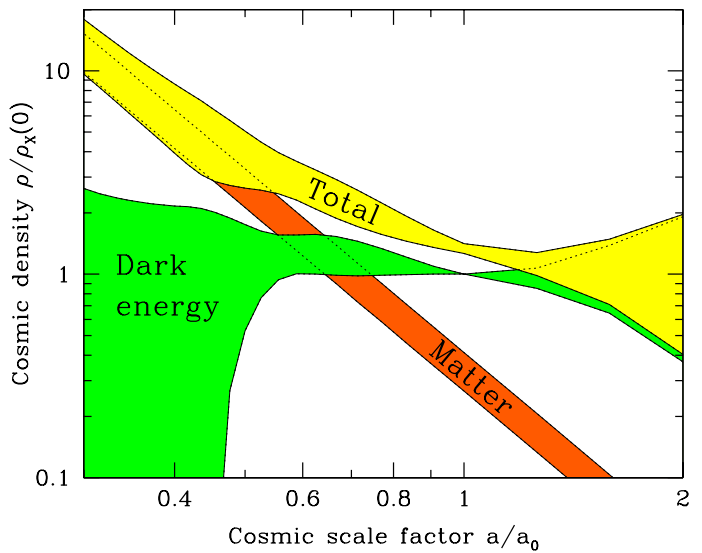}}
%\vskip\smbotskip
\caption[1]{\label{rholsausageFig}\footnotesize%
Zoom of the lower right corner of \fig{arhoFig}: 
$1\sigma$ constraints on the densities of matter and dark energy 
computed from 
SN Ia \cite{Riess04}, CMB and galaxy clustering data
as described in \cite{supernovae}.
}
\end{figure}

The dark energy density is, in practice, defined as the part of the
observed density that we do not understand.
The density contributions that we do understand at least at some level are those from
photons, matter and curvature, $\rhog$, $\rhom$ and $\rhok$:
\beqa{rhogEq}
\rhog(a)&=&{\pi^2\over 15} T^4 = \rhog(\anow)\left({\anow\over a}\right)^4,\\
\label{rhomEq}
\rhom(a)&=& {2\zeta(3)\over\pi^2}(\xib+\xic+\xin) T^3=\rhom(\anow)\left({\anow\over a}\right)^3,\\
%\rhom(a)&=& (\xib+\xic+\xin)n_\gamma = {2\zeta(3)\over\pi^2}(\xib+\xic+\xin) T^3,\\
\label{rhokEq}
\rhok(a)&=&3\mbar^2 H^2(1-\Otot) = \rhok(\anow)\left({\anow\over a}\right)^2,
% = {3\mbar^2 H_0^2(1-\Otot^0)\anow^2\over a^2}
% = {3\mbar^2 k\over a^2},
\eeqa
where $T={a\over a_0}T_0$ is the CMB temperature and $\zeta(3)\approx 1.202$.
%and $k=-1$, $0$ or $1$ for $\Otot>1$, $\Otot=1$ and $\Otot<1$, respectively.
The dark energy density is thus the residual
\beq{rholDefEq}
\rhol(a)\equiv\rho(a)-\rhogmk(a),
\eeq
where $\rhogmk=\rhog+\rhom+\rhok$.
Since all parameters entering into $\rhogmk(a)$ are now fairly well measured 
($\Otot$, $\xib$, $\xic$, $\xin$ as in Table 1, $H_0$, $T_0$) 
\cite{Spergel03,sdsspars,sdsslyaf},
\eq{rholDefEq} determines $\rhol(a)$ directly from the behavior of $\rho(a)$
around the present epoch.
If $\vphi$ settles into a stable potential minimum at $\vphi_0$ long before the present epoch,
then we observe $\rhol=V(\vphi_0)$ independently of time, so $w=-1$.

Dark energy is often discussed in the literature as something separate from inflation, so 
our merging of the two in this paper requires an explanation.

First of all, note that if General Relativity has a true cosmological constant $\rholgr$ hard-wired in,
quite separate from the inflaton-only potential $\tilde V(\vphi)$, then we will only obtain the correct 
predictions from inflation if we absorb this cosmological constant into the potential and insert 
$V(\vphi)\equiv\tilde V(\vphi)+\rhol$ into the equations below. Inserting $\tilde V(\vphi)$ would give
the wrong answer, since the Hubble parameter plays a key roll during inflation and
depends on the {\it total} density via \eq{FriedmanEq}, including $\rholgr$.

Second, note that if there is some other mechanism generating a ``dark energy'' density contribution 
without having any connection to the physics of inflation,
it must be absorbed into $V(\vphi)$ for the same reason.
Moreover, for scalar-field models of dark energy, at least ones with standard kinetic terms, 
it is mathematically straightforward to perform 
this inclusion, either by modifying the dynamics of the inflaton $\vphi$ at very low energies or by including 
the dark energy fields as additional components in the vector $\vphi$.

Third, generic potentials can have many minima, typically with $V(\vphi)\ne 0$,
so they can readily produce nonzero dark energy regardless of whether there is also a non-inflationary source
of dark energy. If inflation is related to the string landscape, say, 
then inflation-based dark energy may not be optional: if there are say $10^{300}$
minima \cite{KKLT03,Susskind03,AshikDouglas04,KKLMMT03,Banks04,WKV04,Denef04,GK04,Freivogel04,Conlon04,DeWolfe04,Nima05,Acharya05}
and $\gg\xi^{-4}Q^{-3}\sim 10^{123}$ of them reproduce our low-energy physics, then it is likely 
that many minima have $\rhol$-values in the range $|\rhol|\simlt\xi^4 Q^3$ 
permitting galaxy formation.

\subsection{Gravitational waves}

During inflation, quantum fluctuations in any massless field $\psi$ will produce effectively classical fluctuations with
power spectrum $P_\psi(k)=(H/2\pi)^2$ at the epoch $a_k$ when the scale $k$ exits the horizon \cite{LiddleLythBook}.
The modes of the spacetime metric corresponding to gravitational waves are examples of such massless fields,
and therefore produce current gravitational wave fluctuations \cite{LiddleLythBook}
\beq{dHtEq}
\deltt(k) = \sqrt{8\rho(a_k)\over 75\pi^2\mbar^4}.
\eeq
Note that this has nothing to do with fluctuations in the inflaton field, depending on 
the cosmic density history $\rho(a)$ alone.

The exit scale $a_k$ is computed by 
solving the equation 
$a_k H(a_k)=a_0 k$ for $a_k$,
where $H=(\rho/3)^{1/2}/\mbar$ as per \eq{FriedmanEq}.
In other words, $a_k$ is determined by following the dotted diagonal in \fig{arhoFig} 
from the point $(a,\rho)=(\anow k/H_0,\rho(\anow))$ up to the left until it
intersects the inflation curve. The current horizon wavenumber $k=H_0$ thus gives 
$a_k=\aexit$ (open triangle) and larger $k$ give larger $a_k$ (open square and circle).
During the observable part ($N\sim 55$) of slow-roll inflation, 
$H$ rarely drops by more than about a percent per e-folding,
so for cosmologically observable scales the approximation $H(a_k)\approx H(\aexit)$ is quite accurate,
giving simply $a_k=\anow k/H(\aexit)$ and thus 
\beq{dHtEq2}
\deltt(k)\approx \sqrt{8\rho\left({\aexit\over Hexit}k\right)\over 75\pi^2\mbar^4}.
\eeq
This has a simple geometric interpretation in terms of \fig{arhoFig}.
Since $\deltt\propto\rho^{1/2}$ and $k\propto a_k$, 
the curve $\deltt(k)$ in a log-log plot is simply the $\rho(a)$ curve from \fig{arhoFig}
compressed by a factor of two vertically. In other words, $\deltt(k)$ 
and $\rho(a)$ are the exact same curve, merely with different axis labeling.

This means that tensor spectral index
\beq{ntEq}
\nt\equiv{d\ln{\deltt}^2\over d\ln k}\approx
{d\ln{\deltt}^2\over d\ln a}={d\ln\rho\over d\ln a}
\eeq
is simply the slope of the $\rho(a)$-curve in \fig{arhoFig},
evaluated at $\aexit$ (open triangle).

\subsection{Density fluctuations}

None of the three above-mentioned observables $\Ok$, $\rhol(a)$ and 
$\deltt(k)$ depended explicitly on the physics of inflation, being computable directly from
the curve $\rho(a)$. We will now see that this is in many cases true even for 
the fourth observable,
the density fluctuations $\delt(k)$.

\subsubsection{Computing $\rho(a)$ and $\delt(k)$ from $V(\vphi)$}

The cosmic density is given by \cite{LiddleLythBook}
\beq{rhoEq}
\rho = V(\vphi) + {1\over 2}|\dot\vphi|^2 + \rhogmk% \approx V(\vphi),
\eeq
and the evolution of the inflation vector is governed by \cite{LiddleLythBook}
\beq{nDphiEq}
\ddot\vphi + 3H\dot\vphi+\nabla V(\vphi) = 0.
\eeq
(For a review of inflation generalizations with non-standard 
kinetic terms, see \cite{LiddleLythBook}.)
We ignore the $\rhogmk$-term in \eq{rhoEq} 
during inflation since it rapidly becomes negligible,
having a steeper slope ($\le -2$) than that of $\rho$ ($> -2$) in \fig{arhoFig}.
%where $\rhogmk$ is the density contribution from radiation ($\propto a^{-4}$), 
%matter ($\propto a^{-3}$),
%spatial curvature ($\propto a^{-2}$) and perhaps other sources.

The slow-roll approximation (SRA) is that the $\dot\vphi$-term in \eq{rhoEq}
and the $\ddot\vphi$-term in \eq{nDphiEq} are negligible, which gives
\beq{SRA_rhoEq}
\rho = V,
\eeq
\beq{nD_SRA_phiEq}
\dot\vphi=-{\nabla\vphi\over 3H}.
\eeq
Using
\eq{FriedmanEq} and the identity 
$\partial/\partial t=H\partial/\partial\ln a = -H\partial/\partial N$, 
this becomes
\beq{nD_SRA_phiprimeEq}
{\partial\vphi\over\partial N}\approx\mbar^2\nabla\ln V.
\eeq
In other words, the slow-rolling inflaton vector evolves just like a particle 
moving in an extremely viscous medium, so that the velocity is proportional to the force, 
and it always rolls in the direction of the gradient, \ie, takes the steepest descent.
The only differences are that the time variable is $-N$ rather than $t$ and 
that the potential is $\ln V$ rather than $V$. 
For the 1-dimensional case, \eq{nD_SRA_phiprimeEq} is usually integrated as
\beq{NintegratedEq}
N\equiv\ln {a\over\aend}={1\over\mbar^2}\int_\phi^\phiend {V(\phi)\over V'(\phi)} d\phi.
\eeq

If valid, the SRA remains valid as long as the slow-roll parameters $\epsilon\ll 1$ and $|\eta|\ll 1$\footnote{For 
the multidimensional case, $|\eta|\ll 1$ is strictly
speaking not a sufficient condition.
The second derivative $\ddot\vphi$ can get large either because the inflaton starts 
rolling faster (large $|\eta|$ so that $|\nabla\ln V|$ grows) or because it changes direction
(a large eigenvalue of the Hessian matrix $(\ln V),_{ij}$ corresponding to an eigenvector misaligned with the gradient).
Generically, such eigenvectors will have a component in the gradient direction as well and cause a large
$|\eta|$, so it requires some fine-tuning to fool the $|\eta|$ and $\epsilon$ tests.
}, 
where
\beq{epsilonetaEq}
\epsilon\equiv{\mbar^2\over 2}|\nabla\ln V|^2,\quad\eta\equiv\mbar^2{V''\over V}.
\eeq
%\beq{epsilonEq}
%\epsilon\equiv{\mbar^2\over 2}|\nabla\ln V|^2,
%\eeq
%\beq{etaEq}
%\eta\equiv\mbar^2{V''\over V}.
%\eeq
For the multidimensional case, $V''$ denotes the second derivative in the direction in which the field is rolling,
\ie, in the direction of the gradient $\nabla\phi$.
For the 1-dimensional case, the third slow-roll parameter %zzz(\eq{xi2Eq2}) 
is defined as 
\beq{xi2Eq}
\xi_2\equiv\mbar^4 {V'(\phi)V'''(\phi)\over V(\phi)^2}  % = \mbar^2\left(y''+y'^2\right).
\eeq
For the $d$-dimensional case (where the vector $\vphi$ has $d$ components),
the power spectrum of density fluctuations $\delt(k)$ is given by 
\cite{Starobinsky85,Salopek95,SasakiStewart96,LythRiotto99,LiddleLythBook}
\beq{nDdHeq}
\delt(k)^2 = {\rho\over 75\pi^2\mbar^2}\sum_{i=1}^d\left({\partial N\over\partial\phi_i}\right)^2,
\eeq
where the right hand side is evaluated at the exit epoch $a_k$ as usual, and 
$N(\vphi)$ denotes the number of e-foldings for the field to roll from $\vphi$ down to some
fixed reference density according to \eq{nD_SRA_phiprimeEq}.
Heuristically, \eq{nDdHeq} is easy to understand:
quantum fluctuations in $\vphi$ are of order $\delta\phi\sim H\sim\rho^{1/2}/\mbar$,
and if such a fluctuation in some particular spatial region 
delays the roll down to our current epoch by an amount 
$\delta N\sim {\partial N\over\partial\phi}\delta\phi$, this will appear an an overdensity
of order $\delta N$. Fluctuations from the different components $\phi_i$ of the inflaton vector $\vphi$ 
add in quadrature to give the total variance $\delt(k)^2$.
% $\vphi$ itself isn't observable today.
% Merely changing $\vphi$ in the roll-direction is equivalent to translating the time variable.
Choosing our coordinates in $\vphi$-space so that the field is momentarily rolling in the $\phi_1$-direction,
we have $\partial N/\partial\phi_1=[\mbar^2\partial\ln V/\partial\phi_1]^{-1}=(2\mbar^2\epsilon)^{-1/2}$, so
\beq{nDdHeq2}
\delt(k)^2 = {\rho\over 75\pi^2\mbar^4}\left[{1\over 2\epsilon} + {1\over\mbar^2}\sum_{2=1}^d\left({\partial N\over\partial\phi_i}\right)^2\right]
\ge {\rho\over 150\pi^2\mbar^4\epsilon},
\eeq
In most studied cases, the walls of the multidimensional gorge in which the inflaton slowly rolls are much steeper than the 
roll direction, rendering the inequality of \eq{nDdHeq2} 
close to an equality \cite{Starobinsky85,Salopek95,SasakiStewart96,LythRiotto99,LiddleLythBook}.

\subsubsection{Computing $\delt(k)$ from $\rho(a)$}

Let us now eliminate $V(\vphi)$ and rewrite the above quantities in terms of the $\rho(a)$ alone
for the one-dimensional case.
%\beq{lnrDefEq}
%\lnr(N)\equiv\ln\rho\left(\aend e^{-N}\right), 
%% $N=-\ln(a/\aend)$ as above.
% a = aend*exp(-N)}\right)
%\eeq
%\ie, $\lnr(N)=\ln\rho(a)$ where $N=-\ln(a/\aend)$ as above.
% a = aend*exp(-N)
Letting primes $'$ denote $\partial/\partial N=-\partial/\partial\ln a$, we obtain
\beq{dNeq}
%\dN=-{1\over H}\dt=-\mbar^2 \nabla_\vphi\ln V \cdot \nabla_\vphi
\eeq

\beq{epsilonEq2b}
\epsilon = {1\over 2}(\ln\rho)'
\eeq

\beq{etaEq2b}
\eta = (\ln\rho)' + {(\ln\rho)''\over 2(\ln\rho)'} 
\eeq

\beq{dHeq2}
\delt(k) = \sqrt{\rho\over 75\pi^2\mbar^4(\ln\rho)'},
\eeq
again evaluated at the exit epoch $a_k$. This gives

\beq{ntEq2}
\nt={\partial\ln{\deltt}^2\over\partial k} = (\ln\rho)' = 2\epsilon
\eeq

\beq{nsEq}
\ns-1={\partial\ln\delt^2\over\partial k} = -[\ln\rho-\ln(\ln\rho)']' 
= {(\ln\rho)''\over(\ln\rho)'} - (\ln\rho)'
\eeq

\beq{alEq}
\al={\partial\ns\over\partial\ln k} 
%= -\left[{(\ln\rho)''\over(\ln\rho)'} - (\ln\rho)'\right]' 
= (\ln\rho)'+{(\ln\rho)''^2\over(\ln\rho)'^2}-{(\ln\rho)'''\over(\ln\rho)'}
%{\partial^2\ln\delt^2\over\partial k^2} 
\eeq
In deriving these and related relations, the following identities are useful
(for brevity, we here --- and only here --- we use the shorthand notation ${\dot{\>}}={d\over d\phi}$ and
and $r=\ln\rho=\ln V$):
\beqa{DiffeventialIdentityEq}
%\dN&=&-{\mpl^2\ovev 8\pi}y'\dphi,\\
%\dN^2&=&\left({\mpl^2\ovev 8\pi}\right)^2\left(y'^2\dphi^2+y'y''\dphi\right),\\
%\dN^3&=&\left({\mpl^2\ovev 8\pi}\right)^3\left(y'^3\dphi^3+3y'^2y''\dphi^2+(y'y''^2+y'^2y''')\dphi\right),\\
\dN&=&-\mbar^2\dot r\dphi,\\
\dN^2&=&\mbar^4\left({\dot r}^2\dphi^2+\dot r \ddot r\dphi\right),\\
\dN^3&=&-\mbar^6\left[{\dot r}^3\dphi^3+3{\dot r}^2\ddot r\dphi^2+(\dot r {\ddot r}^2+{\dot r}^2\dddot r)\dphi\right],
\eeqa
{\etc}
The derivatives of a curve $(\ln a,\ln\rho)=(-N,r)$ therefore satisfy
\beqa{dNeq2}
r'&=&-\mbar^2{\dot r}^2,\\
r''&=&2\mbar^4 {\dot r}^2{\ddot r},\\
r'''&=&-\mbar^6{\dot r}^2\left(2{\ddot r}^2+{\dot r}{\dddot r}\right)\\
r''''&=&-2\mbar^8 {\dot r}^2\left(4{\ddot r}^3+7{\dot r}{\ddot r}{\dddot r}+{\dot r}^2{\ddddot r}\right),\\
%\eeqa
%{\etc}
%\beqa{rderivEq}
\dot r&=&{\sqrt{-r'}\over\mbar}\\
\ddot r&=&-{1\over 2\mbar^2}{r''\over r'}\\
\dddot r&=&-{1\over \mbar^3(-{r'}^{3/2})}{r'''+{r''}^2\over 2r'}
\eeqa

% \xi_2\equiv\mbar^4 {V'(\phi)V'''(\phi)\over V(\phi)^2} 

\section{Inflation near peaks and troughs}

Since the cosmological parameter predictions for inflation with 
$\mh\gg\mbar$ or $\mh\ll\mbar$ depend only on the behavior of the inflaton potential 
near its minima or maxima, respectively, let us review the analytic solutions
for these two important cases and compute the predicted 
cosmological parameter probability distributions.

\subsection{Inflation near a minimum}

Taylor expanding $V(\phi)$ to second order around a minimum $\phi_0$ gives
\beq{quadratic_Veq}
V(\phi) = V(\phi_0) + {1\over 2}V''(\phi_0)(\phi-\phi_0)^2,
\eeq
where, generically, $V''(\phi)>0$.
Since we saw in \Sec{AnthroSec} that only those minima where $|\rhol| = |V(\phi_0)|$ was negligibly small 
contributed significantly to the parameter probability distribution, we need only consider the simple case
$V(\phi_0)=0$, \ie, the standard parabolic potential case \cite{LindeDiffusion86}.
Using equations\eqn{epsilonetaEq}, this gives 
\beq{quadratic_epsilonetaEq}
\epsilon=\eta={2\mbar^2\over (\phi-\phi_0)^2},
\eeq
so inflation ends with $\epsilon=\eta=1$ when $\phi$ has rolled down to 
\beq{pl_phiendEq}
\phiend=\phi_0\pm\sqrt{2}\mbar.
\eeq
\Eq{NintegratedEq} now gives \eq{quadratic_rhoEq}, which in turn 
%zzz(via equations~\ref{parSummaryEq}-\ref{xi2Eq2})
(via equations~\ref{parSummaryEq}-\ref{etaEq2})
gives
%\ref{\ref{xi2Eq}
%% epsilon, eta & xi in terms of N(phi)
%\ref{SRA_nsEq}-\ref{alEq} % give pars in terms of epsilon, eta & xi2 
the well-known results %(taking $\phi\le\phiend<\phi_0$)
\beqa{pl_phiNepsetaxiEq}
%N&=&\ln {a_2\over a_1}={(\phi-\phi_0)^2-(\phiend-\phi_0)^2\over 4\mbar^2},\\
N&=&{(\phi-\phi_0)^2\over 4\mbar^2}-{1\over 2},\\
\phi&=&\phi_0\pm\mbar\sqrt{2(2N+1)}\\
\epsilon&=&\eta={1\over 2N+1},\\
\xi_2&=&0,
\eeqa 
and the cosmological predictions of 
equations\eqn{quadratic_rhoEq}-(\ref{quadratic_ntEq}).

\subsection{Inflation near a maximum}

Taylor expanding $V(\phi)$ to second order around a maximum $\phi_0$ gives
\beq{hill_Veq}
V(\phi) = V(\phi_0)\left[1+{\eta_0\over 2}\left({\phi-\phi_0\over\mbar}\right)^2\right],
\eeq
where $\eta_0$ is the second slow-roll parameter of \eq{epsilonetaEq} evaluated at the maximum.
Generically, $\eta_0<0$.
Using \eq{epsilonetaEq}, this gives 
\beqa{hill_epsilonEq}
\epsilon&=&{\eta_0^2(\phi-\phi_0)^2\over 2\mbar^2}\left[1+{\eta_0\over 2}\left({\phi-\phi_0\over\mbar}\right)^2\right]^{-2},\\
\eta&=&\eta_0\left[1+{\eta_0\over 2}\left({\phi-\phi_0\over\mbar}\right)^2\right]^{-2}.
\eeqa
In the early stages of $\phi$ rolling off the peak, while the term in square brackets remains near unity, 
\eq{NintegratedEq} gives
%\beq{hill_N_Eq}
%N = \eta_0\ln{\phi\over\phiend},
%\eeq
%\ie,
\beq{hill_N_Eq}
(\phi-\phi_0) \propto e^{\eta_0 N},
\eeq
so at the observable epoch $N\sim 55$, the field is exponentially
close to the peak unless $\eta_0$ happens to lie in the small interval
$-0.02\simlt\eta_0<0$ (slow-roll requires $-1<\eta_0<0$).
According to the 2nd order expansion of \eq{hill_Veq}, 
inflation ends only when $|\phi-\phi_0|\simgt\mbar$. 
However, the scaling relations given in
\Sec{MonteSec} show that inflation ends much sooner because
of higher order terms in the Taylor expansion, usually because 
the cubic term causes inflation to end when $|\eta|=1$ with 
$|\phiend-\phi_0|\sim \mh^3/\mbar^2$, which is borne out
by the $\mh\ll\mbar$ simulations in \Sec{MonteSec}.
For the observable regime $N\sim 55$, we thus have
\beqa{hill_phi_Eq}
|\phi-\phi_0| &\sim&{\mh^3\over\mbar^2} e^{\eta_0 N},\\
{V'(\phi)\over V(\phi)}&\sim&{\mh^3\eta_0\over\mbar^4} e^{\eta_0 N}\approx 0,\\
%V'(phi) ~ (V(phi0)eta0/mbar^2) dphi
%        ~ (V(phi0)eta0/mbar^2) (mh^3 exp[]/mbar^2)
%        ~ (V(phi0)/mbar^4) (mh^3 eta0 exp[])
\epsilon&\sim&{\eta_0^2\mh^6\over\mbar^6} e^{2\eta_0 N}\approx 0,\\
\eta&\approx&\eta_0,\\
\xi_2&=&\mbar^4 {V'V'''\over V^2} \sim \eta_0 e^{\eta_0 N} \approx 0,\label{Hillxi2Eq}
% \xi_2\equiv\mbar^4 {V'(\phi)V'''(\phi)\over V(\phi)^2}  
% \xi_2\equiv\mbar^4 (V'/V) (V'''/V)
%        ~ mbar^4 e^{eta0 N} (mh^3 eta0/mbar^4) mh^(-3)
%        ~  eta0 e^{eta0 N} 
% xi2 ~ (mbar/mh)^4 >> 1
%\rho(a)	&\approx&V(\phi_0),\label{hill_rhoeq}\\
%%Q	&=&{Q_t\over\sqrt{\epsilon}}\sim V(\phi_0)^{1/2}{\mbar^3\over\mh^3\eta_0} e^{-\eta_0 N},\label{hill_Qeq}\\
%Q	&=&{Q_t\over 4\sqrt{\epsilon}}\sim {\mv^2\mbar\over\mh^3\eta_0} e^{-\eta_0 N},\label{hill_Qeq}\\
%Q_t	&\approx&\sqrt{8V(\phi_0)\over 75\pi^2\mbar^4}\sim{\mh^2\over\mbar^2} ,\label{hill_QtEq}\\
%\ns	&\approx&1+2\eta_0,\label{hill_nsEq}\\
%\al	&\approx&-2\xi_2\approx 0,\label{hill_alEq}\\
%r	&=&16\epsilon\approx 0,\label{hill_rEq}\\
%\nt	&=&-2\epsilon\approx 0,\label{hill_ntEq}.
\eeqa
and the cosmological predictions of 
equations\eqn{hill_rhoEq}-(\ref{hill_ntEq}).
In \eq{Hillxi2Eq}, we used the scaling relation $V'''/V\sim\mh^3$ which follows from \eq{VunitEq}. 
% Cut, since I get triangular rather than rectangular V''-distribution:
%For the $\mh\ll\mbar$ case, the $\eta_0$-value at a typical peak is of order 
%$(\mbar/\mh)^2\gg 1$, so only a small fraction $\sim(\mh/\mbar)^2$ of all peaks satisfy the SRA requirement
%$\eta\ge-1$ and lead to inflation. 
%Since the intrinsic width of the
%probability distribution for $\eta_0$ at peaks is thus much wider than
%the region $-1\le\eta_0<0$ that contributes to our parameter probability
%distribution, we can make the approximation that $\eta_0$ has a uniform
%distribution on the interval $[-1,0)$, which enables us to calculate the probability 
%distributions for many of the above parameters analytically in the $\mh\ll\mbar$ limit.
%Typically $-\eta_0$ is of order unity, causing $\epsilon$, $r$ and $\nt$ to be exponentially small
%and $Q$ to be exponentially large (for a given vertical energy scale $\mh$).
%This makes $\ns$ uniformly distributed on the interval $[0,2)$ and gives 
%$\alpha$ a scatter of order $(\mbar/\mh)^4$.
%% \xi_2\equiv\mbar^4 {V'(\phi)V'''(\phi)\over V(\phi)^2}  % = \mbar^2\left(y''+y'^2\right)
%% xi2 ~ (mbar/mh)^4 >> 1
%For about $N^{-1}\sim 2\%$ of the cases, we have 
%$-N^{-1}\le\eta_0<0$, giving 
%$\epsilon$ only polynomially small: 
%$\epsilon\sim 10^{-4}(\mh/\mbar)^6.$
%% Explain that I don't see this $\ns$-distribution numerically because of underflow.

%\clearpage
%%%%%%%%%%%%%%%%%%%%%%%%%%%%%%%%%%%%%%%%%%%%%%%%%%%%%%%%%%%%
%%%%%%%%%%%%%%%%%%%%%% REFERENCES: %%%%%%%%%%%%%%%%%%%%%%%%%
%%%%%%%%%%%%%%%%%%%%%%%%%%%%%%%%%%%%%%%%%%%%%%%%%%%%%%%%%%%%

\end{document}